\documentclass{ws-rv9x6}
\usepackage{subfigure}   % required only when side-by-side / subfigures are used
\usepackage{ws-rv-thm}   % comment this line when `amsthm / theorem / ntheorem` package is used
\usepackage{ws-rv-van}   % numbered citation & references (default)
\makeindex
\usepackage[dvipsnames]{xcolor}
\usepackage[colorlinks=true, pdfstartview=FitV, linkcolor=RedOrange, citecolor=Emerald, urlcolor=Cerulean]{hyperref}

\newcommand{\trento}{{\rm T}$\mathrel{\protect\raisebox{-2.1pt}{{\rm R}}}${\rm ENTo}}

\begin{document}

\chapter[Small Systems]{Progress and Challenges in Small Systems\label{ra_ch1}}

\author[J. Noronha, B. Schenke, C. Shen, and W. Zhao]{
Jorge Noronha$^{1}$\footnote{jn0508@illinois.edu},
Bj\"orn Schenke$^{2}$\footnote{bschenke@bnl.gov},
Chun Shen$^{3,4}$\footnote{chunshen@wayne.edu},
and
Wenbin Zhao$^{3,5,6}$\footnote{WenbinZhao@lbl.gov}
}

\address{
$^1$Illinois Center for Advanced Studies of the Universe \& Department of Physics,
University of Illinois Urbana-Champaign, Urbana, IL 61801, USA \\
$^2$Physics Department, Brookhaven National Laboratory, Upton, NY 11973, USA \\
$^3$Department of Physics and Astronomy, Wayne State University, Detroit, Michigan 48201, USA\\
$^4$RIKEN BNL Research Center, Brookhaven National Laboratory, Upton, NY 11973, USA\\
$^5$Nuclear Science Division, Lawrence Berkeley National Laboratory, Berkeley, California 94720, USA\\
$^6$Physics Department, University of California, Berkeley, California 94720, USA\\
}

\begin{abstract}

We present a comprehensive review of the theoretical and experimental progress in the investigation of novel high-temperature quantum chromodynamics phenomena in small systems at both the Relativistic Heavy Ion Collider and the Large Hadron Collider. We highlight the challenges and opportunities associated with studying small systems, by which we generally mean collision systems that involve at least one light ion or even a photon projectile. We discuss perspectives on possible future research directions to better understand the underlying physics at work in the collisions of small systems.

\end{abstract}

\body

\tableofcontents

\section{Introduction}

For over two decades, ultrarelativistic heavy-ion collisions have produced and systematically investigated the properties of the quark-gluon plasma (QGP), an exotic phase of quantum chromodynamics (QCD) in which quarks and gluons are not confined inside protons and neutrons and that existed microseconds after the Big Bang. The current theoretical modeling \cite{Gale:2013da,Romatschke:2017ejr,Shen:2020mgh} of these specks of early universe matter produced in heavy-ion collisions is broadly consistent with the statement that the quark-gluon plasma can flow \cite{Heinz:2013th} like a nearly frictionless, strongly coupled quantum liquid over distance scales not much larger than the proton radius. This makes the quark-gluon plasma formed in colliders the hottest, smallest, densest, most perfect fluid known to humanity. 

Remarkable progress has been achieved over the years regarding the theoretical description of heavy-ion collisions. A general framework consisting of initial state modeling, followed by a pre-equilibrium phase and the subsequent hydrodynamic evolution that is ultimately merged into hadronic transport, forms the core of the so-called standard model of heavy-ion collisions~\cite{Shen:2014vra, Putschke:2019yrg, Nijs:2020roc}. This theoretical framework evolved from original qualitative success to impressive predictive power \cite{Shen:2011eg, Heinz:2013bua, Niemi:2015voa,Noronha-Hostler:2015uye} and quantitative accuracy, currently exemplified by the use of Bayesian inference techniques \cite{Paquet:2023rfd}.

However, there are still many fundamental questions and potential pitfalls in our understanding of heavy-ion collisions and the quark-gluon plasma, driven by our general inability to perform ab initio real-time calculations in quantum chromodynamics beyond weak coupling. In that context, the investigation of the so-called small systems, such as proton-proton ($p$+$p$) and proton-nucleus ($p$+A) collisions, has been particularly illuminating as it requires pushing the boundaries of our understanding of the bulk collective properties of the quark-gluon plasma. 

Traditionally, relativistic hydrodynamics has been primarily employed in studying large collision systems, such as those involving two heavy ions, e.g.~gold or lead. These systems exhibit an abundance of produced particles, a long-lived quark-gluon plasma phase, and definite signatures of collective behavior compatible with hydrodynamics. In the last decade, advances in experimental capabilities have allowed us to explore experimental signatures of collectivity in small collision systems. Those are characterized by more significant pressure gradients and a shorter quark-gluon plasma phase compared to that found in heavy-ion collisions. Nevertheless, despite their size, small systems exhibit intriguing collective behavior, challenging the conventional wisdom that collective effects are exclusive to large collision systems\footnote{We note that some uncertainties appear already at the experimental level, as the reduced number of particles produced in small systems introduces new difficulties in experimental flow measurements \cite{ATLAS:2017rtr, KrizkovaGajdosova:2020pxc}, which do not occur in large systems.}.  
However, one expects the uncertainties in theoretical modeling to increase significantly as one moves down in system size (e.g., from Pb+Pb to $p$+$p$ collisions). 

For example, mirroring what is done in large systems, one may employ relativistic hydrodynamics to investigate the collective properties of the quark-gluon plasma encoded in the anisotropic flow of the final-state particle distributions of small systems. Applying relativistic hydrodynamics to small collision systems, however, presents several challenges. The reduced number of particles produced in the events, and the shorter time for hydrodynamic evolution, necessitate careful consideration of the initial conditions and the freeze-out process. The large spatial gradients in the initial state can drive the system significantly far from equilibrium, which can be challenging to describe using fluid dynamics without violating fundamental physical principles (such as relativistic causality).
Due to their reduced system size, in small systems, there is \emph{a priori} no clear separation between the scales associated with the microscopic physics of QCD processes and the typical scales related to gradients of conserved quantities. In this case, standard arguments relying on the Knudsen number and inverse Reynolds number expansions, widely used in the derivation of fluid dynamical models \cite{Denicol:2012cn, Rocha:2023ilf}, are much less justifiable than in large heavy-ion collision systems (such as central Pb+Pb, or Au+Au, collisions).  
Furthermore, stochastic thermal fluctuations must become more relevant as the system decreases in size \cite{Landau_book}, which introduces an inherent lack of determinism that has not yet been included in state-of-the-art simulations of small systems. 

On the other hand, applying relativistic hydrodynamics to small collision systems represents a unique opportunity for progress in high-energy nuclear physics~\cite{Nagle:2018nvi, Schenke:2021mxx}. By studying flow observables and their hydrodynamic interpretation in such extreme systems, one may uncover novel properties of the quark-gluon plasma near and far from equilibrium, which can shed light on the fundamental nature of strongly interacting matter. Continued theoretical and experimental efforts in this field will deepen our understanding of the quark-gluon plasma and its behavior across collision sizes.

In this chapter, we provide a concise overview of some recent developments and challenges concerning the theoretical modeling of small systems. We intend to be more illustrative than comprehensive, focusing on some theoretical results and experimental measurements, which we believe give a reasonably good idea of the progress achieved through the last decade and the remaining fundamental questions. In this regard, we admit that the presentation is not entirely balanced, with some results receiving more attention than others. 

This paper is structured as follows. In Section \ref{sec:model}, we introduce the models employed to investigate the small quark-gluon plasma droplets formed in small collision systems. This includes a discussion of the initial conditions, the pre-equilibrium dynamics, and the hydrodynamic models. Section \ref{sec:reviewdata} presents the model-to-data comparison, encompassing various observables. Section \ref{sec:opportunities} outlines additional theoretical advances that offer further insights into the physics of small systems. Our final comments are presented in Section \ref{sec:conclusion}.
\emph{Notation}: We use a mostly minus metric signature $(+---)$ %this is horrible :)
and natural units, $\hbar=c=k_B=1$.

\section{Multi-stage modeling for small system dynamics}
\label{sec:model}
\subsection{Initial state models}\label{sec:initial}

Hydrodynamic simulations rely on models for the initial conditions. In principle, these should provide the entire energy-momentum tensor and conserved currents over the entire space at the initial time. However, many models provide only the spatial energy density distribution, which for heavy ion collisions with a long lifetime may be a good approximation but, for small systems, is likely insufficient, as initial velocity and off-equilibrium corrections can significantly affect the produced particle distributions. 

Several studies have employed Monte Carlo Glauber type models \cite{Miller:2007ri,dEnterria:2020dwq} with nucleon degrees of freedom without substructure to initialize hydrodynamic simulations of small systems \cite{Bozek:2011if,Bozek:2012gr,Bozek:2013df,Bozek:2013uha,Bozek:2013uha,Bozek:2013ska,Bzdak:2013zma,Qin:2013bha,Werner:2013ipa,Kozlov:2014fqa,Romatschke:2015gxa,Shen:2016zpp,Weller:2017tsr,Giacalone:2017uqx,Sievert:2019zjr,Summerfield:2021oex,Katz:2019qwv}. These calculations generally reproduce azimuthal anisotropies measured in experiments at Relativistic Heavy Ion Collider (RHIC) and the Large Hadron Collider (LHC). 

The success of these simulations depends on the choice of energy or entropy deposition, which can be proportional to the sum of the thickness functions~\footnote{The thickness function is the integral of the nuclear density function over the direction of the beamline.} of the projectile and target or other combinations, such as the product of thickness functions. Only the former option yields good agreement with experimental data for nucleon-based initial state models. For example, in $p$+A collisions, the proton projectile is perfectly round when assuming no substructure, and fluctuations can only originate from the fluctuating positions of nucleons in the target. In this case, only the sum of thickness functions produces enough fluctuations to explain the data~\cite{Shen:2014vra,Moreland:2014oya}, while the product leads to overly smooth energy density distributions~\cite{Schenke:2014zha}.

Nevertheless, there are strong arguments for why the sum should be disfavored. First, all Bayesian analyses using the \trento{} model \cite{Moreland:2014oya}, which parametrizes the dependence of the initial entropy or energy density distribution on the thickness functions, prefer an entropy density proportional to the square root of the product of the two thickness functions \cite{Bernhard:2016tnd,Giacalone:2017uqx,JETSCAPE:2020shq,JETSCAPE:2020mzn,Nijs:2020roc}, rather than the sum, which is also included in \trento{}'s parameter space. 
Further, the effective theories and models that allow for calculating the energy deposition without assuming the dependence on the thickness function all find proportionality to (some power of) their product rather than the sum.
This includes the weak coupling calculations based on the color glass condensate (CGC) effective theory \cite{McLerran:1993ni,McLerran:1993ka,Kovner:1995ja,Iancu:2003xm,Lappi:2006hq} and 
strong coupling holographic calculations \cite{Romatschke:2013re,Romatschke:2017ejr}. 

By adhering to an energy or entropy deposition description that utilizes the product of thickness functions or a power thereof, an accurate representation of anisotropic flow data in small systems can only be achieved by incorporating subnucleon fluctuations. This has been explicitly demonstrated in the CGC-based IP-Glasma \cite{Schenke:2014zha} and \trento{} \cite{Moreland:2018gsh} models.

The fluctuating spatial structure of a nucleon has not been computed from first principles. To make progress, one can parametrize the substructure as a collection of hot spots with certain sizes and then constrain the parameters through Bayesian analyses of heavy-ion and small system collisions.\cite{Moreland:2018gsh,Nijs:2020roc} Although the precise number of hot spots could not be well constrained, a finite number of hot spots was favored over a smooth nucleon. The width of the subnucleonic hot spots was constrained to approximately $0.4\,\rm{fm}$, similar to the typical width used for the \emph{nucleon} width in the IP-Glasma model.

Another approach involves utilizing input from electron-proton ($e$+$p$) scattering experiments. It has been demonstrated that the cross-sections for incoherent diffractive vector meson production are sensitive to the proton substructure and its fluctuations \cite{Mantysaari:2016ykx, Mantysaari:2016jaz,Mantysaari:2022ffw,Mantysaari:2023qsq}. By constraining the parameters of a hot spot model using data from HERA \cite{Chekanov:2002rm,Aktas:2003zi,Aktas:2005xu,Chekanov:2002xi,Alexa:2013xxa}, and applying this model in proton-nucleus ($p$+A) collisions, successful reproduction of flow observables has been achieved \cite{Mantysaari:2017cni,Schenke:2019pmk,Schenke:2020mbo}.
The most sophisticated extraction of nucleon substructure parameters from HERA data used Bayesian techniques \cite{Mantysaari:2022ffw}. While some parameters could be well constrained, the number of hot spots, for example, was not well determined. A global analysis using both $e$+$p$ and hadronic collisions could potentially improve this situation.

As mentioned above, the initial state description should provide more than the spatial distributions of energy density. For example, the spatial distribution of the entire initial energy-momentum tensor can be computed within the IP-Glasma model with the caveat that the system under consideration is purely gluonic. In the classical approximation, the coupling is fixed, and the system is conformally invariant. This means that the equation of state from this calculation differs from the QCD equation of state \cite{Borsanyi:2013bia,HotQCD:2014kol}. Furthermore, conformal invariance implies that this system's bulk viscosity transport coefficient vanishes. Such differences introduce challenges when matching to hydrodynamics\cite{NunesdaSilva:2020bfs,daSilva:2022xwu}, for which one usually employs an equation of state extracted from lattice QCD. For more on the early time dynamics that transition the initial state to hydrodynamics, we refer the reader to the next section. 

The IP-Glasma initial energy-momentum tensor includes (shear) viscous corrections and an initial flow profile. Interestingly, the initial transverse flow velocity distribution is already anisotropic, a consequence of correlations among the color fields in the colliding nuclei \cite{Schenke:2015aqa,Lappi:2015vta}. Assuming boost invariance of these correlations and performing a (2+1)D (IP-Glasma-like) Yang-Mills evolution, such correlations were shown to survive even after hydrodynamic evolution for low multiplicity events \cite{Giacalone:2020byk}. However, performing small $x$-evolution to obtain a rapidity-dependent initial state shows that the correlations leading to such initial state momentum anisotropies decay rapidly with the rapidity difference \cite{Schenke:2022mjv}. One may also expect that full (3+1)D Yang-Mills simulations, which include rapid growth of small fluctuations \cite{Berges:2014yta,Ipp:2020igo,McDonald:2023qwc} and lead to at least partial pressure isotropization, will further reduce the initial state anisotropies. 

The extension to full 3D initial conditions has usually been done by introducing a parametric longitudinal envelope folded with the transverse energy density profile~\cite{Hirano:2005xf, Bozek:2010vz, Bozek:2013uha, Ke:2016jrd, Barej:2019xef, Wu:2021fjf, Du:2022yok, Jiang:2023vxp}. In this family of initial state models, one can further impose energy-momentum conservation when mapping the initial-state density profiles to hydrodynamic fields~\cite{Shen:2020jwv, Soeder:2023vdn}. Different implementations result in different amounts of longitudinal fluctuations in the initial energy density profiles, which can be constrained by the longitudinal decorrelation measurements. 

In Section \ref{sec:long} we will discuss in more detail the need for simulations that explicitly consider all three spatial dimensions, i.e., that do not assume boost invariance. That means that the initial state model needs to provide the energy-momentum tensor as a function of both transverse space and space-time rapidity. Several models that achieve this have been employed in the description of small systems.\cite{Lin:2004en, Bozek:2014cya, Barej:2019xef, Shen:2022oyg, Zhao:2022ugy}.

\subsection{The need for pre-equilibrium dynamics models}\label{sec:pre_eq}

Immediately after the collisions between the nuclei, the matter is expected to find itself in a gluon-dominated phase \cite{McLerran:1993ka}. In principle, this idea can be used to determine the spatial dependence of all the components of the system's energy-momentum tensor \cite{Schenke:2014zha} and other conserved currents \cite{Carzon:2019qja,Shen:2022oyg,Du:2023msv,Pihan:2023dsb} (in this section, for simplicity, we focus on the energy-momentum tensor). However, the system is usually still in a far-from-equilibrium state at such early times, which implies that matching the initial state model directly to hydrodynamics cannot be done without drastic, unjustified assumptions. This can be better understood as follows.

Assume that some initial state model, say IP-Glasma, gives an initial $T^{\mu\nu}$ at some early time $\tau_0$, which we intend to use to define the initial conditions for hydrodynamic evolution directly. One can then use Landau matching \cite{Landau_book} to define a flow 4-velocity $u^\mu$ and energy density $\varepsilon$ at $\tau_0$ as an eigenvector and its corresponding eigenvalue of $T^{\mu\nu}$, i.e., $T^{\mu\nu}u_\nu = \varepsilon u^\mu$. We note that one usually assumes that this procedure gives a flow velocity that obeys $u_\mu u^\mu = 1$ and $\varepsilon$ is non-negative. We note that these properties are not guaranteed to hold in general\cite{Arnold:2014jva,Rougemont:2021qyk,Rougemont:2021gjm}. Putting such issues aside, one can unambiguously decompose the initial energy-momentum tensor as
\begin{equation}
    T^{\mu\nu} = \varepsilon u^\mu u^\nu - [ p(\varepsilon) + \Pi ]\Delta^{\mu\nu} + \pi^{\mu\nu} \,, 
    \label{eqn:LandauDecomp}    
\end{equation} 
where $p(\varepsilon)$ is the thermodynamic pressure related to the energy density through the equation of state, $\Pi$ is the bulk pressure, $\Delta^{\mu\nu}=g^{\mu\nu}-u^\mu u^\nu$, and $\pi^{\mu\nu}$ is the shear stress tensor defined by $\pi^{\mu\nu} = \Delta^{\mu\nu}_{\alpha\beta}T^{\alpha\beta}$, where $\Delta^{\mu\nu}_{\alpha\beta} = \frac{1}{2}\left( \Delta^\mu_\alpha \Delta^\nu_\beta + \Delta^\mu_\beta \Delta^\nu_\alpha \right) - \frac{1}{3}\Delta^{\mu\nu}\Delta_{\alpha\beta}$. 

If hydrodynamics is a good approximation for the initial $T^{\mu\nu}$, then one would expect that the deviations from equilibrium, quantified by the inverse Reynolds numbers $\sim |\Pi|/p, \sqrt{|\pi^{\mu\nu}\pi_{\mu\nu}|}/p$, are locally \emph{small} at the initial time $\tau_0$. It turns out that this is generally not the case even in large heavy-ion collision systems \cite{Niemi:2014wta,Noronha-Hostler:2015coa,Schenke:2019pmk,Inghirami:2022afu}. Therefore, one generally expects that, at sufficiently early times, direct matching of the energy-momentum tensor from initial state models to hydrodynamics is a very poor approximation. This should be even more pressing in small systems, because of their reduced size compared to large heavy-ion collisions.  

In recent years, pre-equilibrium models have been used to bridge the gap between the initial state energy-momentum tensor and the energy-momentum tensor at the start of the hydrodynamics phase. Such models are used to describe the evolution of the system from $\tau_0$ up to $\tau_\text{hydro}$, the time at which a hydrodynamic description is valid. %so that $\tau_0 < \tau < \tau_\text{hydro}$.

The simplest model one can consider is free streaming. In this case, one assumes that on-shell non-interacting massless partons emerge isotropically from an initial hard scattering at the initial time $\tau_0$ \cite{Broniowski:2008qk, Liu:2015nwa}. Assuming longitudinal boost invariance, the energy-momentum tensor of such a system can be obtained by integrating the initial number density of partons in the transverse plane, effectively smoothing out the energy density of the system. This free streaming dynamics is suddenly interrupted at $\tau_\text{fs} > \tau_0$, where the system is suddenly assumed to attain local thermal equilibrium. Clearly, this sudden transition from a zero to finite (strong) coupling is unphysical. Therefore, free streaming pre-equilibrium dynamics is necessarily a drastic assumption that, though practical, is theoretically unjustified.

Progress toward a more realistic scenario was explored in \cite{Kurkela:2018vqr, Kurkela:2018wud} using the K{\o}MP{\o}ST framework. There, the system's evolution from $\tau_0$ to $\tau_\text{hydro}$ is described by an effective kinetic theory model of weakly coupled QCD \cite{Arnold:2002zm}. In this approach, the system's dynamics at early times after the collision are assumed to be determined by classical Yang-Mills equations. The system expands, its energy density drops, and at a time $\tau_\text{EKT}$ the subsequent evolution of the system is modeled by effective kinetic processes that drive the plasma towards a state where inverse Reynolds numbers have sufficiently decreased such that relativistic viscous fluid dynamics models may become a good approximation.

While this represents significant progress, current approximations in such pre-equilibrium models can still be improved. For example, one can take the trace of Eq.~\eqref{eqn:LandauDecomp} to find that the total isotropic pressure is $p(\varepsilon) + \Pi = \frac{ \varepsilon - T^\mu_\mu }{3}$. One can see that for conformal pre-equilibrium models, such as the original K{\o}MP{\o}ST framework where $T^\mu_\mu = 0$, a simple calculation gives $p(\varepsilon) + \Pi= \frac{\varepsilon}{3}$. This implies that an artificial bulk viscous pressure component must be added at the beginning of the hydrodynamic evolution because the conformal pressure $\epsilon/3$ differs from the QCD pressure. Uncertainties associated with this issue have been investigated~\cite{Schenke:2019pmk,NunesdaSilva:2020bfs,daSilva:2022xwu,Liyanage:2022nua, Jaiswal:2021uvv}.
Secondly, the dilute cold corona of the collision requires a non-perturbative description of QCD confinement and parton-hadron interactions~\cite{Kanakubo:2019ogh, Werner:2023jps}. 

Pre-equilibrium models are needed in self-consistent hydrodynamic simulations of large and small collision systems. Their inclusion in the standard modeling of heavy-ion collisions emphasizes the need to investigate how the far-from-equilibrium initial QCD fields decohere towards a state amenable to hydrodynamic description. In fact, investigations into the physics of the pre-equilibrium phase must consider the inherent non-Abelian physics of QCD in out-of-equilibrium conditions, which is bound to lead to several new theoretical developments. For example, the pre-equilibrium phase should be crucial to understanding the onset of hydrodynamics and the presence of hydrodynamic attractors \cite{Heller:2015dha,Strickland:2017kux,Giacalone:2019ldn}, which have been proposed as a way to extend the domain of applicability of hydrodynamics to the far-from-equilibrium regime \cite{Jankowski:2023fdz}. 

\subsection{Relativistic hydrodynamics}\label{sec:hydro}

Fluid dynamic behavior emerges in vastly different systems in nature. Applications can be found across many scales, ranging from cosmology to the fluid dynamic phenomena observed in everyday life. The ubiquitousness of hydrodynamics stems from the presence of conservation laws and the assumption of a significant separation between different characteristic length scales in the system. In general, this hierarchy is quantified by the Knudsen number $\mathrm{Kn} \sim \ell/L$, which broadly denotes the ratio between the relevant microscopic scale $\ell$ and a macroscopic scale $L$ associated with the variations of conserved quantities (such as energy, momentum, and baryon number). Fluid dynamics is expected to be a good approximation to describe the long-time, large wavelength physics of many-body systems in situations where $\mathrm{Kn} \ll 1$ \cite{Landau_book}.

Relativistic hydrodynamics plays a crucial role in understanding the behavior of matter and energy under extreme conditions \cite{Romatschke:2017ejr}. The equations of motion of relativistic hydrodynamics encode the conservation of energy, momentum, and conserved charges:
\begin{equation}
    \partial_\mu T^{\mu \nu} = 0, \qquad \partial_\mu J^\mu = 0.
\end{equation}
Here, $T^{\mu\nu}$ is the system's energy-momentum tensor, and the $J^\mu$ represents the 4-current of conserved charges (such as baryon number in the case of heavy-ion collisions). In its simplest incarnation, known as ideal fluid dynamics, dissipative processes are neglected, and these equations are closed using the matter's equation of state, e.g., $p(\varepsilon, n_B)$, where $n_B$ is the baryon density~\cite{Monnai:2021kgu}.

In many aspects, one may argue that the quark-gluon plasma produced in ultrarelativistic collisions lies at the edge of the applicability of fluid dynamics. In small systems, this is even more pressing since ``macroscopic scales" are, at the most, comparable to the proton radius (of the order of $1$ fm), and the relevant microscopic scales can be estimated in terms of the inverse temperature $\ell \sim 1/T$. This exercise shows that for the temperature range expected in such systems, $T \sim 150 - 450$ MeV, the Knudsen number $\mathrm{Kn}$ is not necessarily small \cite{Niemi:2014wta, Noronha-Hostler:2015coa, Shen:2016zpp, Schenke:2019pmk, Shen:2020mgh}. This has important consequences for the hydrodynamic modeling of the quark-gluon plasma, as discussed below.

In high-energy nuclear collisions, relativistic hydrodynamics has proven to be an indispensable tool for investigating the intricate dynamics of the quark-gluon plasma. Because the underlying QCD system is relativistic, and dissipative processes may not be neglected (as $\mathrm{Kn}$ is not generally small), a \emph{causal} formulation of relativistic \emph{viscous} hydrodynamics is essential for its phenomenological applications in heavy-ion collisions. Given that the standard relativistic generalization of Navier-Stokes equations, derived by Eckart \cite{EckartViscous} and Landau and Lifshitz \cite{Landau_book}, lead to acausal evolution \cite{PichonViscous} where small disturbances around the equilibrium state grow exponentially \cite{Hiscock_Lindblom_instability_1985}, the development of other frameworks where dissipation is not fundamentally at odds with relativistic causality became necessary. It is now understood that the issues found in Eckart, and Landau and Lifshitz's theories follow from very general arguments, as proved recently\cite{Bemfica:2020zjp,Gavassino:2021owo}. Basically, the main insight is that causality is necessary for the stability of the equilibrium state in relativistic fluids. The previous formulations from Eckart, and Landau and Lifshitz give rise to non-hyperbolic partial differential equations (PDE)s \cite{PichonViscous} that necessarily violate relativistic causality, creating unphysical instabilities of the equilibrium state.   

The issues found in previous formulations can be fixed using the new approach put forward by Bemfica, Disconzi, Noronha, and Kovtun (BDNK) \cite{BemficaDisconziNoronha,Kovtun:2019hdm,Bemfica:2019knx,Bemfica:2020zjp,Hoult:2020eho}. Following standard effective theory reasoning \cite{Kovtun:2012rj}, in BDNK non-equilibrium corrections are taken into account in constitutive relations using the most general expressions compatible with the symmetries involving combinations of first-order spacetime derivatives of the standard hydrodynamic variables (which vanish for the equilibrium state). This procedure creates several new terms in the equations of motion, which were not considered before and are crucial to restoring causality and stability. Though at this point, this new formalism has not been applied to realistic quark-gluon plasma simulations, nontrivial numerical solutions have already been developed in other contexts\cite{Pandya:2021ief,Pandya:2022pif,Pandya:2022sff,Bantilan:2022ech,Rocha:2022ind}. 

Current hydrodynamic simulations of the quark-gluon plasma employ equations of motion that can lead to causal and stable evolution within the framework originally devised by Israel and Stewart \cite{MIS-6}. This approach differs from the BDNK reasoning mentioned above because in Israel-Stewart theories \cite{MIS-6,Baier:2007ix,Denicol:2012cn} the dissipative fluxes (such as the shear-stress tensor, $\pi_{\mu\nu}$, and the bulk scalar, $\Pi$) are not defined via constitutive relations, i.e., they are not expressed only using the hydrodynamic variables and their derivatives. Rather, the dissipative fluxes evolve according to additional equations of motion of relaxation type (derived either from entropic considerations \cite{MIS-6}, kinetic theory \cite{Denicol:2012cn}, effective theory arguments \cite{Baier:2007ix}, or anisotropic hydrodynamics \cite{Alqahtani:2017mhy}), which introduce a relaxation time that characterizes how the dissipative fluxes relax towards their Navier-Stokes values. The relaxation time effectively restores causality as, for example, flow gradients are not automatically converted into acceleration as in the relativistic Navier-Stokes equations. In-depth discussions of the different theories of relativistic viscous fluid dynamics can be found in the literature\cite{Rocha:2023ilf}.

All state-of-the-art hydrodynamic simulations of the quark-gluon plasma formed in heavy-ion collisions have been performed using Israel-Stewart-like approaches. The equations of motion are now solved by many different groups, using either Eulerian \cite{Luzum:2008cw, Schenke:2010nt, Schenke:2010rr, Bozek:2013uha, Shen:2016zpp, Weller:2017tsr, Mantysaari:2017cni, Schenke:2020mbo} or Lagrangian numerical algorithms \cite{Noronha-Hostler:2013gga,Noronha-Hostler:2014dqa,Noronha-Hostler:2015coa}, which pass standard tests of accuracy \cite{Marrochio:2013wla}. However, despite this significant progress in solving these equations of motion in the context of heavy-ion collisions, there are still uncertainties in the modeling coming from the hydrodynamic description. This is especially relevant in the case of simulations of small collision systems, as discussed below. 

\subsubsection{Large Knudsen numbers and inverse Reynold's numbers in small systems}

Relativistic hydrodynamics describes the macroscopic dynamics of many-body systems. The system's entropy is packed in a small volume in space for high multiplicity events of ultrarelativistic proton-nucleus and proton-proton collisions. Despite small particle multiplicity, the local energy density can be large enough such that, at least in principle, the system may behave as a fluid.

In small collision systems, the limited size and short lifetime of the created QGP pose unique challenges in the hydrodynamic description of the system's dynamics. One significant aspect relevant to the overall applicability of hydrodynamics in small systems is that, in such systems, Knudsen numbers and inverse Reynolds numbers can become large.

Let us first make a few general comments about the Knudsen number in small systems. The initial transverse volume in small systems is about 50 times smaller than in central Pb+Pb collisions. The reduction of the macroscopic size significantly increases the Knudsen number at early times. This indicates that the underlying collision rate is small, meaning that the collision time is larger than the characteristic timescale of the hydrodynamic expansion. As a result, the assumptions of local thermal equilibrium and continuous fluid behavior, crucial to our standard understanding of the domain of applicability of hydrodynamics, are progressively challenged.

The inverse Reynolds number ($\mathrm{Re}^{-1}$) serves as a complementary measure, characterizing the importance of viscous effects in the system. It describes the ratio of viscous forces to inertial forces, and it can be estimated, for example, as the ratio of the viscous pressure contributions to the fluid cell's local equilibrium enthalpy. In practice, one may define the inverse Reynolds number associated with shear stress as \cite{Denicol:2012cn,Rocha:2023ilf}
\begin{equation}
\mathrm{Re}^{-1}_\pi = \frac{\sqrt{\vert \pi_{\mu\nu}\pi^{\mu\nu}\vert}}{\varepsilon+p}\,.
\label{defineReynolds}
\end{equation}
Near equilibrium, this quantity becomes small, as viscous stresses are then a small contribution to the pressure/energy density. 

We note that while $\mathrm{Re}^{-1}$ and $\mathrm{Kn}$ are certainly related, in general they are not the same. To illustrate this point, consider the case of shear-viscous effects. In that case, one may estimate the shear Knudsen number as
\begin{equation}
\mathrm{Kn}_{\sigma} \sim \frac{\sqrt{\vert \sigma_{\mu\nu}\sigma^{\mu\nu} \vert}}{T},
\label{defineKnudsen}
\end{equation}
where the shear tensor is $\sigma^{\mu\nu} = \Delta^{\mu\nu\alpha\beta}\nabla_\alpha u_\beta$. 
The $1/T$ factor denotes the microscopic scale\footnote{For a dilute gas, one may use the mean free path as the microscopic scale $\ell \sim 1/n \sigma_0$, where $n$ is the density and $\sigma_0$ the cross-section. For a strongly coupled system, $\ell\sim 1/T$ is typically used in such estimates.}, while the other factor describes the scale associated with gradients of the flow velocity. Now one may compare Eq.~\eqref{defineReynolds} with Eq.~\eqref{defineKnudsen}. In the Navier-Stokes regime, where $\pi_{\mu\nu} = 2 \eta \sigma_{\mu\nu}$, these quantities are completely equivalent. However, in Israel-Stewart theories $\pi_{\mu\nu}$ evolves according to a differential equation  
\begin{equation}
\tau_\pi\Delta^{\mu\nu}_{\alpha\beta}u^\lambda \partial_\lambda \pi^{\alpha\beta}+\pi^{\mu\nu} = 2 \eta \sigma^{\mu\nu}+\ldots\,,
\end{equation}
where $\tau_\pi$ is the shear relaxation time and $\ldots$ denote other terms usually employed in simulations \cite{Rocha:2023ilf}. Thus, depending on the initial conditions, the state of the system, and the values of the transport coefficients, in general, $\pi^{\mu\nu}$ is not well approximated by $2\eta\sigma^{\mu\nu}$. In that \emph{transient} case, $\mathrm{Re}^{-1}$ and $\mathrm{Kn}$ should be treated as separate parameters that may be used in a series expansion \cite{Denicol:2012cn}. In small collision systems, the large pressure gradients lead to a fast expansion rate, which drives the system significantly out of equilibrium. Consequently, the viscous effects can dominate the dynamics, leading to a large inverse Reynolds number, which is not reflected by the corresponding $\mathrm{Kn}$.

The presence of large Knudsen and inverse Reynolds numbers in small systems challenges the applicability of conventional relativistic viscous hydrodynamic models that assume perturbative viscous corrections around local thermal equilibrium. To address these challenges, efforts have been made to develop hybrid approaches that combine relativistic hydrodynamics with effective kinetic theory frameworks (which were discussed in Sec.\ \ref{sec:pre_eq}). These hybrid models aim to capture the non-equilibrium dynamics and viscous corrections that become increasingly significant in small systems.

\subsubsection{Possible cavitation under rapid expansion}

The rapid expansion observed in small collision systems suggests that cavitation may occur and play an important role in the hydrodynamic description. Cavitation, commonly observed in fluid dynamics \cite{brennen_2013}, refers to the formation and subsequent collapse of vapor or gas-filled voids within a liquid medium. It arises when the local pressure drops below the vapor pressure, leading to the nucleation and growth of bubbles. 

In heavy-ion collisions, the hydrodynamic evolution is characterized by an intense and rapid expansion of the QGP, leading to a drop in pressure and temperature. Under certain conditions, this rapid expansion may give rise to a local total pressure that falls close to zero, potentially leading to cavitation within the QGP \cite{Torrieri:2007fb,Rajagopal:2009yw,Habich:2014tpa,Denicol:2015bpa,Byres:2019xld,Schenke:2019pmk}. The idea behind it is simple: the system's total pressure has equilibrium and non-equilibrium contributions, i.e., $p_{\mathrm{total}}=p+\Pi$. In Navier-Stokes theory $\Pi = -\zeta \partial_\mu u^\mu$ \cite{Rocha:2023ilf}, where $\zeta$ is the bulk viscosity. For an expanding system $\partial_\mu u^\mu>0$, hence $\Pi<0$, and, if the expansion is sufficiently large, there could be regions where the total pressure vanishes. In practice, such arguments are qualitative at best, given that in the transient hydrodynamic regime, $\Pi$ is not well approximated by Navier-Stokes values, and consequently, the bulk inverse Reynolds number does not equal the associated Knudsen number. In any case, if cavitation can occur in heavy-ion collisions, it is reasonable to expect that it is more likely to take place in small systems because of their large expansion rate/out-of-equilibrium corrections. It is interesting to note that having zero local pressure does not necessarily lead to fundamental problems. For example, causality can still be preserved even if $p+\Pi <0$ as long as $\varepsilon+p+\Pi>0$ \cite{Bemfica:2019cop}. 

Since the transition from the QGP phase to the hadron gas phase is a smooth cross-over, cavitation in small collision systems may or may not have significant consequences for interpreting experimental observables. Nevertheless, these low-pressure voids can introduce non-equilibrium effects and modify the system's evolution. The expansion dynamics of the QGP, which is typically described by viscous hydrodynamics, may be affected by the presence of these low-pressure voids.

The possibility of cavitation under rapid expansion in small systems presents an intriguing avenue for research in relativistic hydrodynamics. Understanding and quantifying the possible cavitation effects in small systems is challenging. The dynamics of bubble formation and collapse in the QGP require a detailed understanding of the equation of state, the thermodynamic properties of the medium, and the transport coefficients. Furthermore, the evolution of the cavitation bubbles must be considered in the context of the overall expansion and freeze-out processes. Incorporating these complexities into hydrodynamic models and analyzing their implications on experimental observables represent active research areas.

\subsubsection{Causality violations when simulating small systems}

Causality is, in a broad sense, the statement that no information in physical systems can be propagated faster than the speed of light. As mentioned before, incorporating dissipative effects in a way that is consistent with relativistic causality is still a matter of intense investigation \cite{Gavassino:2023mad,Wang:2023csj,Hoult:2023clg}. Causality is a necessary requirement for thermodynamic stability in relativistic many-body systems \cite{Gavassino:2021owo}. For small disturbances around equilibrium, it is possible to obtain conditions involving the equation of state and the transport coefficients of the system under which causality and stability can be established. This is the case in all the theories employed to investigate relativistic viscous fluids \cite{Rocha:2023ilf}, such as BDNK \cite{BemficaDisconziNoronha,Kovtun:2019hdm,Bemfica:2019knx,Bemfica:2020zjp,Hoult:2020eho} and Israel-Stewart theories \cite{Hiscock_Lindblom_stability_1983,Olson:1989ey,Denicol:2008ha,Pu:2009fj,Gavassino:2021cli,Gavassino:2021kjm}. Once nonlinear corrections are included, the equations of motion of hydrodynamics become a nonlinear set of PDEs, and establishing causality (or hyperbolicity) is a highly challenging task. In BDNK theories, where viscous corrections do not require the introduction of additional fields besides those in equilibrium, causality in the full nonlinear regime has been rigorously established \cite{BemficaDisconziNoronha,Kovtun:2019hdm,Bemfica:2019knx,Bemfica:2020zjp,Hoult:2020eho,Abboud:2023hos}. For Israel-Stewart theory, however, the analysis is much more complicated because the causality conditions now explicitly include the dissipative fluxes, in addition to transport coefficients and the equation of state. This implies that the constraints needed to ensure causality in the linear regime around equilibrium are not sufficiently powerful to guarantee causality in the nonlinear regime. In that case, only a few general results, valid for general equations of state, in 3+1 dimensions, etc, are known \cite{Bemfica:2019cop, Bemfica:2020xym}.    

When performing numerical simulations of heavy-ion collisions, it is crucial to ensure that the evolution preserves causality. Violating causality conditions can lead to significant consequences that impact the reliability and interpretation of the results \cite{Plumberg:2021bme,Chiu:2021muk}. This is particularly relevant for small collision systems \cite{Krupczak:2023jpa}. Causality violations can occur due to several factors. One common issue is the choice of relaxation times for the shear and bulk viscous tensors. Suppose they are too close to the causality bounds derived in the linear perturbative region. In that case, there is little room for the system to stay out of equilibrium, according to the causality conditions recently derived for the Israel-Stewart hydrodynamic theory in the full non-equilibrium regime \cite{Bemfica:2020xym}. In some cases, serious violations of causality conditions can generate numerically unstable modes that grow exponentially. However, in other cases, the signs of causality violations are hard to trace, and the numerical simulations remain stable. 

The violation of causality conditions in numerical simulations of small systems demands careful consideration, because it can undermine the reliability and accuracy of the obtained results. Unphysical artifacts introduced by causality violations can distort the interpretation of the system's dynamics and compromise the comparison with experimental data \cite{Plumberg:2021bme,Chiu:2021muk,Krupczak:2023jpa}. This hampers the extraction of meaningful insights about the underlying physics and, in particular, the transport properties of the QGP in small collision systems.
Moreover, violating causality conditions can hinder developing and improving theoretical models and numerical algorithms. Identifying and addressing the sources of unphysical behavior becomes challenging when causality violations are present. Resolving these issues requires careful investigations within the complex numerical simulations.

To mitigate the consequences of causality violations, ongoing research focuses on developing more robust theories and implementing numerical algorithms that explicitly account for the nonlinear causality conditions \cite{Bemfica:2019cop,Bemfica:2020xym} in small system simulations \cite{Plumberg:2021bme,Chiu:2021muk,Krupczak:2023jpa}. 

\subsubsection{Sensitivity to second-order transport coefficients}

In large collision systems, the effects of viscosity on the collective behavior of the QGP are well understood \cite{Gale:2013da, Romatschke:2017ejr}. They can be characterized by the first-order transport coefficients, namely the shear viscosity ($\eta$) and the bulk viscosity ($\zeta$). Recent studies have revealed that small collision systems exhibit a heightened sensitivity to both the first-order and second-order transport coefficients, such as the viscous relaxation times $\tau_{\pi}$ and $\tau_\Pi$ \cite{Schenke:2019pmk,Shen:2020gef,Chiu:2021muk}.

In large collision systems, the collective behavior is dominated by the long-lived QGP phase, reducing the sensitivity to the specific values of the second-order transport coefficients. However, small collision systems have a reduced QGP lifetime, leading to shorter hydrodynamic evolution timescales close to the viscous relaxation times.

The sensitivity of small collision systems to the transport coefficients manifests in various ways. First, the magnitude of the viscous effects becomes more pronounced in small systems, leading to enhanced sensitivity to the values of shear and bulk viscosity and their temperature dependence \cite{Schenke:2019pmk}.
Second, the hydrodynamic evolution in small systems is more susceptible to non-equilibrium and transient effects~\cite{Strickland:2014pga, Strickland:2018exs}. The shorter duration of the QGP phase reduces the time available for the system to relax towards local thermal equilibrium, amplifying the influence of dissipative processes. In this context, accurately determining the transport coefficients becomes crucial, as their values directly affect the system's ability to reach thermal equilibrium and influence the subsequent particle production and flow patterns.

The sensitivity of small collision systems to the second-order transport coefficients poses challenges to theoretical modeling and interpretation of experimental data. Precise knowledge of the transport coefficients is required to accurately describe the dynamics and extract meaningful information from experimental observables. This necessitates a deeper understanding of the underlying microscopic processes and improved constraints on the transport coefficients from experimental measurements.

To address these challenges, efforts are underway to systematically study the second-order transport coefficients and their temperature dependence within different theoretical frameworks \cite{Denicol:2014vaa,Rocha:2021zcw,Rocha:2022fqz,deBrito:2023tgb}. These studies involve comparisons between theoretical calculations, hydrodynamic simulations, and experimental data \cite{Nijs:2020roc, Nijs:2021clz, Chiu:2021muk, JETSCAPE:2020mzn}.

\begin{figure*}[htbp]
\begin{center}
\includegraphics[width=0.7\textwidth]{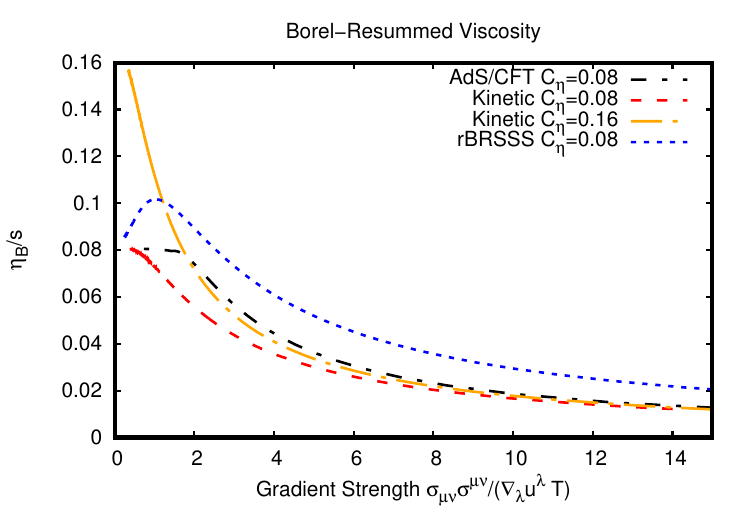}
\end{center}
\caption{Effective shear viscosity versus gradient strength
 for hydrodynamic attractors in different theories (rBRSSS, kinetic theory and AdS/CFT). For large gradient strength, $\frac{\eta_B}{s}\rightarrow 0$.
This figure is taken from published work. \cite{Romatschke:2017vte}.
}
\label{fig:effectiveetas}
\end{figure*} 

\subsubsection{Resummed transport coefficients and attractors}

Hydrodynamics is traditionally based on the assumption that the system is near local equilibrium. A trivial rearrangement of this statement in terms of the Knudsen number is that hydrodynamics is under control when $\mathrm{Kn} \ll 1$. 
However, there is growing evidence supporting the capability of hydrodynamics to describe small colliding systems that are not close to local equilibrium and in situations where $\mathrm{Kn}\sim 1$.
Recently, it has been argued that part of this success may be attributed to the existence of hydrodynamic attractors \cite{Heller:2015dha,Romatschke:2017vte,Strickland:2017kux}. According to this view \cite{Jankowski:2023fdz}, the solution of the dynamical equations converges quickly to this attractor through non-hydrodynamic mode decay.
Despite the fact that the nonlinear hydrodynamic gradient series can diverge \cite{Heller:2013fn,Buchel:2016cbj}, the series may be resummed to give rise to the hydrodynamic attractor, which may be approximated by second-order hydrodynamic solutions (at least, in very simplified cases). 

Resummation of the gradient series may also provide a way to define transport coefficients far from equilibrium~\cite{Romatschke:2017vte,Denicol:2020eij}. Figure \ref{fig:effectiveetas} shows this effective shear viscosity $\eta_B/s$ versus gradient strength~\cite{Romatschke:2017vte}. For small gradients (system near equilibrium), the effective viscosity reduces to the viscosity computed using linear response theory. For large gradient strength (system far from equilibrium), the effective viscosity $\eta_B/s$ tends to zero. Thus, in systems out-of-equilibrium, it is possible that higher-order viscous corrections effectively reduce the value of the viscosity entering the viscous hydrodynamic equations \cite{Romatschke:2017vte,Blaizot:2017ucy}. 

The idea that the far-from-equilibrium properties of the system may be taken into account in new resummed transport coefficients can be potentially relevant for phenomenology. 
For example, a unified hydrodynamic description of the quark-gluon plasma generated in ultrarelativistic $p$+Pb and Pb+Pb collisions at $\sqrt{s_{\rm NN}}=5.02$~TeV has been proposed.\cite{Moreland:2018gsh, Schenke:2020mbo} The authors employ a Bayesian parameter estimation model calibration framework to conduct a global analysis of bulk observables in both $p$+Pb and Pb+Pb collisions.
Nevertheless, it is crucial to acknowledge that, as previously discussed, smaller systems exhibit larger gradient strengths, leading to a comparatively smaller effective viscosity than that observed in heavy-ion collisions.
A potential avenue for future research involves a more comprehensive approach that integrates small and large systems. 
This framework could involve determining the viscosity in small systems in the context of gradient series resummation and resummed transport coefficients.  The anticipated outcome is a smaller effective viscosity and/or viscous effects in small systems than it would be naively expected from standard bare $\mathrm{Kn}$ arguments. This innovative approach holds promise for refining our understanding of the nuanced dynamics within different collision systems.

\subsubsection{Impacts from stochastic noise on the hydrodynamic description in small systems}

Dissipative systems near equilibrium must display thermal fluctuations -- this is the physics behind the celebrated fluctuation-dissipation theorem \cite{Callen:1951vq, Kubo:1957mj}. Therefore, stochastic noise arises from the inherent fluctuations present in the microscopic dissipative dynamics of the heavy-ion collision system. In small systems, the reduced size and short lifetime are expected to amplify the relative importance of these stochastic effects. The fluctuations effectively contribute as a stochastic source term to the hydrodynamic equations, converting the system to a nonlinear set of stochastic PDEs, which is very challenging to directly evolve numerically \cite{Young:2014pka,Singh:2018dpk}.

The impacts of stochastic noise on the hydrodynamic description in small systems are expected to be twofold. First, the fluctuations can introduce an additional source of event-by-event variations in the system's evolution (for the same initial state), resulting in broadened distributions for many observables. In particular, stochastic noise can lead to additional event-by-event fluctuations in the azimuthal anisotropy coefficients ($v_n$) and other flow-related observables, which are not currently considered in state-of-the-art simulations. 

Second, the need for the inclusion of stochastic noise in small systems provides another source of uncertainty in our current hydrodynamic models that assume smooth and deterministic evolution. The fluctuations can induce effective non-equilibrium behavior, which is particularly relevant in small systems, where the QGP lifetime is shorter and the hydrodynamic evolution timescale is comparable to or shorter than the characteristic relaxation timescales.

Understanding and quantifying the impacts of stochastic noise on the hydrodynamic description of relativistic systems is a very active area of research. Several approaches and applications have been pursued in recent years, which include work on hydro-kinetics \cite{Akamatsu:2016llw, Akamatsu:2017rdu, An:2019osr, Martinez:2018wia, De:2022tkb}, fluctuations near a critical point  \cite{Stephanov:2017ghc, Nahrgang:2018afz, Martinez:2019bsn, Rajagopal:2019xwg, An:2019csj, An:2020vri, Nahrgang:2020yxm, Dore:2020jye, Du:2020bxp, Pradeep:2022mkf, An:2022jgc}, general studies of fluctuations in the context of heavy-ion collisions \cite{Kapusta:2011gt, Young:2013fka, Sakai:2017rfi, Singh:2018dpk, De:2020yyx, Sakai:2020pjw, Kuroki:2023ebq}, and more \cite{Calzetta:1997aj, Kovtun:2003vj, Dunkel:2008ngc, Kovtun:2011np, Kovtun:2012rj, Kumar:2013twa, Young:2014pka, Murase:2016rhl, Kapusta:2014dja, Martinez:2017jjf, Murase:2019cwc, Calzetta:2020wzr, Torrieri:2020ezm, Dore:2021xqq, Petrosyan:2021lqi, Abbasi:2022rum, Chen:2022ryi}. Many of the theoretical developments have focused on the Schwinger-Keldysh action on the closed time path \cite{Crossley:2015evo, Sieberer:2015hba, Haehl:2016pec, Glorioso:2017fpd, Liu:2018kfw} where actions are constructed as an effective field theory from the underlying quantum mechanical system using the dynamical Kubo-Martin-Schwinger symmetry \cite{Kubo:1957mj, Martin:1959jp,Grozdanov:2013dba, Kovtun:2014hpa, Harder:2015nxa,Haehl:2015pja, Haehl:2015uoc, Jensen:2017kzi, Jensen:2018hse, Haehl:2018lcu, Jain:2020zhu,Jain:2023obu}. New works have explicitly considered the recent developments concerning causality and thermodynamic stability in relativistic fluids \cite{Mullins:2023tjg,Mullins:2023ott}.

Experimental measurements as functions of the rapidity gap \cite{STAR:2016vqt} can provide valuable insights into the impacts of stochastic noise on small systems. Measurements of higher-order cumulants and flow fluctuations, as well as the study of event-by-event correlations in small systems with stochastic fluctuations, are expected to be crucial for unraveling the underlying physics and extracting more accurate information about the transport properties of the QGP~\cite{CMS:2023bvg}.

\subsubsection{Particlization and freeze-out in small systems}

To compute experimental observables, the fluid needs to be converted into particles. This particlization stage is usually performed by the Cooper-Frye procedure~\cite{Cooper:1974mv}, which imposes continuity requirements for the energy-momentum current across the conversion hyper-surface under the grand-canonical ensemble. Each fluid cell on the conversion hyper-surface thermally emits hadrons at its temperature and chemical potential, which are then boosted according to the underlying flow velocity~\cite{Shen:2014vra}. This prescription produces particle thermal spectra, which also include perturbative out-of-equilibrium corrections from the viscous part of the fluid's energy-momentum tensor~\cite{Teaney:2003kp, Dusling:2009df, Romatschke:2009im}. The details of such corrections to the particle spectra depend on the underlying theory and, in practice, are an additional source of theoretical uncertainty~\cite{Wolff:2016vcm, JETSCAPE:2020avt, JETSCAPE:2020mzn, JETSCAPE:2020shq}. Especially if the viscous corrections are large at particlization, observables can be affected by the underlying assumptions, especially for experimental observables with $p_T \gtrsim 2$\,GeV.

For small systems, the available energy in the system could be limited. As the particle multiplicity becomes small, one would need to consider the effects coming from the conservation of energy, momentum, and conserved charges~\cite{Oliinychenko:2019zfk, Vovchenko:2020kwg, Oliinychenko:2020cmr}.

Because the fluid cells in small systems particlize at a similar local energy density compared to that in large collision systems~\cite{Schenke:2020mbo}, the system's final size is comparable with those in large collision systems at the same final-state hadron multiplicity~\cite{Heinz:2019dbd}. Due to the faster local expansion rate in small systems, the collision rate in the hadronic scatterings was found to be lower than that in the large systems at the same particle multiplicity~\cite{Shen:2015qba, Shen:2016zpp}. The hadronic rescatterings effects are not significant in small systems~\cite{Shen:2016zpp}.

\subsubsection{The role of longitudinal structure in full (3+1)D simulations for small systems}
\label{sec:long}

Many small systems are asymmetric light+heavy collisions, which explicitly break the longitudinal boost-invariance at the collision geometry level. Because of longitudinal momentum conservation, the initial-state energy density profiles have non-trivial longitudinal structures, which result in rapidity-dependent particle production and anisotropic flow coefficients in the final state~\cite{Shen:2017fnn, Shen:2017ruz, Zhao:2022ayk, Shen:2022oyg, Zhao:2022ugy, Shen:2022daw, Shen:2023awv, Shen:2023pgb}. We will discuss the phenomenological studies concerning these effects in Sec.~\ref{sec:vnRap}.

Measurements of rapidity decorrelation of flow observables study how the anisotropic flow vector fluctuates and evolves as a function of the particle rapidity~\cite{CMS:2015xmx, ATLAS:2017rij}
\begin{equation}
    r_n(\eta) = \frac{\Re \{\langle Q_n(-\eta) (Q_n^\mathrm{ref})^* \rangle\}}{\Re\{ \langle Q_n(\eta) (Q_n^\mathrm{ref})^* \rangle \}},
\end{equation}
where the $n$-th order anisotropic flow vector $Q_n$ is defined as
\begin{equation}
    Q_n \equiv \sum_{j=1}^M e^{i n \phi_j},
    \label{eqQnDef}
\end{equation}
and the sum runs over all $M$ final state particles and their transverse momenta's azimuthal angles $\{\phi_j\}$.
The slope of $r_n$ as a function of $\eta$ measures how fast the anisotropic flow vector $Q_n(-\eta)$ decorrelates from $Q_n(\eta)$. The reference flow vector $Q_n^\mathrm{ref}$ is usually chosen at forward or backward rapidity with at least one unit of rapidity gap from both the $Q_n(\pm \eta)$ vectors to suppress non-flow correlations. In light+heavy collisions, the asymmetric collision geometry and event-by-event longitudinal fluctuations contribute to the decorrelation of the flow vectors in $r_n(\eta)$. Full (3+1)D dynamic evolution is essential to study the flow rapidity decorrelation measurements and constrain the longitudinal dynamics of the collision systems~\cite{Bozek:2017qir, Pang:2018zzo, Zhao:2022ayk}.

The (3+1)D dynamical evolution of the relativistic heavy-ion collisions allows us to interpret the final-state anisotropic flow vector $Q_n(\eta)$ as a convolution of the initial state eccentricities $\varepsilon_n(\eta_s)$ with the hydrodynamic response function $G_n(\eta - \eta_s)$~\cite{Li:2019eni, Franco:2019ihq}
\begin{equation}
    Q_n(\eta) = \int d \eta_s G_n(\eta - \eta_s) \varepsilon_n(\eta_s).
\end{equation}
This equation is a generalization of the simple linear response $Q_n = k_n \varepsilon_n$ observed in boost-invariant simulations (for the harmonic order $n = 2, 3$). Because the hydrodynamic response function $G(\eta - \eta_s)$ is a two-point function, thermal fluctuations in the hydrodynamic phase also contribute to the hydrodynamic response function and the rapidity decorrelation. A large flow rapidity decorrelation would reduce the sensitivity of the final-state flow measurements to the initial geometry.

Performing (3+1)D dynamics is crucial when studying experimental observables related to the evolution of longitudinal flow velocity, for example, the fluid vorticity and $\Lambda$ polarization~\cite{Ryu:2021lnx, Alzhrani:2022dpi}. In asymmetric light+heavy collision systems, the light nucleus drills through the heavy nucleus, resulting in a nontrivial longitudinal flow profile. The transverse coordinate dependence of the longitudinal flow is directly related to the size of initial-state flow vorticity in the collision system, which imprints itself to the final-state polarization of the $\Lambda$ hyperons. Theoretical work predicted the existence of ``smoke ring'' patterns for flow vorticity in asymmetric collision systems~\cite{Lisa:2021zkj}. Future measurements of such a pattern in the $\Lambda$ polarization would verify and probe the fluid dynamic nature of the produced collision system at the length scale of velocity gradients~\cite{Serenone:2021zef, Ribeiro:2023waz}.

\section{Phenomenological studies in small systems}
\label{sec:reviewdata}

In this section, we will review the diverse signals associated with quark-gluon plasma formation in the soft sector of small systems. We will conduct a thorough comparison between the model calculations and experimental data.

\subsection{Collective flow}
Collective flow is one of the most important observables in relativistic nuclear collisions. Its study provides valuable information on the initial state and several QGP properties. Collective flow observables determine the anisotropy in particle transverse momentum distributions correlated with the flow symmetry plane $\Psi_{n}$~\cite{Ollitrault:1992bk}. 
The various characteristic patterns of the anisotropic flow can be obtained from a Fourier expansion of the azimuthal particle distribution in a given event~\cite{Voloshin:1994mz}:
\begin{equation}
\frac{{\rm d} N}{{\rm d} \varphi} \propto 1+ 2 \sum_{n=1}^{\infty} v_{n} \, e^{in(\varphi-\Psi_{n})}
\label{eq:vnDef}
\end{equation}
where $v_{n} = \langle \cos \, [n(\varphi - \Psi_n)] \rangle$ is the anisotropic flow coefficient and $\Psi_{n}$ the corresponding flow symmetry plane, both fluctuating event by event.
Here, the angular average $\langle O \rangle$ is defined as
\begin{equation}
    \langle O \rangle \equiv \frac{\int d \phi \, O \, dN/d\phi}{\int d \phi\, dN/d\phi}.
\end{equation}
In general, the first coefficient, $v_1$, is called {\emph{directed flow}},  the second coefficient, $v_2$, is called {\emph{elliptic flow}} and the third coefficient $v_3$, is called {\emph{triangular flow}}. For $n \geq 3$, we refer to the $v_n$ as higher-order flow harmonics.
Because the global rotation of the collision orientation introduces an arbitrary random phase event by event, the flow symmetry plane cannot be directly measured. Therefore, the anisotropic flow coefficient $v_{n}$ cannot be measured using single-particle information.
A popular approach is the event-plane method~\cite{Poskanzer:1998yz}, where the azimuthal correlation of emitted particles is measured with respect to an event-plane. However, it was found that the results from the event-plane method strongly depend on the resolution of the event-plane (limited mainly by the finite number of measured particles), which introduces an uncontrolled bias in the measurement~\cite{Luzum:2012da}.
As an alternative approach, multi-particle azimuthal correlation measurements \cite{Bilandzic:2010jr,Bilandzic:2013kga} have been employed as they allow for a more robust measurement of the underlying anisotropic flow.

The $Q$-cumulant method measures the flow harmonics $v_n$ from multi-particle correlations without knowing the event plane. With the flow $Q_n$-vector defined in Eq.~\eqref{eqQnDef}, the 2-and 4-particle azimuthal correlations in a single event can be calculated as~\cite{Bilandzic:2010jr}:
\begin{equation}
\begin{aligned}
\langle 2 \rangle_{n,-n} & = \frac{|Q_{n}|^{2} - M} {M(M-1)}, \\
\langle 4 \rangle_{n,n,-n,-n} & = \frac{|Q_{n}|^{4} + |Q_{2n}|^{2} - 2 \cdot {\rm{Re}}[Q_{2n}Q_{n}^{*}Q_{n}^{*}]  } { M(M-1)(M-2)(M-3) }  \\
& \ \  ~ ~ - 2 \frac{ 2(M-2) \cdot |Q_{n}|^{2} - M(M-3) } { M(M-1)(M-2)(M-3) },
\end{aligned}
\label{Eq:Mean24}
\end{equation}
where $M$ is the particle multiplicity for the flow vector.
Here, we have used the general notation of the single-event $k$-particle correlators $\langle k \rangle_{n_{1}, n_{2}, ..., n_{k}} \equiv \langle \cos(n_1\varphi_1\! + \!n_2\varphi_2\!+\!\cdots\!+\!n_k\varphi_k) \rangle  \,(n_1\geq n_2 \geq \cdots \geq n_k)$ and $\langle ... \rangle$ means an average over all the particles in a single event. After averaging over all events within the selected centrality bin, the obtained 2- and 4-particle cumulants are:
\begin{equation}
\begin{aligned}
c_{n}\{2\} & = \langle \langle 2 \rangle \rangle_{n,-n}, \\
c_{n}\{4\} & = \langle \langle 4 \rangle \rangle_{n,n,-n,-n} - 2 \cdot \langle \langle 2 \rangle \rangle^{2}_{n,-n},
\end{aligned}
\label{Eq:c24}
\end{equation}
Then, the 2- and 4-particle integrated flow harmonics can be calculated as~\cite{Bilandzic:2010jr}:
\begin{eqnarray}
v_n\{2\} &= \sqrt{c_n\{2\}}, \quad  v_n\{4\} &= \sqrt[4]{-c_n\{4\}}.
\label{Eq:v24}
\end{eqnarray}

In general, the 4-particle correlations used to determine the flow harmonics $v_n\{4\}$ can largely suppress
the non-flow effects from jets, resonance decays, etc. Non-flow effects significantly influence $v_n\{2\}$ obtained from the 2-particle correlations. To suppress such non-flow effects, one divides the whole event into sub-events with a certain pseudorapidity gap $|\Delta \eta|$ and then calculates the 2-particle azimuthal correlations to further suppress the non-flow contamination. 

The high energy proton–lead ($p$+Pb)  and proton–proton ($p$+$p$) collisions at the LHC were originally aimed to provide the reference data (the ``no-QGP'' reference) for high energy nucleus-nucleus collisions.  However, various unexpected phenomena have been observed in these small systems, especially in the high multiplicity region. One surprising discovery was the long-range ``ridge'' structure in two-particle azimuthal correlations with a large pseudo-rapidity separation in high multiplicity $p$+Pb and $p$+$p$ collisions. Such long-range correlation structures were first discovered in Au+Au and Pb+Pb collisions and interpreted as a signature of the collective expansion. This has also triggered the study of collectivity in other small systems, such as $p$+Au, $d$+Au, and $^3$He+Au  at RHIC~\cite{PHENIX:2017xrm,PHENIX:2018lia,Schenke:2019pmk, Zhao:2022ugy}, as well as the system size scan of various collision systems at the LHC~\cite{Citron:2018lsq}.
Keeping in mind the caveats discussed in Section \ref{sec:model}, the observed flow-like signals in the small systems can be quantitatively or semi-quantitatively described by hydrodynamic calculations,  which translate initial spatial anisotropies into final momentum anisotropies of produced hadrons with the collective expansion of the bulk matter.  

\begin{figure}[tb]
  \centering
  \includegraphics[width=0.48\textwidth]{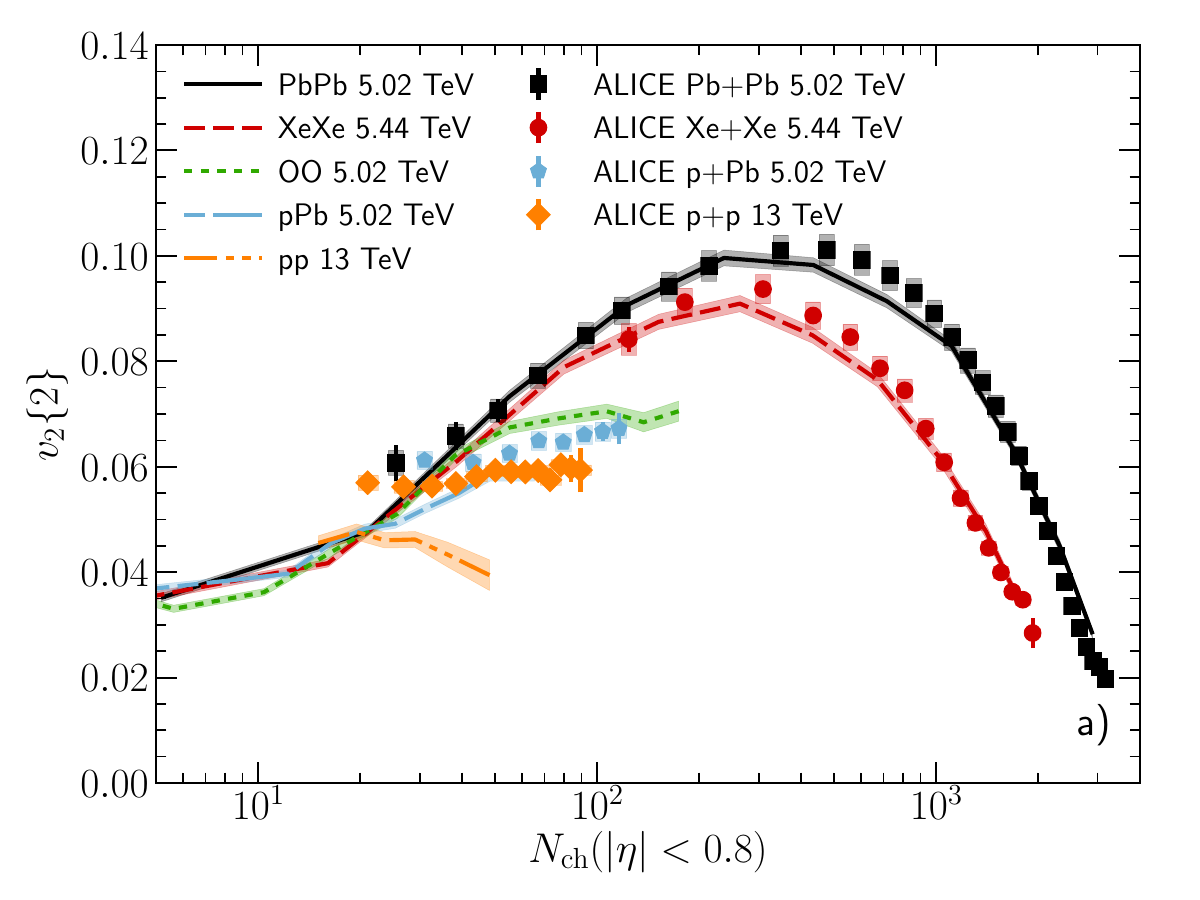}\includegraphics[width=0.48\textwidth]{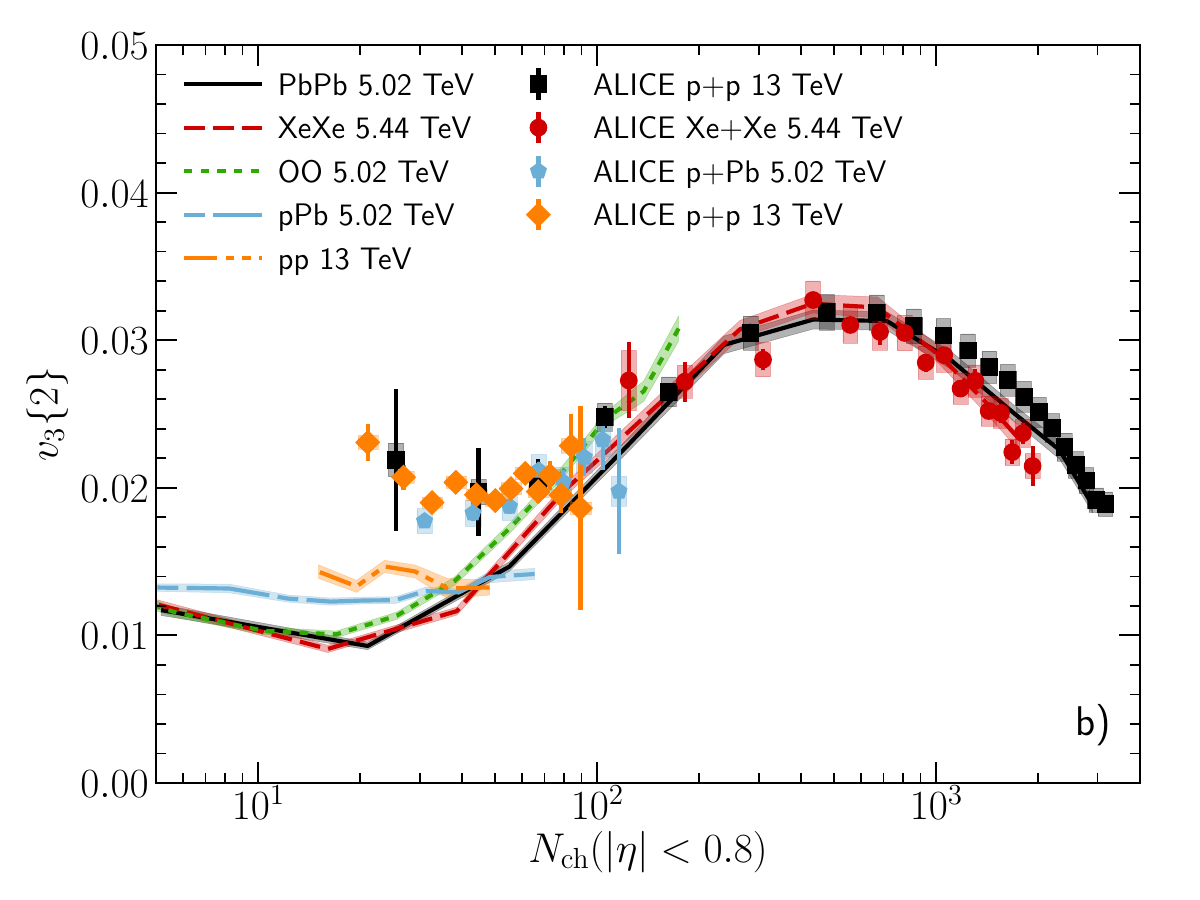}
  \caption{Anisotropy coefficients $v_2\{2\}$ (a) and $v_3\{2\}$ (b) for charged hadrons vs.~charged hadron multiplicity in various collision systems at LHC, compared to experimental data from the ALICE Collaboration \cite{ALICE:2019zfl}. The figures are taken from published work.\cite{Schenke:2020mbo}. \label{fig:v2v3}}
\end{figure}

Fig.~\ref{fig:v2v3} shows hydrodynamic calculations with IP-Glasma initial conditions~\cite{Schenke:2020mbo} of the charged hadron $v_2\{2\}$ (left) and  $v_3\{2\}$ (right) as a function of charged hadron multiplicity from small to large collision systems at LHC and compares to experimental data \cite{ALICE:2019zfl}. Overall, the hydrodynamic model reproduces the multiplicity dependence of the integrated $v_2\{2\}$  and  $v_3\{2\}$ from $p$+Pb to Pb+Pb collisions well.
In particular, hydrodynamics seems to reproduce well the $v_2\{2\}$  in Xe+Xe and Pb+Pb collisions from peripheral to central collisions.
For $p$+Pb collisions, the agreement becomes worse, and for $p$+$p$ collisions, this hydrodynamic simulation misses the experimentally observed magnitude and multiplicity dependence.
The situation is similar for $v_3\{2\}$. The agreement is best for more central larger systems. The $v_3\{2\}$ is rather insensitive to the system's average geometry, as it is driven solely by fluctuations. Similar to $v_2\{2\}$, hydrodynamic calculations underestimate the experimental data for $v_3\{2\}$ both in $p$+$p$ and $p$+Pb collisions, which requires further study in the future. 

\begin{figure}[tb]
  \centering
  \includegraphics[width=0.7\textwidth]{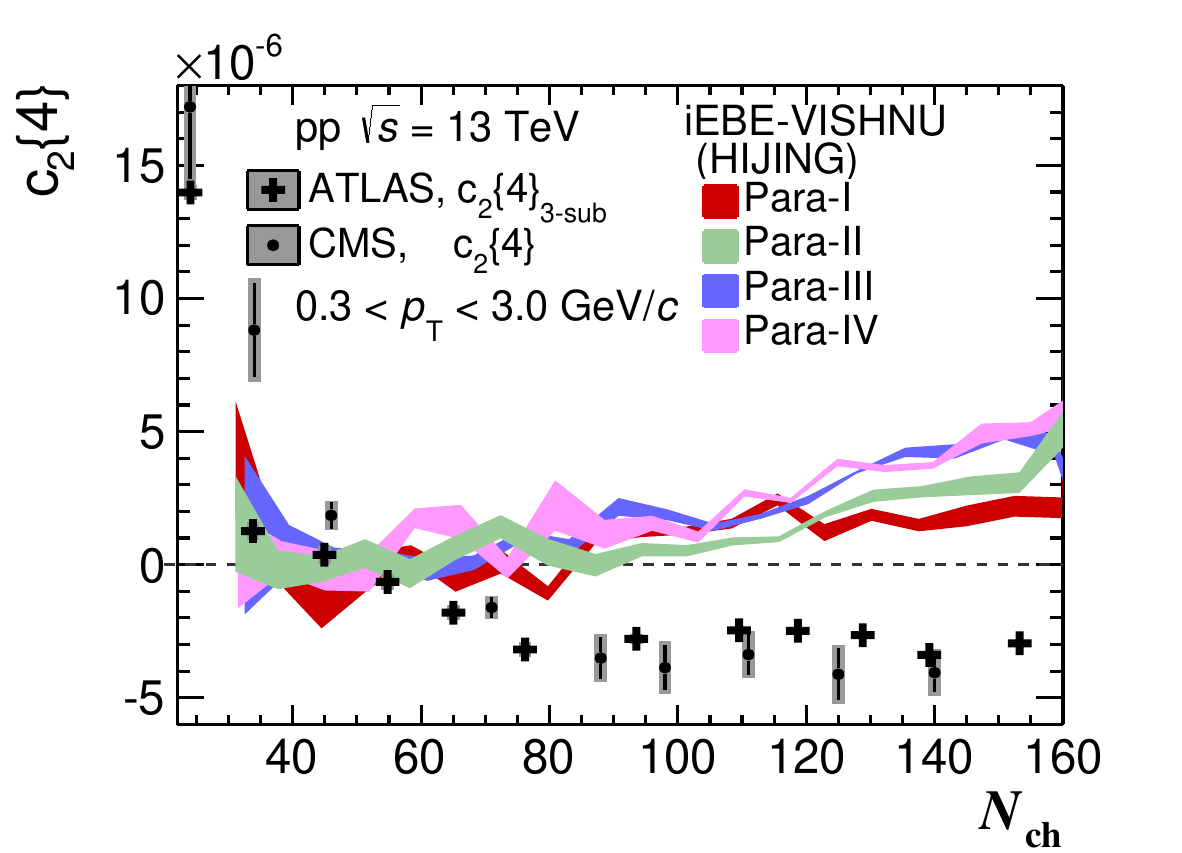}
  \caption{Four-particle correlations, $c_2\{4\}$,  as a function of $N_{ch}$ for charged hadrons in pp collisions 13 TeV, calculated by  hydrodynamic model. The CMS data and the ATLAS data are taken from~\cite{CMS:2016fnw} and~\cite{Zhou:2017wls}, respectively. This figure is taken from published work.~\cite{Zhao:2017rgg}. \label{fig:c24pp}}
\end{figure}

\subsection{Multi-particle correlations}
Compared with the 2-particle correlations, multi-particle cumulants are less influenced by the non-flow effects and thus are key observables to evaluate the anisotropic collectivity in small systems. Additionally, the multi-particle and 2-particle cumulants show different sensitivities to event-by-event flow fluctuations~\cite{Voloshin:2008dg,Borghini:2000sa,Jia:2014pza}. An extensive measurement of these different cumulants could provide tight constraints on the initial state fluctuations. To extract real values of the flow coefficients, the  2-, 4-, 6-, and 8-particle cumulants are expected to carry positive, negative, positive, and negative signs, respectively. Such ``changing sign pattern'' has been observed in the measured 2- and multi-particle cumulants in Pb+Pb collisions at the LHC~\cite{ALICE:2011ab, Adam:2016izf}. Similarly, 2- and multi-particle cumulants have been measured in high multiplicity $p$+$p$ and $p$+Pb collisions at the LHC. However, it was found that the standard multi-particle cumulants in the small systems are still largely affected by the residual short-range non-flow and its fluctuations,  which even presents fake flow signals with the ``right sign''.  To further remove the residual non-flow from jets, 2- and 3-subevent methods for the multi-particle cumulants were developed.\cite{Jia:2017hbm, Huo:2017nms}  With this technique, several experimental collaborations have confirmed the observations of positive 2-particle cumulants and negative 4-particle cumulants in high multiplicity $p$+$p$ and $p$+Pb collisions at the LHC~\cite{ATLAS:2017rtr}.

With properly tuned parameters, hydrodynamic simulations can quantitatively or at least semi-quantitatively describe the two-particle correlations in $p$+Pb and $p$+$p$ and multi-particle correlations in $p$+Pb collisions at the LHC and RHIC~(see reviews \cite{Dusling:2015gta,Loizides:2016tew,Schlichting:2016sqo,Nagle:2018nvi,Schenke:2019pmk,Shen:2020mgh,Schenke:2021mxx}). 
However, the measured negative $c_2\{4\}$ in $p$+$p$ collisions, which could naively be interpreted as evidence of hydrodynamic flow, could not be reproduced by any of the hydrodynamic calculations on the market~\cite{Zhao:2017rgg,Schenke:2020mbo,Zhao:2020pty}. As shown in Fig.\,\ref{fig:c24pp}, hydrodynamic model calculations with different initial conditions yield positive values for $c_2\{4\}$ in $p$+$p$ collisions. It is still unknown if the wrong sign of $c_2\{4\}$ is because of the incorrect fluctuation spectrum in initial conditions or due to issues with the applicability of the hydrodynamic model to $p$+$p$ collisions.

\subsection{Partonic collectivity and strangeness enhancement in small systems}
The origins of the observed collective behavior  in small systems are still under
debate. For the soft hadron data measured in high-multiplicity $p$+Pb collisions,  hydrodynamics or transport models based on final-state effects can describe many of the flow-like signals. On the other hand, color glass condensate  models~\cite{Kovner:2012jm,Dusling:2017dqg,Kovchegov:2012nd,Lappi:2015vta} including the IP-Glasma model~\cite{Schenke:2015aqa,Schenke:2016lrs}, which by themselves produce momentum anisotropies as initial-state effects, have also been used to explain some features of the observed collectivity. However, generally, the initial state models produce the wrong multiplicity dependence of the anisotropic flow. As discussed in Section~\ref{sec:initial}, the usual assumption of boost-invariance of the signal has been challenged, as has the probability of its survival if 3+1 dimensional early time evolution is included. 

In large systems, hard probes have been used as signatures for QGP formation. In small systems, due to their limited sizes and lifetime, the energy lost by energetic partons no longer leads to obvious signatures at high $p_T$ to discern if the QGP is formed. The relatively small nuclear modification effects for large $p_T$ light and heavy flavor hadrons and jets measured in small systems are consistent with the expectations of cold nuclear matter effects~\cite{ALICE:2012mj,CMS:2015ved,ALICE:2018vuu,Eskola:2016oht,Dong:2019byy,ALICE:2016yta}. For more information about the hard probes in small systems, we refer the reader to subsection \ref{sec:hardprobe}.

Besides collective flow and the quenching of energetic jets, the number of constituent quark (NCQ) scaling of elliptic flow and strangeness enhancement was proposed long ago as signals of QGP formation in high-energy nuclear collisions. 
The ATLAS, CMS, and ALICE Collaborations have measured the $v_2(p_T)$ of charged and identified hadrons with high precision in high multiplicity events of small systems~\cite{CMS:2018loe,ATLAS:2016yzd,Pacik:2018gix}. The resulting data show a similar approximate NCQ scaling of $v_2$ at intermediate $p_T$ as observed in heavy ion collisions. Also, strangeness enhancement in small colliding systems was reported by the ALICE Collaboration \cite{ALICE:2016fzo}.
Yield ratios of (multi)strange hadrons to charged pions exhibit a monotonic and continuous increase as functions of multiplicity across collision systems.
In high-multiplicity events of small systems, the strange hadron yield ratios almost reach similar values to those in heavy-ion collisions.

\begin{figure*}[ht]
\centering
\includegraphics[width=1.0\textwidth]{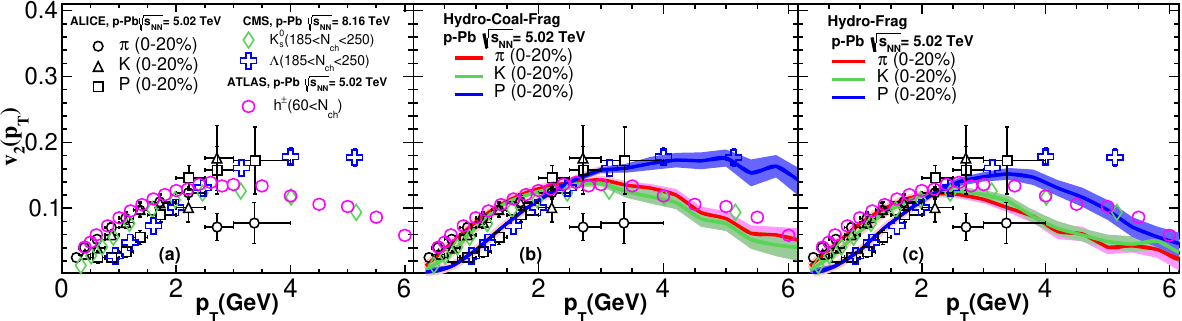}
\caption{Differential elliptic flow $v_2(p_T)$ of pions, kaons and protons in 0-20\%   p-Pb at $\sqrt{s_{NN}}=5.02$ TeV measured in experiments (left window) and calculated with the {\tt Hydro-Coal-Frag} model (middle window) and the {\tt Hydro-Frag} model (right window).  The ALICE, CMS, and ATLAS data are taken from~\cite{ALICE:2013snk}, ~\cite{CMS:2018loe} and~\cite{ATLAS:2016yzd}, respectively. This figure is taken from published work. \cite{Zhao:2020wcd} }
\label{fig:v2ptpPb}
\end{figure*}

The role  of  partonic  degrees  of  freedom  in  high  multiplicity  $p$+Pb  collisions was investigated in the quark coalescence model \cite{Zhao:2020wcd}. The partonic collectivity will manifest itself at intermediate $p_T$ of hadrons after the quark coalescence process. 
In this model, the mesons and baryons are coalesced from the quark and antiquark employing the overlaps of the quarks' phase-space distributions and the hadrons' Wigner functions~\cite{Greco:2003xt,Han:2016uhh}.
In the calculation shown in Fig.\,\ref{fig:v2ptpPb} \cite{Zhao:2020wcd}, the coalescence process includes thermal-thermal, thermal-hard and hard-hard parton recombinations, where the thermal quarks are taken from the hydrodynamic medium and hard partons are taken from the linear Boltzmann transport energy loss model~\cite{Wang:2013cia,He:2015pra,Cao:2017hhk}.
The remnant hard partons are grouped into strings and fragmented to hard hadrons using the ``hadron standalone mode" of PYTHIA8~\cite{Sjostrand:2007gs}.

As shown in Fig.~\ref{fig:v2ptpPb}, the newly developed hybrid hadronization model  nicely  describes  the  measured proton-to-pion ratio and the $v_2(p_T)$  of  pions,  kaons  and  protons  from  0  to  6 GeV in high multiplicity $p$+Pb collisions at 5.02 TeV. Specifically, the low $p_T$ mass ordering of $v_2(p_T)$ of identified hadrons is reproduced by the hydrodynamic part. For $p_T>2.5$ GeV,  the $v_2(p_T)$  of protons becomes larger than that of pions and kaons, which is attributed to the quark coalescence process. In contrast,  without  the quark coalescence  process, the model fails  to describe the $v_2(p_T)$ of identified hadrons at intermediate $p_T$. This demonstrates that including the quark coalescence contribution to the production of  hadrons  is  essential  in  reproducing  the  measured $v_2(p_T)$  of identified  hadrons  at  intermediate $p_T$. It thus provides a strong indication for the existence of partonic degrees of freedom and the  possible  formation  of  the  QGP  in  high  multiplicity $p$+Pb collisions at 5.02 TeV.

\begin{figure*}[htbp]
\begin{center}
\includegraphics[bb=0 0 468 540,width=0.45\textwidth]{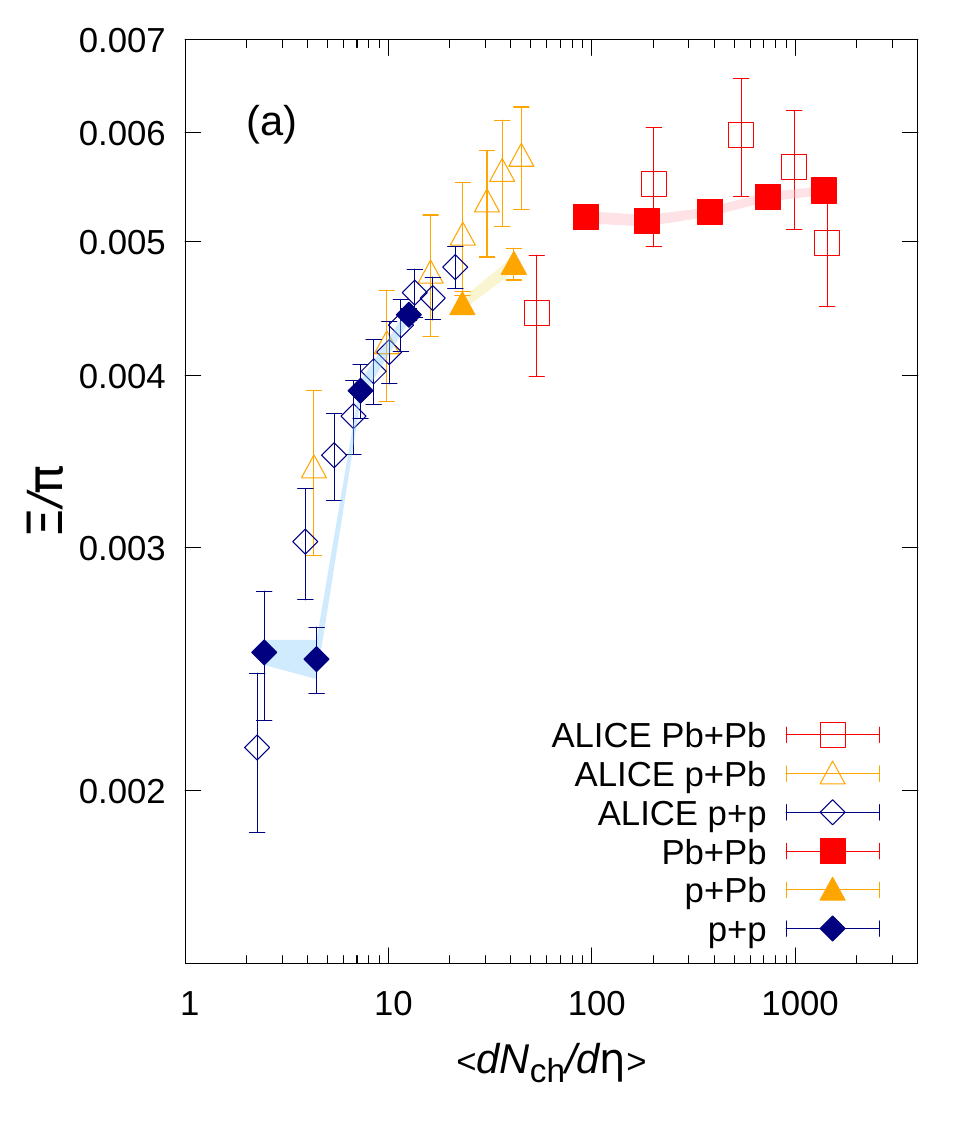}
\hspace{25pt}
\includegraphics[bb=0 0 468 540,width=0.45\textwidth]{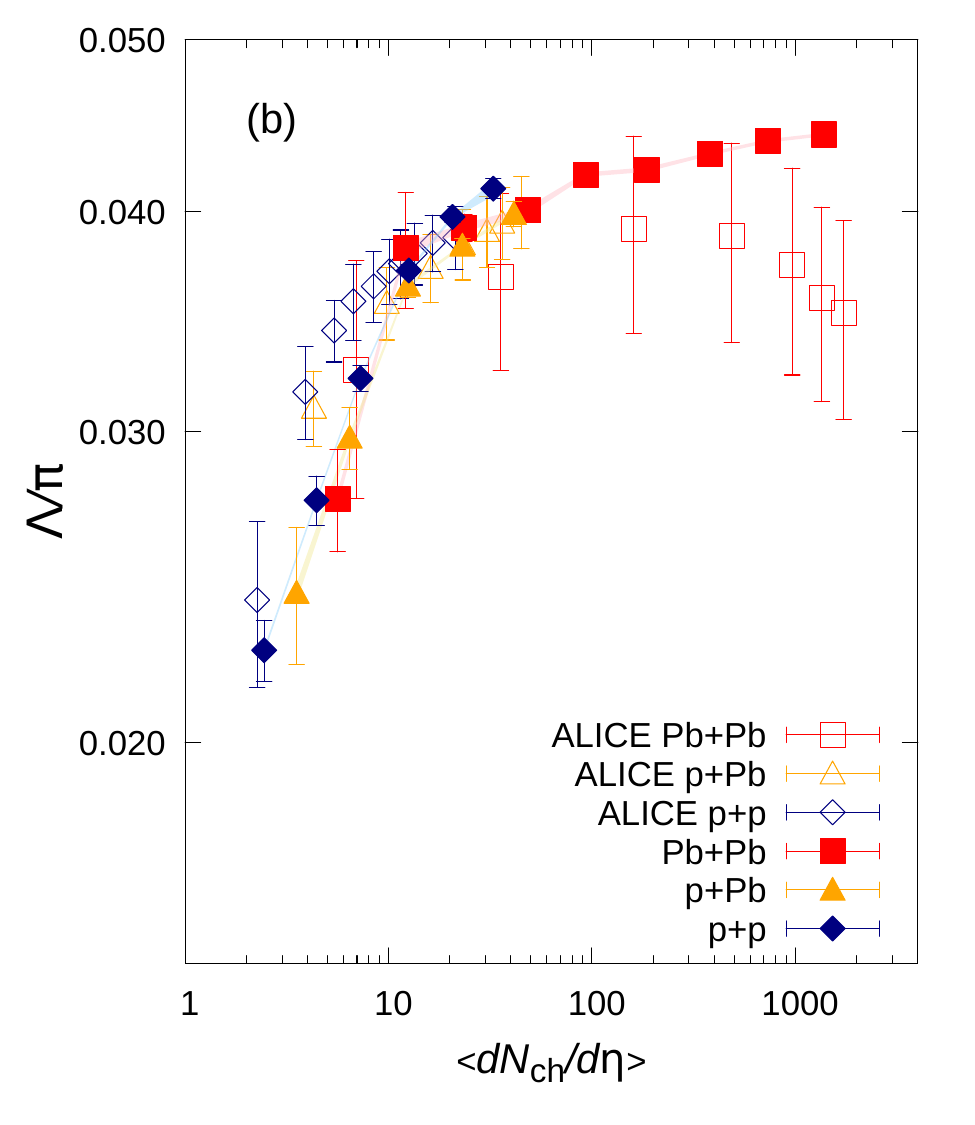}
\end{center}
\caption{Hadron yield ratios of (a) cascades ($\Xi^{-}$ and $\bar{\Xi}^{+}$) and
(b) lambdas ($\Lambda$ and $\bar{\Lambda}$) to charged pions ($\pi^-$ and $\pi^{+}$)
as functions of multiplicity at midrapidity, $\left<dN_{\mathrm{ch}}/d\eta\right>$,
from $p$+$p$ to Pb+Pb  collisions at LHC energies.
The ALICE data is taken from \cite{ALICE:2014jbq,ALICE:2016sak,ALICE:2018pal,ALICE:2013mez,ALICE:2013wgn,ALICE:2015mpp,ALICE:2016fzo}.  This figure is taken from published work. \cite{Kanakubo:2019ogh}.
}
\label{fig:yield-ratios}
\end{figure*} 

Concerning strangeness  enhancement,  since the colliding  nuclei do not  contain  a  strange  valence quark, yields  of  strange  hadrons  should  be  sensitive  to  details of the reaction dynamics. A dynamical core–corona framework to study the event activity dependence of the hadron yield ratios in various collision systems and over a wide range of collision energies was developed.\cite{Kanakubo:2019ogh} In this approach, the system generated in high-energy nuclear collisions is described with two components:  equilibrated matter (core) and non-equilibrated matter (corona). Non-equilibrated partons can act as sources of QGP fluid as they lose energy and momentum. The core/fluid is described by hydrodynamics. The low-density corona, where partons suffer only a few collisions, is treated microscopically, and hadrons are produced via string fragmentation. Fig. \ref{fig:yield-ratios} shows that, because of the interplay between core and corona components, the model reasonably describes the strangeness of hadron yield ratios to charged pions as functions of multiplicity ranging from $p$+$p$ to $p$+Pb to Pb+Pb collisions. The ratios increase up to $\left< dN_{ch}/d\eta \right>\sim 100$ and saturate in high multiplicity events. This tendency implies that the contribution of the fluid becomes large and dominant in high multiplicity events. Importantly, at a given multiplicity value, the high multiplicity small systems have similar values of the strange hadron ratios to large systems, where the system reaches the chemically equilibrated fluid limit. This result provides a strong indication of partial QGP generation in high multiplicity small colliding systems.

\subsection{The $v_n$-$p_T$ correlations}
An interesting observable involving the correlation between the elliptic momentum anisotropy $v_2$ and the average transverse momentum $[p_T]$ has been proposed to help constrain initial state models.\cite{Bozek:2016yoj} This observable has been used to extract information on the nucleon size and on the origin of the observed momentum anisotropy \cite{Schenke:2020uqq,Giacalone:2021clp}.
The correlation between $v_n^2$ and $[p_t]$ expressed by the Pearson correlation coefficient is \cite{Bozek:2016yoj}
\begin{equation}
\label{eq:pcc}
    \hat{\rho}_n \equiv \hat{\rho}(v_n^2,[p_t]) = \frac{\langle \delta v_n^2 \delta [p_t] \rangle}{\sqrt{\langle (\delta v_n^2 )^2 \rangle \langle (\delta [p_t])^2 \rangle } },
\end{equation}
where $\delta O = O - \langle O \rangle$ for any observable $O$ and the flow anisotropic coefficient $v_n$ is defined in Eq.~(\ref{eq:vnDef}).

\begin{figure*}[htbp]
\begin{center}
\includegraphics[width=0.6\textwidth]{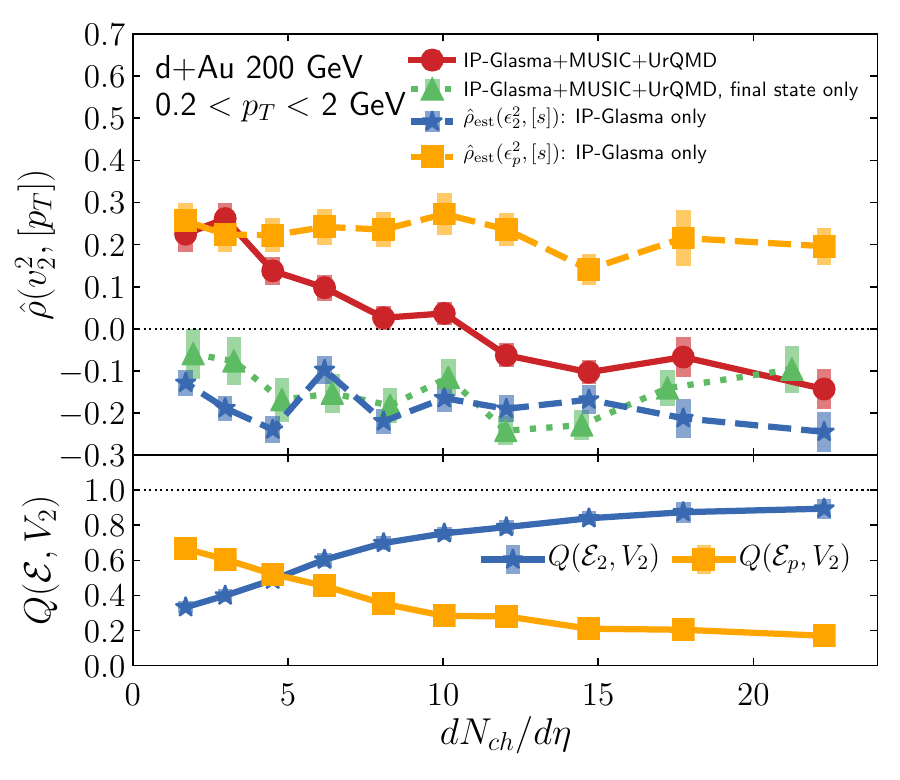}
\includegraphics[width=0.95\textwidth]{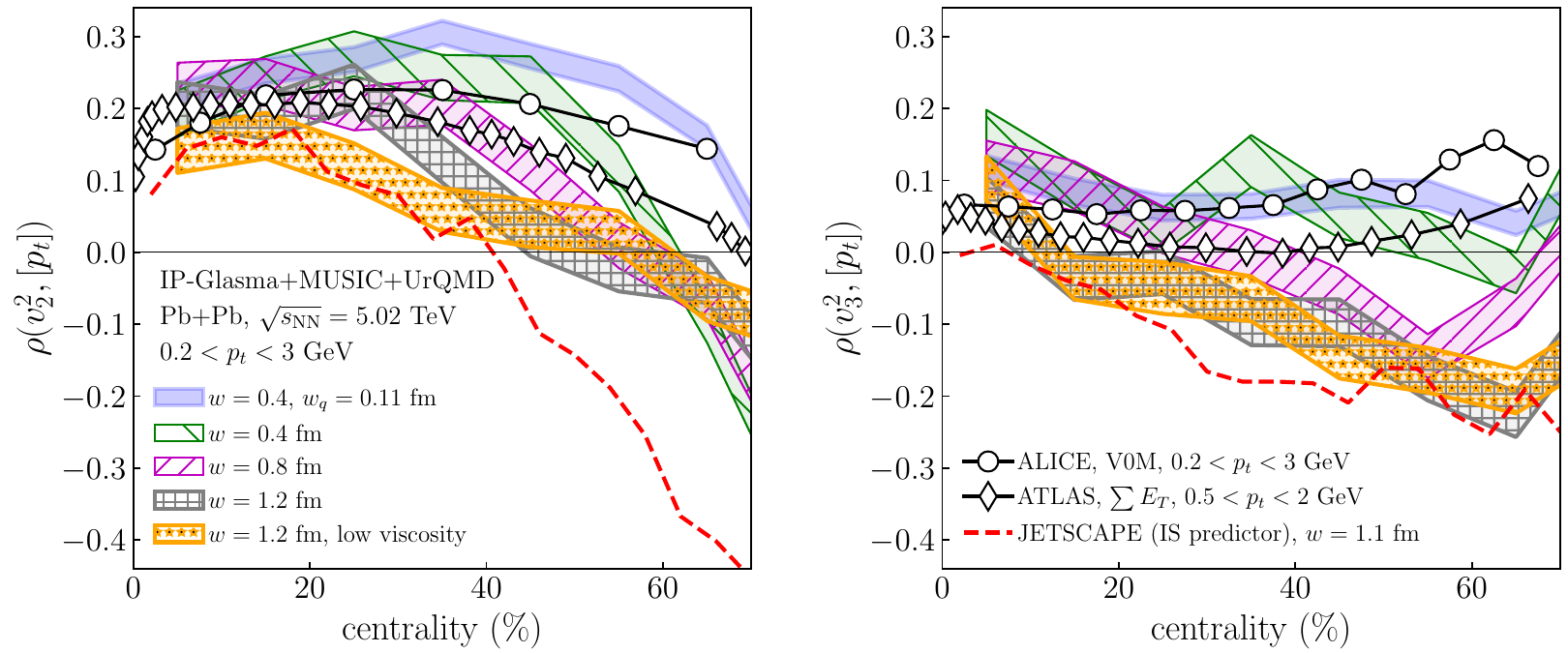}
\end{center}
\caption{Top: The correlator $\hat{\rho}(v_2^2,[p_t])$ (circles, solid lines) together with estimators based on the initial geometry ($\hat{\rho}_{\rm est}(\varepsilon_2^2,[s])$, stars) and the initial momentum anisotropy ($\hat{\rho}_{\rm est}(\varepsilon_p^2,[s])$, squares) in d+Au collisions at $\sqrt{s}=200\,{\rm GeV}$, and the Pearson coefficients between $v_2$ and the initial ellipticity (stars) and the initial momentum anisotropy (squares), respectively.  Bottom:  Results of the IP-Glasma+MUSIC+UrQMD framework for $\rho(v_2^2,[p_t])$ with different types of shaded bands correspond to different values of the nucleon width in Pb+Pb collisions. 
These figures are taken from published works. \cite{Schenke:2020uqq,Giacalone:2021clp}.
}
\label{fig:Pearson-rho}
\end{figure*} 

Two predictors for the $\hat{\rho}$-correlator as a function of multiplicities are presented in the top panel of Fig.~\ref{fig:Pearson-rho} for $200\,{\rm GeV}$ $d$+Au collisions. The $\hat{\rho}_{\rm est}(\epsilon_2^2,[s])$, is based entirely on the initial geometry using the initial spatial eccentricity $\epsilon_2$,
which is a good estimator for $v_2$. The  $\hat{\rho}_{\rm est}(\epsilon_p^2,[s])$ uses the initial momentum anisotropy $\epsilon_p$ as estimator for $v_2$. It is computed from the initial energy-momentum tensor of the IP-Glasma model as \cite{Schenke:2019pmk}
\begin{equation}\label{eq:epsilonp}
    \mathcal{E}_p \equiv \epsilon_p e^{i 2 \psi_2^p} \equiv \frac{\langle T^{xx}-T^{yy}\rangle + i\langle 2 T^{xy}\rangle}{\langle T^{xx}+T^{yy}\rangle}\,,
\end{equation}
evaluated at $\tau =\,0.1\,{\rm fm}$, where here $\langle\cdot\rangle$ is defined as average over the transverse plane~\cite{Schenke:2019pmk}.  The $[p_T]$ is estimated from the average initial entropy density $[s]$, where $[s]=[e^{3/4}]$ for ideal parton gas at $\tau =\,0.1\,{\rm fm}$.
The top panel in Fig.~\ref{fig:Pearson-rho} also displays the Pearson coefficients of $\mathcal{E}_2$ with $V_2$ and $\mathcal{E}_p$ with $V_2$, which is defined as \cite{Gardim:2011xv,Gardim:2014tya,Betz:2016ayq}
\begin{equation}
    Q(\mathcal{E}, V_2) = \frac{{\rm Re} \langle\mathcal{E} V_2^*\rangle}{\sqrt{\langle |\mathcal{E}|^2\rangle \langle |V_2|^2\rangle}}\,,
\end{equation}
where $V_2$ is the complex valued $2^{\rm nd}$ order flow harmonic.

One can see that for higher multiplicities $\hat{\rho}(v_2^2,[p_T])$ approaches the geometric estimator, while at lower multiplicities, the initial momentum anisotropy predicts $\hat{\rho}(v_2^2,[p_T])$ better.
Based on the color domain interpretation of the initial state momentum anisotropy \cite{Kovner:2010xk,Kovner:2011pe,Dumitru:2014dra,Dumitru:2014vka,Lappi:2015vta}, the $\hat{\rho}_{\rm est}(\epsilon_p^2,[s])$ is expected to be positive, because at fixed multiplicity, a larger $[p_T]$ selects events with smaller transverse size. This reduces the number of color domains with an average size of $1/Q_s$, which enhances the magnitude of initial momentum anisotropy in the CGC description \cite{Lappi:2015vta}.

The Pearson coefficients $Q(\mathcal{E}, V_2)$ show that the behavior of $\hat{\rho}(v_2^2,[p_T])$ is a result of the geometry dominating the elliptic flow in high multiplicity events, and the initial momentum anisotropy driving the final $v_2$ at low multiplicity.
The Pearson coefficients were studied already before \cite{Schenke:2019pmk}, but they are not experimentally observable. In contrast, the $\hat{\rho}(v_2^2,[p_T])$ is an observable whose sign change as a function of multiplicity is an indicator of the origin of the elliptic flow in small systems, and the presence of initial state momentum anisotropies as predicted from the color glass condensate. 

We note, however, that as discussed earlier, the above study assumed boost invariance and that recent studies going beyond that assumption indicate that the initial state momentum anisotropies are short-range in rapidity ($|\Delta \eta| <1$) \cite{Schenke:2022mjv}. Given the rapidity gaps used in the experimental determination of $\hat{\rho}(v_2^2,[p_T])$, this means that the initial state momentum contribution should be much smaller than shown in Fig.\,\ref{fig:Pearson-rho}. Furthermore, a complete CGC calculation beyond the classical limit, including the rapid growth of small fluctuations \cite{Berges:2014yta}, is also expected to wash out initial momentum anisotropies.

The bottom panel of Fig.\,\ref{fig:Pearson-rho} shows the $v_n^2$-$[p_t]$ correlations, modeled using different nucleon sizes in Pb+Pb collisions at 5.02 TeV. Note that $\hat{\rho}_2$ for a smooth nucleon with $w=0.4$ fm differs above 50\% centrality from the result with sub-nucleon constituents, showing that the details of the sub-nucleon structure are important. The figure reports as well experimental data from the ATLAS \cite{ATLAS-CONF-2021-001} and the ALICE \cite{ALICE:2021gxt} collaborations at LHC. Data is in qualitative disagreement with calculations implementing $w=0.8$ fm or larger and suggests that it is indeed possible to constrain the size of nucleons (or their constituents) from such observations in heavy-ion collisions. It also shows the results from a low-viscosity run, which demonstrates that $\hat{\rho}_n$ are driven by initial-state properties and are largely insensitive to medium effects. The sensitivity of $\hat{\rho}(v_3^2$-$[p_t])$ to variations in $w$ is also noteworthy. Nucleon structure properties significantly influence the observable of final-state two-particle correlations, particularly in peripheral collisions. Notably, the sign of $\hat{\rho}(v_3^2$-$[p_t])$ can flip depending on the value of $w$  in peripheral collisions.

\subsection{Rapidity dependent anisotropic flow in small systems}
\label{sec:vnRap}

\begin{figure*}[htbp]
\begin{center}
\includegraphics[width=1.\textwidth]{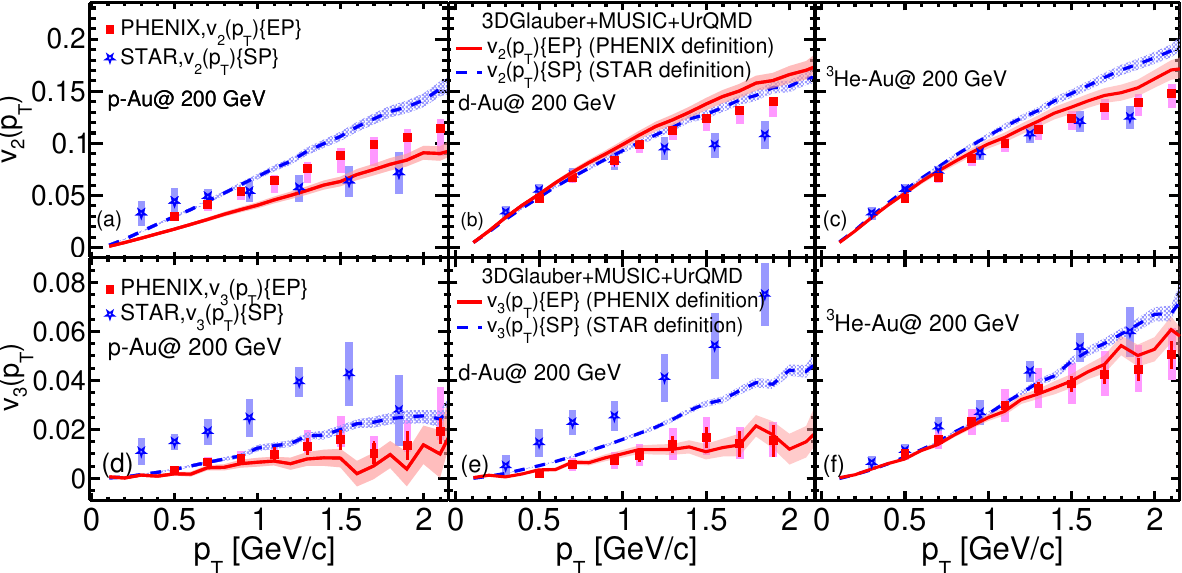}
\end{center}
\caption{  The anisotropic flow $v_n(p_T)$ as a function of $p_T$ in central p+Au, d+Au and $^3$He+Au collisions computed from the (3+1)D hydrodynamic model. The PHENIX and STAR data are from~\cite{PHENIX:2018lia,STAR:2022pfn}.
This figure is taken from published work. \cite{Zhao:2022ugy}.
}
\label{fig:vnptrhic}
\end{figure*} 

\begin{figure*}[htbp]
\begin{center}
\includegraphics[width=0.9\textwidth]{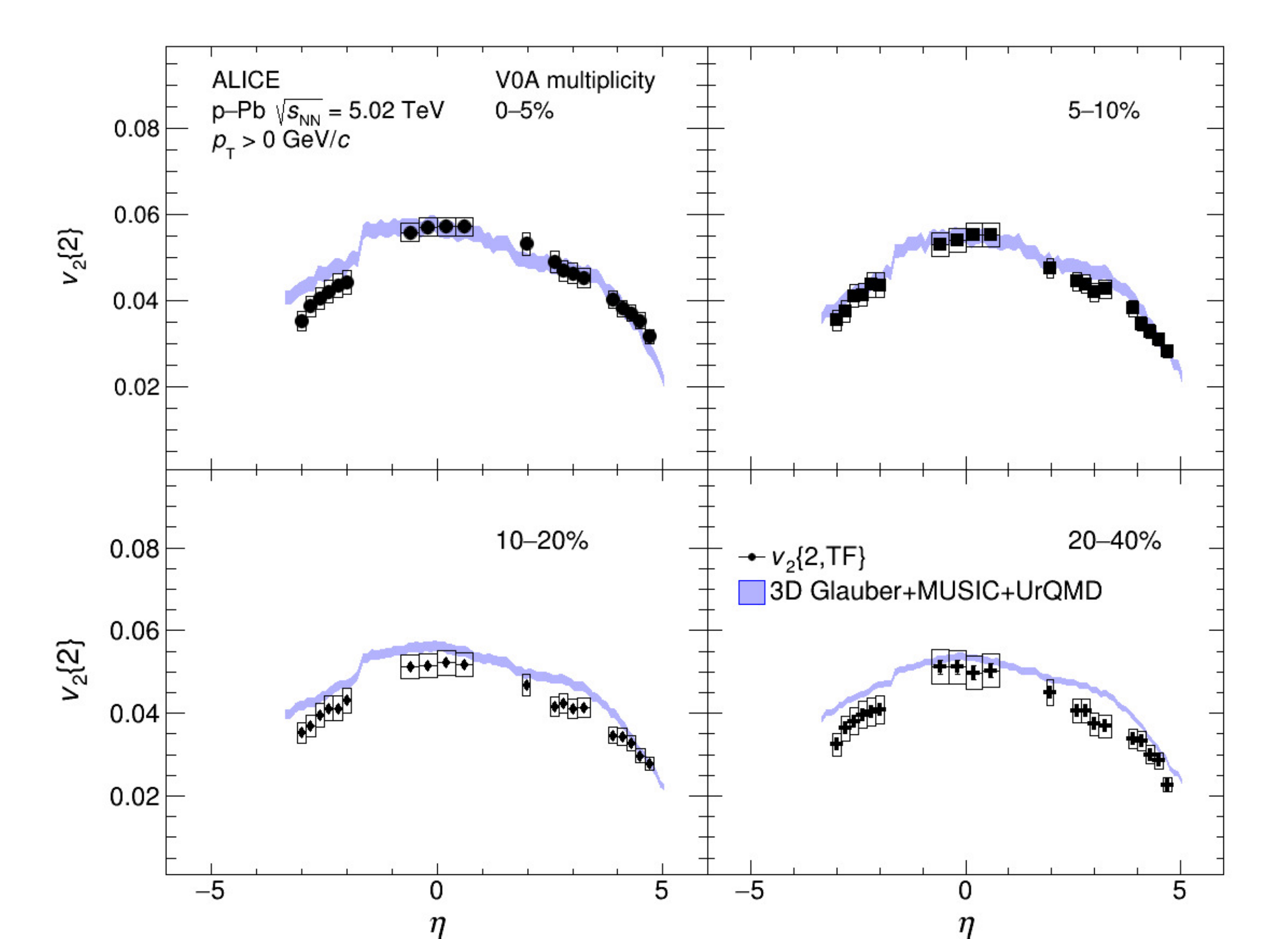}
\end{center}
\caption{  The pseudorapidity dependence of $p_{T}$ integrated $v_2$ in $p$+Pb at $\sqrt{s_{NN}}=5.02$ TeV. Comparison of the ALICE data~\cite{ALICE:2023gyf} (black circles) with a calculation within a (3+1)D hydrodynamical model (blue band)~\cite{Zhao:2022ayk} for the 0--5\% (top left), 5--10\% (top right), 10--20\% (bottom left), and 20--40\% (bottom right) centrality classes.
This figure is taken from published work \cite{ALICE:2023gyf}.
}
\label{fig:vnetappb}
\end{figure*}

The (3+1)D dynamics is also crucial to explore the collectivity in central p+Au, d+Au and $\rm {^{3}He}$+Au collisions at RHIC and high-multiplicity $\gamma^*$-nucleus collisions in ultra-peripheral collisions (UPC) at the LHC~\cite{Zhao:2022ayk, Zhao:2022ugy,Ryu:2023bmx, STAR:2023wmd}. Figure \ref{fig:vnptrhic}  shows the $v_n(p_T)$ ($n = 2, 3$) for charged hadrons compared between  (3+1)D  hydrodynamic model simulations and the experimental data from the PHENIX and STAR Collaborations. The (3+1)D  model gives an overall good description of the PHENIX data for d+Au and $\rm {^{3}He}$+Au collisions. The model has some tension with the STAR data for $v_3(p_T)$ in p+Au and d+Au collisions.~\cite{Zhao:2022ugy,Ryu:2023bmx}

It is crucial to note that both in the model calculations and experimental data, the $v_3(p_T)$ with the STAR definition are systemically larger than those determined using the PHENIX definition. In (3+1)D hydrodynamic calculations, this difference is mainly caused by the different magnitudes of the longitudinal decorrelation of flow vectors of $v_3$ between the different pseudo-rapidity bins used by the two collaborations. Figure \ref{fig:vnetappb} shows the pseudorapidity dependence of $p_T$ integrated $v_2(\eta)$ in $p$+Pb at 5.02 TeV.  In the (3+1)D model calculations, the longitudinal decorrelation of flow vectors plays a significant role in shaping the pseudorapidity dependence of $v_2$. 
For UPCs, the incoming energies between the virtual photons and the nucleus are highly unbalanced and fluctuate event by event. The large global rapidity shift between the center-of-mass frame and the lab frame introduces the non-trivial longitudinal decorrelations of flow vectors. This underscores the significance of employing (3+1)D model simulations in small systems.

\subsection{Electromagnetic radiation in small systems}

Black-body radiation is a characteristic phenomenon for any finite-temperature system in equilibrium. Electromagnetic (EM) radiation from relativistic heavy-ion collisions is clean because photons and dileptons suffer negligible final-state interactions after production~\cite{Shen:2016odt, David:2019wpt, Geurts:2022xmk}. An observation of thermal EM radiation is a smoking gun for producing thermalized hot matter in collisions. Furthermore, their transverse momentum spectra encode information on local temperature and blue shift from flow velocity. Therefore, searching for thermal EM radiation in small systems can provide direct evidence of producing a nearly equilibrated QCD medium in these collision systems~\cite{Shen:2015qba, Shen:2016zpp, Shen:2016egw, Shen:2016mmv}. 

Detecting thermal photon production from relativistic nuclear collisions has been challenging experimentally because of overwhelming background radiation from hadronic decays and prompt photon production. Nevertheless, the PHENIX Collaboration measured the direct (including thermal) photon radiation in small systems, such as ($p$, $d$, $^3$He)+Au collisions at 200 GeV. A hint of thermal photon enhancement above the prompt photon background was observed in 0-5\% most central collision events~\cite{Esha:2023ooh}. While the measurements were consistent with the theoretical predictions based on hydrodynamic evolution and thermal photon emission rates, they still contain considerable uncertainty, preventing any decisive conclusion. 

Future systematic studies of thermal photon radiation as a function of charged hadron multiplicity across multiple collision systems~\cite{Shen:2023aeg} would shed additional light on characterizing the nature of the QCD matter produced in small systems.

\subsection{Hard probes and the $R_{pA}$ and $v_2(p_T)$ puzzle in small systems}
\label{sec:hardprobe}

Motivated by the significant azimuthal anisotropies measured in small systems for low-$p_T$ hadrons with patterns similar to those observed in heavy-ion collisions, measurements aimed at detecting signs of jet quenching in these small
collision systems were performed. However, they have found no such effect. Analyses of high $p_T$ hadron and jet $p_T$ spectra indicate production yields consistent with those in $p$+$p$ collisions scaled by the expected nuclear thickness in $p$+Pb~\cite{ATLAS:2014qaj,ALICE:2012mj,ATLAS:2019vcm} and
$d$+Au collisions~\cite{PHENIX:2015fgy}. On the other hand, the experimental observations of the hadron azimuthal anisotropy up to $p_T \approx 12$~GeV indicate a non-zero anisotropy extending into the region beyond the usual hydrodynamic interpretation and into the regime of jet quenching~\cite{ATLAS:2014qaj,ATLAS:2019vcm}. In heavy ion collisions, this effect was interpreted as resulting from directionally dependent jet quenching, driven by the anisotropic shape of the underlying medium. However, it is unlikely that there can be differential jet quenching as a function of orientation relative to the QGP geometry if there is no jet quenching in $p$+Pb collisions in the first place.\cite{Park:2016jap}

Thus, there are two related outstanding puzzles, one being the lack of jet quenching observed in the spectra, if indeed small droplets of QGP are formed, and the other being the mechanism that leads to high-$p_T$ hadron anisotropies other than differential jet quenching. In recent developments, some progress has been achieved in addressing this issue \cite{JETSCAPE:2023xbc, Soudi:2023epi}.

\section{Further opportunities with small systems}
\label{sec:opportunities}
\begin{figure*}[htbp]
\begin{center}
\includegraphics[width=0.7\textwidth]{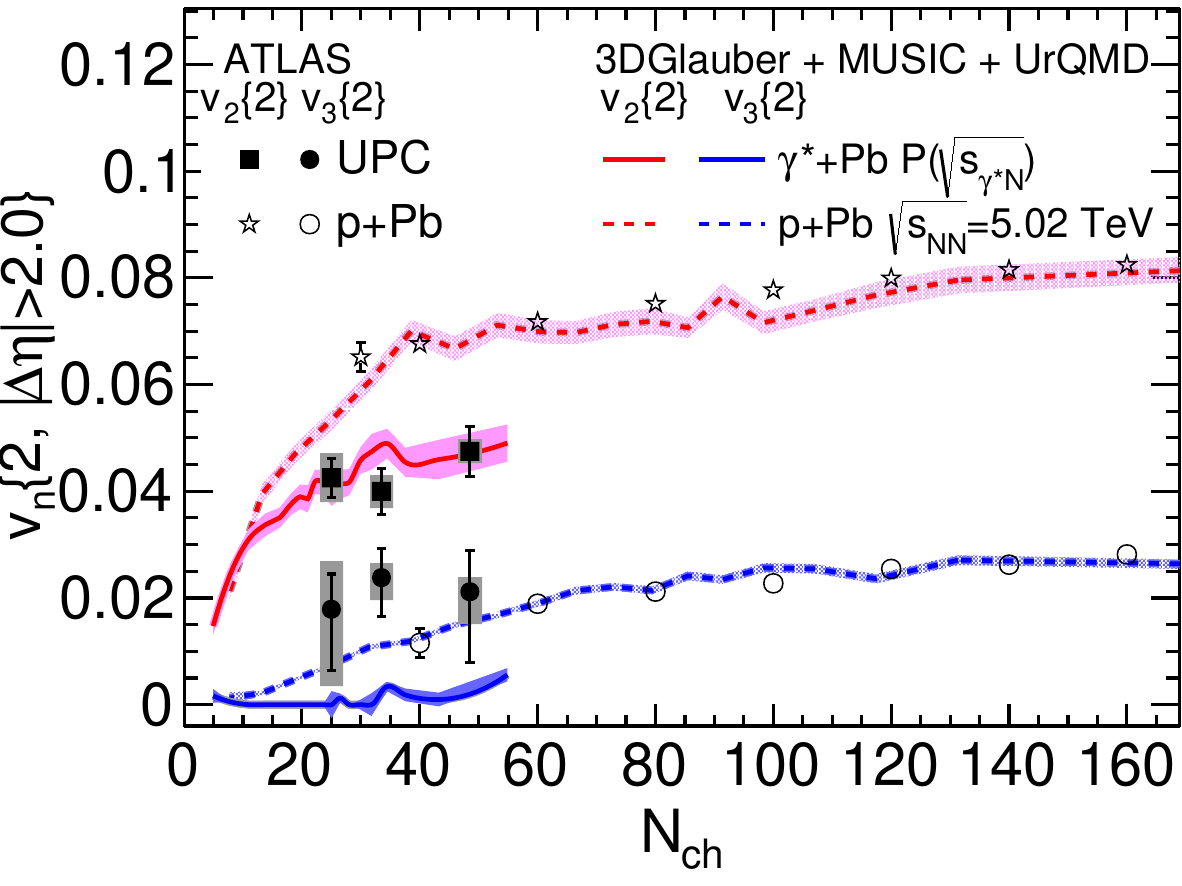}
\end{center}
\caption{ Charged hadron anisotropic flow coefficients $v_2\{2\}$ and $v_3\{2\}$ as functions of charged hadron multiplicity $N_{\rm ch}$ in $p$+Pb (dashed lines) and $\gamma^*$+Pb (solid lines) collisions at LHC energies from (3+1)D hydrodynamics simulations. The ATLAS data is from~\cite{ATLAS:2016yzd,ATLAS:2021jhn}.
This figure is taken from published work. \cite{Zhao:2022ayk}.
}
\label{fig:vnppbupc}
\end{figure*} 

\subsection{Confronting ab initio nuclear structure with small system measurements}

The collision geometry of small systems is dominated by the shape of light nuclei, which can be computed using ab initio methods in nuclear structure theory~\cite{Bender:2003jk}. Therefore, small collision systems, especially light+heavy collisions, can serve as excellent cameras for the light colliding nuclei~\cite{Niel:2023zij}. Unlike low-energy nuclear experiments, relativistic nuclear collisions can probe the multi-nucleon wavefunction at the collision impact~\cite{Giacalone:2023hwk, Ryssens:2023fkv, Mantysaari:2023qsq}. Therefore, precision measurements on anisotropic flow coefficients can set complementary constraints on the multi-particle correlations inside the nuclei. On the other hand, the precision nuclear structure inputs to relativistic nuclear collisions can significantly reduce the theoretical uncertainty in modeling the collisions of small systems. 

\subsection{Accessing features of saturated QCD systems}

An alternative approach to unraveling the nature of long-range azimuthal anisotropies in small systems emerges within the framework of the color glass condensate effective theory.  In this framework, the effective degrees of freedom are color sources at large $x$ and gauge fields at small $x$. At high energies, the former are frozen color configurations, randomly distributed event-by-event due to time dilation. The latter are dynamic fields coupled to the static color sources. %It is the stochastic nature of the sources, combined with the separation of time scales. 
In the CGC, small $x$ gluons saturate with large occupation numbers, and their typical momenta peak at the saturation scale $Q_s$. These gluon fields exhibit correlations within a characteristic length of $1/Q_s$, which can be interpreted as color field domains of that size.
Within this picture, given an initial non-perturbative distribution of sources at an initial scale, one can compute systematically $n$-point gluon correlation functions and their evolution with $x$ order by order in perturbation theory \cite{Gelis:2010nm}.

For the anisotropic flow in small systems, it was demonstrated that  multi-gluon production from the CGC gives rise to notable correlations in azimuthal anisotropies across long-range rapidities in momentum space without the need for final state interactions.\cite{Krasnitz:2002ng,Gelis:2008ad,Dumitru:2008wn}  Recent calculations also have demonstrated that CGC effective theory can capture certain collective features in small systems, such as multiparticle cumulants \cite{Dusling:2018hsg} and the mass ordering of anisotropy coefficients \cite{Schenke:2016lrs,Greif:2020rhi}. However, as alluded to earlier, the CGC calculations struggle with reproducing the magnitudes or the correct systematics of the experimental data with multiplicity or collision system (e.g., $p$+Au, $d$+Au, $^3$He+Au at RHIC \cite{PHENIX:2018lia}). This suggests that final state effects are necessary to describe the data, at least for high multiplicities regions in small systems.

\subsection{Probing the boundaries of collective behavior}

Intriguing experimental results on two-particle azimuthal correlations in ultra-peripheral Pb+Pb collisions (UPCs) at the LHC have recently appeared~\cite{ATLAS:2021jhn}. These UPCs involve appreciable rates of photo-nuclear interactions~\cite{Bertulani:2005ru,Baltz:2007kq}, and the ATLAS measurements of such photo-nuclear ($\gamma^*$+Pb) interactions in Pb+Pb UPCs indicate the persistence of collective phenomena with the strength of correlations comparable to that observed in proton-proton and proton-lead collisions in similar multiplicity ranges~\cite{ATLAS:2021jhn}.

On the theoretical front, hydrodynamic simulations with the appropriate initial state model have proven effective in describing the systematic behavior of measured azimuthal anisotropies. These simulations span a wide range of systems, from high-energy heavy-ion collisions to small hadronic collision systems producing only a limited number of charged hadrons per unit of rapidity. 
Collective QGP signatures in $\gamma^*$+Pb collisions at the LHC were explored using a full (3+1)D dynamical framework\cite{Shen:2022oyg} with hydrodynamics and hadronic transport.\cite{Zhao:2022ayk} Fig. \ref{fig:vnppbupc} shows the multiplicity dependence of the $p_T$-integrated anisotropic flow coefficients $v_2\{2\}$ and $v_3\{2\}$ computed with two subevents with the $|\Delta\eta| > 2.0$ for $\gamma^{*}$+Pb and $p$+Pb collisions.
Extrapolating from $p$+Pb to $\gamma^{*}+$Pb collisions,  hydrodynamic calculations,  assuming strong final-state interactions, reproduce the hierarchy observed for the elliptic flow coefficient $v_2$ in $N_\mathrm{ch} \in [20, 60]$ in the ATLAS data.  Within this framework, the elliptic flow hierarchy between $p$+Pb and $\gamma^*$+Pb collisions is dominated by the difference in longitudinal flow decorrelations. The hydrodynamic model predicts  triangular flow in $\gamma^{*}+$Pb collisions  smaller than that in $p$+Pb collisions at the same charged hadron multiplicity, again because of the larger longitudinal decorrelation, which does not agree with the ATLAS data. The magnitude of $v_3\{2\}$ in $\gamma^{*}+$Pb collisions may be sensitive to the vector meson's detailed substructure fluctuations.

The color glass condensate effective theory also generates significant correlations in the initial state of $\gamma^*$+Pb collision systems. It has also been shown to mimic collective behavior to a certain degree \cite{Shi:2020djm}. 
Notably, the hydrodynamic model and the CGC effective theory predict the opposite dependence of the collectivity on the $\gamma^*$'s virtuality $Q^2$. 
In $\gamma^*$+Pb collisions, the average size of the 
projectile vector meson decreases with increasing $Q^2$. Thus, hydrodynamics predicts that vector mesons with smaller virtuality lead to greater elliptic flow coefficients. This is attributed to the increased transverse space for geometric fluctuations, resulting in larger average ellipticities. In contrast, the CGC model predicts that vector mesons with smaller virtuality result in weaker elliptic flow coefficients. This is because the increased number of independent color domains in the transverse space reduces two-particle correlations on average.
Future experiments at an Electron-Ion Collider (EIC) will provide direct access to the photon's virtuality, allowing for systematic tests of predictions from both the hydrodynamic and CGC frameworks. 

More recently, the CMS Collaboration has reported a new measurement of two-particle correlations of particles within a jet cone in $p$+$p$ collisions at 13 TeV \cite{CMS:2023iam}. For low multiplicity events, the extracted $v_2\{2\}$ was found to be consistent with the PYTHIA8 or SHERPA model results,  where the observed correlations arise from short-range correlations associated with the presence of jets or mini-jets. Surprisingly, for very high multiplicity jets, the CMS data reveal an increasing trend in the value of $v_2$.  The PYTHIA8 and SHERPA models, which do not include the long-range collective effects, fail to describe this data. This discrepancy raises intriguing questions about QGP droplets within a jet and calls for further investigations in the future. 

\section{Instead of Conclusions}
\label{sec:conclusion}

Research on the physics of small collision systems is ongoing, with several open theoretical and phenomenological questions. Therefore, instead of conclusions, in this section, we briefly summarize the progress made to date and list some important open questions that need to be studied in the future. 

In small systems, hydrodynamic + hadronic transport models, together with the proper initial conditions and early time evolution, can describe most of the features of the experimental data at a largely quantitative level. 
Nonetheless, challenges persist in hydrodynamic simulations of small systems. Large Knudsen and inverse Reynolds numbers pose significant challenges to applying hydrodynamics in small systems. Additionally, driven by far-from-equilibrium effects, numerical simulations may violate causality, potentially introducing unphysical artifacts. 

Furthermore, a multitude of open questions remains. For example, the sign of the four-particle correlation from hydrodynamic simulations ($c_2\{4\}>0$) conflicts with experimental data in proton-proton collisions. Jet quenching does not seem to be observed in small systems, and the $R_{pA}$ and $v_2(p_T)$ cannot simultaneously be described within any model framework in small systems.
Clear evidence of enhanced production of electromagnetic probes, indicating thermal production, is also still outstanding.

Hydrodynamics is being pushed to the edge of validity in these small colliding systems. This has led to probably the most significant advances in relativistic fluid dynamic theory in recent decades. Yet, more theoretical work is still needed to provide a complete understanding of how collectivity emerges in small systems and if and how it can be consistently described within a hydrodynamic theory. Experimentally, we are moving to even more extreme cases, such as ultraperipheral collisions and even the interior of jets, and the future EIC is likely also to provide new insights into the origins of collective behavior in multi-particle systems governed by quantum chromodynamics.

\section{Acknowledgments}
This work is supported by the U.S. Department of Energy, Office of Science, Office of Nuclear Physics, under DOE Contract No.~DE-SC0012704 (B.S.) and Award No.~DE-SC0021969 (C.S.).
J.N. is partially supported by the U.S. Department of Energy, Office of Science, Office for Nuclear Physics under Award No. DE-SC0023861. C.S. acknowledges a DOE Office of Science Early Career Award. 
W.B.Z. is supported by the National Science Foundation (NSF) under grant numbers ACI-2004571 within the framework of the XSCAPE project of the JETSCAPE collaboration and US DOE under Contract No.~DE-AC02-05CH11231, and within the framework of the Saturated Glue (SURGE) Topical Theory Collaboration.

\bibliographystyle{ws-rv-van}
\bibliography{referencesInspires,References_Jorge,references,References_Jorge1}

\def\cprime{$'$} \def\cprime{$'$}
\begin{thebibliography}{351}
\providecommand{\natexlab}[1]{#1}
\providecommand{\url}[1]{\texttt{#1}}
\expandafter\ifx\csname urlstyle\endcsname\relax
  \providecommand{\doi}[1]{doi: #1}\else
  \providecommand{\doi}{doi: \begingroup \urlstyle{rm}\Url}\fi

\bibitem{Gale:2013da}
C.~Gale, S.~Jeon, and B.~Schenke, {Hydrodynamic Modeling of Heavy-Ion
  Collisions}, \emph{Int. J. Mod. Phys. A}. {\bf 28}, \penalty0 1340011
  (2013).
\newblock \doi{10.1142/S0217751X13400113}.

\bibitem{Romatschke:2017ejr}
P.~Romatschke and U.~Romatschke, \emph{{Relativistic Fluid Dynamics In and Out
  of Equilibrium}}. Cambridge Monographs on Mathematical Physics, Cambridge
  University Press  (5, 2019).
\newblock ISBN 978-1-108-48368-1, 978-1-108-75002-8.
\newblock \doi{10.1017/9781108651998}.

\bibitem{Shen:2020mgh}
C.~Shen and L.~Yan, {Recent development of hydrodynamic modeling in heavy-ion
  collisions}, \emph{Nucl. Sci. Tech.} {\bf 31}\penalty0 (12), \penalty0 122
  (10, 2020).
\newblock \doi{10.1007/s41365-020-00829-z}.

\bibitem{Heinz:2013th}
U.~Heinz and R.~Snellings, {Collective flow and viscosity in relativistic
  heavy-ion collisions}, \emph{Ann. Rev. Nucl. Part. Sci.} {\bf 63}, \penalty0
  123--151  (2013).
\newblock \doi{10.1146/annurev-nucl-102212-170540}.

\bibitem{Shen:2014vra}
C.~Shen, Z.~Qiu, H.~Song, J.~Bernhard, S.~Bass, and U.~Heinz, The iebe-vishnu
  code package for relativistic heavy-ion collisions, \emph{Comput. Phys.
  Commun.} {\bf 199}, \penalty0 61--85  (2016).
\newblock \doi{10.1016/j.cpc.2015.08.039}.

\bibitem{Putschke:2019yrg}
J.~H. Putschke et~al., {The JETSCAPE framework}  (3, 2019).

\bibitem{Nijs:2020roc}
G.~Nijs, W.~van~der Schee, U.~G\"ursoy, and R.~Snellings, {Bayesian analysis of
  heavy ion collisions with the heavy ion computational framework Trajectum},
  \emph{Phys. Rev. C}. {\bf 103}\penalty0 (5), \penalty0 054909  (2021).
\newblock \doi{10.1103/PhysRevC.103.054909}.

\bibitem{Shen:2011eg}
C.~Shen, U.~Heinz, P.~Huovinen, and H.~Song, {Radial and elliptic flow in Pb+Pb
  collisions at the Large Hadron Collider from viscous hydrodynamic},
  \emph{Phys. Rev. C}. {\bf 84}, \penalty0 044903  (2011).
\newblock \doi{10.1103/PhysRevC.84.044903}.

\bibitem{Heinz:2013bua}
U.~Heinz, Z.~Qiu, and C.~Shen, {Fluctuating flow angles and anisotropic flow
  measurements}, \emph{Phys. Rev. C}. {\bf 87}\penalty0 (3), \penalty0 034913
  (2013).
\newblock \doi{10.1103/PhysRevC.87.034913}.

\bibitem{Niemi:2015voa}
H.~Niemi, K.~J. Eskola, R.~Paatelainen, and K.~Tuominen, {Predictions for 5.023
  TeV Pb + Pb collisions at the CERN Large Hadron Collider}, \emph{Phys. Rev.
  C}. {\bf 93}\penalty0 (1), \penalty0 014912  (2016).
\newblock \doi{10.1103/PhysRevC.93.014912}.

\bibitem{Noronha-Hostler:2015uye}
J.~Noronha-Hostler, M.~Luzum, and J.-Y. Ollitrault, {Hydrodynamic predictions
  for 5.02 TeV Pb-Pb collisions}, \emph{Phys. Rev. C}. {\bf 93}\penalty0 (3),
  \penalty0 034912  (2016).
\newblock \doi{10.1103/PhysRevC.93.034912}.

\bibitem{Paquet:2023rfd}
J.-F. Paquet, {Applications of emulation and Bayesian methods in heavy-ion
  physics}  (10, 2023).

\bibitem{ATLAS:2017rtr}
M.~Aaboud et~al., {Measurement of long-range multiparticle azimuthal
  correlations with the subevent cumulant method in $pp$ and $p + Pb$
  collisions with the ATLAS detector at the CERN Large Hadron Collider},
  \emph{Phys. Rev. C}. {\bf 97}\penalty0 (2), \penalty0 024904  (2018).
\newblock \doi{10.1103/PhysRevC.97.024904}.

\bibitem{KrizkovaGajdosova:2020pxc}
K.~K\v{r}\'\i{}\v{z}kov\'a~Gajdo\v{s}ov\'a, {Probing QGP with flow: An
  experimental overview}, \emph{Nucl. Phys. A}. {\bf 1005}, \penalty0 121802
  (2021).
\newblock \doi{10.1016/j.nuclphysa.2020.121802}.

\bibitem{Denicol:2012cn}
G.~S. Denicol, H.~Niemi, E.~Molnar, and D.~H. Rischke, Derivation of transient
  relativistic fluid dynamics from the {B}oltzmann equation, \emph{Phys. Rev.}
  {\bf D85}, \penalty0 114047  (2012).
\newblock \doi{10.1103/PhysRevD.85.114047, 10.1103/PhysRevD.91.039902}.
\newblock [Erratum: Phys. Rev.D91,no.3,039902(2015)].

\bibitem{Rocha:2023ilf}
G.~S. Rocha, D.~Wagner, G.~S. Denicol, J.~Noronha, and D.~H. Rischke, {Theories
  of Relativistic Dissipative Fluid Dynamics}  (11, 2023).

\bibitem{Landau_book}
L.~D. Landau and E.~M. Lifshitz, \emph{Fluid Mechanics}. Pergamon  (1987).
\newblock \doi{https://doi.org/10.1016/B978-0-08-033933-7.50001-5}.

\bibitem{Nagle:2018nvi}
J.~L. Nagle and W.~A. Zajc, {Small System Collectivity in Relativistic Hadronic
  and Nuclear Collisions}, \emph{Ann. Rev. Nucl. Part. Sci.} {\bf 68},
  \penalty0 211--235  (2018).
\newblock \doi{10.1146/annurev-nucl-101916-123209}.

\bibitem{Schenke:2021mxx}
B.~Schenke, {The smallest fluid on Earth}, \emph{Rept. Prog. Phys.} {\bf
  84}\penalty0 (8), \penalty0 082301  (2021).
\newblock \doi{10.1088/1361-6633/ac14c9}.

\bibitem{Miller:2007ri}
M.~L. Miller, K.~Reygers, S.~J. Sanders, and P.~Steinberg, {Glauber modeling in
  high energy nuclear collisions}, \emph{Ann. Rev. Nucl. Part. Sci.} {\bf 57},
  \penalty0 205--243  (2007).
\newblock \doi{10.1146/annurev.nucl.57.090506.123020}.

\bibitem{dEnterria:2020dwq}
D.~d'Enterria and C.~Loizides, {Progress in the Glauber Model at Collider
  Energies}, \emph{Ann. Rev. Nucl. Part. Sci.} {\bf 71}, \penalty0 315--344
  (2021).
\newblock \doi{10.1146/annurev-nucl-102419-060007}.

\bibitem{Bozek:2011if}
P.~Bozek, {Collective flow in p-Pb and d-Pd collisions at TeV energies},
  \emph{Phys. Rev. C}. {\bf 85}, \penalty0 014911  (2012).
\newblock \doi{10.1103/PhysRevC.85.014911}.

\bibitem{Bozek:2012gr}
P.~Bozek and W.~Broniowski, {Correlations from hydrodynamic flow in p-Pb
  collisions}, \emph{Phys. Lett. B}. {\bf 718}, \penalty0 1557--1561  (2013).
\newblock \doi{10.1016/j.physletb.2012.12.051}.

\bibitem{Bozek:2013df}
P.~Bozek and W.~Broniowski, {Size of the emission source and collectivity in
  ultra-relativistic p-Pb collisions}, \emph{Phys. Lett. B}. {\bf 720},
  \penalty0 250--253  (2013).
\newblock \doi{10.1016/j.physletb.2013.02.014}.

\bibitem{Bozek:2013uha}
P.~Bozek and W.~Broniowski, {Collective dynamics in high-energy proton-nucleus
  collisions}, \emph{Phys. Rev. C}. {\bf 88}\penalty0 (1), \penalty0 014903
  (2013).
\newblock \doi{10.1103/PhysRevC.88.014903}.

\bibitem{Bozek:2013ska}
P.~Bozek, W.~Broniowski, and G.~Torrieri, {Mass hierarchy in identified
  particle distributions in proton-lead collisions}, \emph{Phys. Rev. Lett.}
  {\bf 111}, \penalty0 172303  (2013).
\newblock \doi{10.1103/PhysRevLett.111.172303}.

\bibitem{Bzdak:2013zma}
A.~Bzdak, B.~Schenke, P.~Tribedy, and R.~Venugopalan, {Initial state geometry
  and the role of hydrodynamics in proton-proton, proton-nucleus and
  deuteron-nucleus collisions}, \emph{Phys. Rev. C}. {\bf 87}\penalty0 (6),
  \penalty0 064906  (2013).
\newblock \doi{10.1103/PhysRevC.87.064906}.

\bibitem{Qin:2013bha}
G.-Y. Qin and B.~M\"uller, {Elliptic and triangular flow anisotropy in
  deuteron-gold collisions at $\sqrt{s_{NN}}=200$ GeV at RHIC and in
  proton-lead collisions at $\sqrt{s_{NN}}=5.02$ TeV at the LHC}, \emph{Phys.
  Rev. C}. {\bf 89}\penalty0 (4), \penalty0 044902  (2014).
\newblock \doi{10.1103/PhysRevC.89.044902}.

\bibitem{Werner:2013ipa}
K.~Werner, M.~Bleicher, B.~Guiot, I.~Karpenko, and T.~Pierog, {Evidence for
  Flow from Hydrodynamic Simulations of $p$-Pb Collisions at 5.02 TeV from
  $\nu_2$ Mass Splitting}, \emph{Phys. Rev. Lett.} {\bf 112}\penalty0 (23),
  \penalty0 232301  (2014).
\newblock \doi{10.1103/PhysRevLett.112.232301}.

\bibitem{Kozlov:2014fqa}
I.~Kozlov, M.~Luzum, G.~Denicol, S.~Jeon, and C.~Gale, {Transverse momentum
  structure of pair correlations as a signature of collective behavior in small
  collision systems}  (5, 2014).

\bibitem{Romatschke:2015gxa}
P.~Romatschke, {Light-Heavy Ion Collisions: A window into pre-equilibrium QCD
  dynamics?}, \emph{Eur. Phys. J. C}. {\bf 75}\penalty0 (7), \penalty0 305
  (2015).
\newblock \doi{10.1140/epjc/s10052-015-3509-3}.

\bibitem{Shen:2016zpp}
C.~Shen, J.-F. Paquet, G.~S. Denicol, S.~Jeon, and C.~Gale, {Collectivity and
  electromagnetic radiation in small systems}, \emph{Phys. Rev. C}. {\bf
  95}\penalty0 (1), \penalty0 014906  (2017).
\newblock \doi{10.1103/PhysRevC.95.014906}.

\bibitem{Weller:2017tsr}
R.~D. Weller and P.~Romatschke, {One fluid to rule them all: viscous
  hydrodynamic description of event-by-event central p+p, p+Pb and Pb+Pb
  collisions at $\sqrt{s}=5.02$ TeV}, \emph{Phys. Lett. B}. {\bf 774},
  \penalty0 351--356  (2017).
\newblock \doi{10.1016/j.physletb.2017.09.077}.

\bibitem{Giacalone:2017uqx}
G.~Giacalone, J.~Noronha-Hostler, and J.-Y. Ollitrault, {Relative flow
  fluctuations as a probe of initial state fluctuations}, \emph{Phys. Rev. C}.
  {\bf 95}\penalty0 (5), \penalty0 054910  (2017).
\newblock \doi{10.1103/PhysRevC.95.054910}.

\bibitem{Sievert:2019zjr}
M.~D. Sievert and J.~Noronha-Hostler, {CERN Large Hadron Collider system size
  scan predictions for PbPb, XeXe, ArAr, and OO with relativistic
  hydrodynamics}, \emph{Phys. Rev. C}. {\bf 100}\penalty0 (2), \penalty0 024904
   (2019).
\newblock \doi{10.1103/PhysRevC.100.024904}.

\bibitem{Summerfield:2021oex}
N.~Summerfield, B.-N. Lu, C.~Plumberg, D.~Lee, J.~Noronha-Hostler, and
  A.~Timmins, {$^{16}$O $^{16}$O collisions at energies available at the BNL
  Relativistic Heavy Ion Collider and at the CERN Large Hadron Collider
  comparing $\alpha$ clustering versus substructure}, \emph{Phys. Rev. C}. {\bf
  104}\penalty0 (4), \penalty0 L041901  (2021).
\newblock \doi{10.1103/PhysRevC.104.L041901}.

\bibitem{Katz:2019qwv}
R.~Katz, C.~A.~G. Prado, J.~Noronha-Hostler, and A.~A.~P. Suaide, {System-size
  scan of $D$ meson $R_{AA}$ and $v_n$ using PbPb, XeXe, ArAr, and OO
  collisions at energies available at the CERN Large Hadron Collider},
  \emph{Phys. Rev. C}. {\bf 102}\penalty0 (4), \penalty0 041901  (2020).
\newblock \doi{10.1103/PhysRevC.102.041901}.

\bibitem{Moreland:2014oya}
J.~S. Moreland, J.~E. Bernhard, and S.~A. Bass, {Alternative ansatz to wounded
  nucleon and binary collision scaling in high-energy nuclear collisions},
  \emph{Phys. Rev. C}. {\bf 92}\penalty0 (1), \penalty0 011901  (2015).
\newblock \doi{10.1103/PhysRevC.92.011901}.

\bibitem{Schenke:2014zha}
B.~Schenke and R.~Venugopalan, {Eccentric protons? Sensitivity of flow to
  system size and shape in p+p, p+Pb and Pb+Pb collisions}, \emph{Phys. Rev.
  Lett.} {\bf 113}, \penalty0 102301  (2014).
\newblock \doi{10.1103/PhysRevLett.113.102301}.

\bibitem{Bernhard:2016tnd}
J.~E. Bernhard, J.~S. Moreland, S.~A. Bass, J.~Liu, and U.~Heinz, {Applying
  Bayesian parameter estimation to relativistic heavy-ion collisions:
  simultaneous characterization of the initial state and quark-gluon plasma
  medium}, \emph{Phys. Rev. C}. {\bf 94}\penalty0 (2), \penalty0 024907
  (2016).
\newblock \doi{10.1103/PhysRevC.94.024907}.

\bibitem{JETSCAPE:2020shq}
D.~Everett et~al., {Phenomenological constraints on the transport properties of
  QCD matter with data-driven model averaging}, \emph{Phys. Rev. Lett.} {\bf
  126}\penalty0 (24), \penalty0 242301  (2021).
\newblock \doi{10.1103/PhysRevLett.126.242301}.

\bibitem{JETSCAPE:2020mzn}
D.~Everett et~al., {Multisystem Bayesian constraints on the transport
  coefficients of QCD matter}, \emph{Phys. Rev. C}. {\bf 103}\penalty0 (5),
  \penalty0 054904  (2021).
\newblock \doi{10.1103/PhysRevC.103.054904}.

\bibitem{McLerran:1993ni}
L.~D. McLerran and R.~Venugopalan, {Computing quark and gluon distribution
  functions for very large nuclei}, \emph{Phys. Rev. D}. {\bf 49}, \penalty0
  2233--2241  (1994).
\newblock \doi{10.1103/PhysRevD.49.2233}.

\bibitem{McLerran:1993ka}
L.~D. McLerran and R.~Venugopalan, {Gluon distribution functions for very large
  nuclei at small transverse momentum}, \emph{Phys. Rev. D}. {\bf 49},
  \penalty0 3352--3355  (1994).
\newblock \doi{10.1103/PhysRevD.49.3352}.

\bibitem{Kovner:1995ja}
A.~Kovner, L.~D. McLerran, and H.~Weigert, {Gluon production from nonAbelian
  Weizsacker-Williams fields in nucleus-nucleus collisions}, \emph{Phys. Rev.
  D}. {\bf 52}, \penalty0 6231--6237  (1995).
\newblock \doi{10.1103/PhysRevD.52.6231}.

\bibitem{Iancu:2003xm}
E.~Iancu and R.~Venugopalan, \emph{{The Color glass condensate and high-energy
  scattering in QCD}}, In eds. R.~C. Hwa and X.-N. Wang, \emph{{Quark-gluon
  plasma 4}}, pp. 249--3363 (3.
\newblock , 2003).
\newblock \doi{10.1142/9789812795533_0005}.

\bibitem{Lappi:2006hq}
T.~Lappi, {Energy density of the glasma}, \emph{Phys. Lett. B}. {\bf 643},
  \penalty0 11--16  (2006).
\newblock \doi{10.1016/j.physletb.2006.10.017}.

\bibitem{Romatschke:2013re}
P.~Romatschke and J.~D. Hogg, {Pre-Equilibrium Radial Flow from Central
  Shock-Wave Collisions in AdS5}, \emph{JHEP}. {\bf 04}, \penalty0 048  (2013).
\newblock \doi{10.1007/JHEP04(2013)048}.

\bibitem{Moreland:2018gsh}
J.~S. Moreland, J.~E. Bernhard, and S.~A. Bass, {Bayesian calibration of a
  hybrid nuclear collision model using p-Pb and Pb-Pb data at energies
  available at the CERN Large Hadron Collider}, \emph{Phys. Rev. C}. {\bf
  101}\penalty0 (2), \penalty0 024911  (2020).
\newblock \doi{10.1103/PhysRevC.101.024911}.

\bibitem{Mantysaari:2016ykx}
H.~M\"antysaari and B.~Schenke, {Evidence of strong proton shape fluctuations
  from incoherent diffraction}, \emph{Phys. Rev. Lett.} {\bf 117}\penalty0 (5),
  \penalty0 052301  (2016).
\newblock \doi{10.1103/PhysRevLett.117.052301}.

\bibitem{Mantysaari:2016jaz}
H.~M\"antysaari and B.~Schenke, {Revealing proton shape fluctuations with
  incoherent diffraction at high energy}, \emph{Phys. Rev. D}. {\bf
  94}\penalty0 (3), \penalty0 034042  (2016).
\newblock \doi{10.1103/PhysRevD.94.034042}.

\bibitem{Mantysaari:2022ffw}
H.~M\"antysaari, B.~Schenke, C.~Shen, and W.~Zhao, {Bayesian inference of the
  fluctuating proton shape}, \emph{Phys. Lett. B}. {\bf 833}, \penalty0 137348
  (2022).
\newblock \doi{10.1016/j.physletb.2022.137348}.

\bibitem{Mantysaari:2023qsq}
H.~M\"antysaari, B.~Schenke, C.~Shen, and W.~Zhao, {Multiscale Imaging of
  Nuclear Deformation at the Electron-Ion Collider}, \emph{Phys. Rev. Lett.}
  {\bf 131}\penalty0 (6), \penalty0 062301  (2023).
\newblock \doi{10.1103/PhysRevLett.131.062301}.

\bibitem{Chekanov:2002rm}
S.~Chekanov et~al., {Measurement of proton dissociative diffractive
  photoproduction of vector mesons at large momentum transfer at HERA},
  \emph{Eur. Phys. J. C}. {\bf 26}, \penalty0 389--409  (2003).
\newblock \doi{10.1140/epjc/s2002-01079-0}.

\bibitem{Aktas:2003zi}
A.~Aktas et~al., {Diffractive photoproduction of $J/\psi$ mesons with large
  momentum transfer at HERA}, \emph{Phys. Lett. B}. {\bf 568}, \penalty0
  205--218  (2003).
\newblock \doi{10.1016/j.physletb.2003.06.056}.

\bibitem{Aktas:2005xu}
A.~Aktas et~al., {Elastic J/psi production at HERA}, \emph{Eur. Phys. J. C}.
  {\bf 46}, \penalty0 585--603  (2006).
\newblock \doi{10.1140/epjc/s2006-02519-5}.

\bibitem{Chekanov:2002xi}
S.~Chekanov et~al., {Exclusive photoproduction of J / psi mesons at HERA},
  \emph{Eur. Phys. J. C}. {\bf 24}, \penalty0 345--360  (2002).
\newblock \doi{10.1007/s10052-002-0953-7}.

\bibitem{Alexa:2013xxa}
C.~Alexa et~al., {Elastic and Proton-Dissociative Photoproduction of J/psi
  Mesons at HERA}, \emph{Eur. Phys. J. C}. {\bf 73}\penalty0 (6), \penalty0
  2466  (2013).
\newblock \doi{10.1140/epjc/s10052-013-2466-y}.

\bibitem{Mantysaari:2017cni}
H.~M\"antysaari, B.~Schenke, C.~Shen, and P.~Tribedy, {Imprints of fluctuating
  proton shapes on flow in proton-lead collisions at the LHC}, \emph{Phys.
  Lett. B}. {\bf 772}, \penalty0 681--686  (2017).
\newblock \doi{10.1016/j.physletb.2017.07.038}.

\bibitem{Schenke:2019pmk}
B.~Schenke, C.~Shen, and P.~Tribedy, {Hybrid Color Glass Condensate and
  hydrodynamic description of the Relativistic Heavy Ion Collider small system
  scan}, \emph{Phys. Lett. B}. {\bf 803}, \penalty0 135322  (2020).
\newblock \doi{10.1016/j.physletb.2020.135322}.

\bibitem{Schenke:2020mbo}
B.~Schenke, C.~Shen, and P.~Tribedy, {Running the gamut of high energy nuclear
  collisions}, \emph{Phys. Rev. C}. {\bf 102}\penalty0 (4), \penalty0 044905
  (2020).
\newblock \doi{10.1103/PhysRevC.102.044905}.

\bibitem{Borsanyi:2013bia}
S.~Borsanyi, Z.~Fodor, C.~Hoelbling, S.~D. Katz, S.~Krieg, and K.~K. Szabo,
  {Full result for the QCD equation of state with 2+1 flavors}, \emph{Phys.
  Lett. B}. {\bf 730}, \penalty0 99--104  (2014).
\newblock \doi{10.1016/j.physletb.2014.01.007}.

\bibitem{HotQCD:2014kol}
A.~Bazavov et~al., {Equation of state in ( 2+1 )-flavor QCD}, \emph{Phys. Rev.
  D}. {\bf 90}, \penalty0 094503  (2014).
\newblock \doi{10.1103/PhysRevD.90.094503}.

\bibitem{NunesdaSilva:2020bfs}
T.~Nunes~da Silva, D.~Chinellato, M.~Hippert, W.~Serenone, J.~Takahashi, G.~S.
  Denicol, M.~Luzum, and J.~Noronha, {Pre-hydrodynamic evolution and its
  signatures in final-state heavy-ion observables}, \emph{Phys. Rev. C}. {\bf
  103}, \penalty0 054906  (2021).
\newblock \doi{10.1103/PhysRevC.103.054906}.

\bibitem{daSilva:2022xwu}
T.~N. da~Silva, D.~D. Chinellato, A.~V. Giannini, M.~N. Ferreira, G.~S.
  Denicol, M.~Hippert, M.~Luzum, J.~Noronha, and J.~Takahashi, {Prehydrodynamic
  evolution in large and small systems}, \emph{Phys. Rev. C}. {\bf
  107}\penalty0 (4), \penalty0 044901  (2023).
\newblock \doi{10.1103/PhysRevC.107.044901}.

\bibitem{Schenke:2015aqa}
B.~Schenke, S.~Schlichting, and R.~Venugopalan, {Azimuthal anisotropies in
  p$+$Pb collisions from classical Yang\textendash{}Mills dynamics},
  \emph{Phys. Lett. B}. {\bf 747}, \penalty0 76--82  (2015).
\newblock \doi{10.1016/j.physletb.2015.05.051}.

\bibitem{Lappi:2015vta}
T.~Lappi, B.~Schenke, S.~Schlichting, and R.~Venugopalan, {Tracing the origin
  of azimuthal gluon correlations in the color glass condensate}, \emph{JHEP}.
  {\bf 01}, \penalty0 061  (2016).
\newblock \doi{10.1007/JHEP01(2016)061}.

\bibitem{Giacalone:2020byk}
G.~Giacalone, B.~Schenke, and C.~Shen, {Observable signatures of initial state
  momentum anisotropies in nuclear collisions}, \emph{Phys. Rev. Lett.} {\bf
  125}\penalty0 (19), \penalty0 192301  (2020).
\newblock \doi{10.1103/PhysRevLett.125.192301}.

\bibitem{Schenke:2022mjv}
B.~Schenke, S.~Schlichting, and P.~Singh, {Rapidity dependence of initial state
  geometry and momentum correlations in p+Pb collisions}, \emph{Phys. Rev. D}.
  {\bf 105}\penalty0 (9), \penalty0 094023  (2022).
\newblock \doi{10.1103/PhysRevD.105.094023}.

\bibitem{Berges:2014yta}
J.~Berges, B.~Schenke, S.~Schlichting, and R.~Venugopalan, {Turbulent
  thermalization process in high-energy heavy-ion collisions}, \emph{Nucl.
  Phys. A}. {\bf 931}, \penalty0 348--353  (2014).
\newblock \doi{10.1016/j.nuclphysa.2014.08.103}.

\bibitem{Ipp:2020igo}
A.~Ipp and D.~I. M\"uller, {Progress on 3+1D Glasma simulations}, \emph{Eur.
  Phys. J. A}. {\bf 56}\penalty0 (9), \penalty0 243  (2020).
\newblock \doi{10.1140/epja/s10050-020-00241-6}.

\bibitem{McDonald:2023qwc}
S.~McDonald, S.~Jeon, and C.~Gale, {3+1D initialization and evolution of the
  glasma}, \emph{Phys. Rev. C}. {\bf 108}\penalty0 (6), \penalty0 064910
  (2023).
\newblock \doi{10.1103/PhysRevC.108.064910}.

\bibitem{Hirano:2005xf}
T.~Hirano, U.~W. Heinz, D.~Kharzeev, R.~Lacey, and Y.~Nara, {Hadronic
  dissipative effects on elliptic flow in ultrarelativistic heavy-ion
  collisions}, \emph{Phys. Lett. B}. {\bf 636}, \penalty0 299--304  (2006).
\newblock \doi{10.1016/j.physletb.2006.03.060}.

\bibitem{Bozek:2010vz}
P.~Bozek, W.~Broniowski, and J.~Moreira, {Torqued fireballs in relativistic
  heavy-ion collisions}, \emph{Phys. Rev. C}. {\bf 83}, \penalty0 034911
  (2011).
\newblock \doi{10.1103/PhysRevC.83.034911}.

\bibitem{Ke:2016jrd}
W.~Ke, J.~S. Moreland, J.~E. Bernhard, and S.~A. Bass, {Constraints on
  rapidity-dependent initial conditions from charged particle pseudorapidity
  densities and two-particle correlations}, \emph{Phys. Rev. C}. {\bf
  96}\penalty0 (4), \penalty0 044912  (2017).
\newblock \doi{10.1103/PhysRevC.96.044912}.

\bibitem{Barej:2019xef}
M.~Barej, A.~Bzdak, and P.~Gutowski, {Wounded nucleon, quark, and quark-diquark
  emission functions versus experimental results from the BNL Relativistic
  Heavy Ion Collider at $\sqrt {s_{NN}}$ =200 GeV}, \emph{Phys. Rev. C}. {\bf
  100}\penalty0 (6), \penalty0 064902  (2019).
\newblock \doi{10.1103/PhysRevC.100.064902}.

\bibitem{Wu:2021fjf}
X.-Y. Wu, G.-Y. Qin, L.-G. Pang, and X.-N. Wang, {(3+1)-D viscous hydrodynamics
  at finite net baryon density: Identified particle spectra, anisotropic flows,
  and flow fluctuations across energies relevant to the beam-energy scan at
  RHIC}, \emph{Phys. Rev. C}. {\bf 105}\penalty0 (3), \penalty0 034909  (2022).
\newblock \doi{10.1103/PhysRevC.105.034909}.

\bibitem{Du:2022yok}
L.~Du, C.~Shen, S.~Jeon, and C.~Gale, {Probing initial baryon stopping and
  equation~of state with rapidity-dependent directed flow of identified
  particles}, \emph{Phys. Rev. C}. {\bf 108}\penalty0 (4), \penalty0 L041901
  (2023).
\newblock \doi{10.1103/PhysRevC.108.L041901}.

\bibitem{Jiang:2023vxp}
Z.-F. Jiang, X.-Y. Wu, S.~Cao, and B.-W. Zhang, {Hyperon polarization and its
  relation with directed flow in high-energy nuclear collisions}, \emph{Phys.
  Rev. C}. {\bf 108}\penalty0 (6), \penalty0 064904  (2023).
\newblock \doi{10.1103/PhysRevC.108.064904}.

\bibitem{Shen:2020jwv}
C.~Shen and S.~Alzhrani, {Collision-geometry-based 3D initial condition for
  relativistic heavy-ion collisions}, \emph{Phys. Rev. C}. {\bf 102}\penalty0
  (1), \penalty0 014909  (2020).
\newblock \doi{10.1103/PhysRevC.102.014909}.

\bibitem{Soeder:2023vdn}
D.~Soeder, W.~Ke, J.~F. Paquet, and S.~A. Bass, {Bayesian parameter estimation
  with a new three-dimensional initial-conditions model for ultrarelativistic
  heavy-ion collisions}  (6, 2023).

\bibitem{Lin:2004en}
Z.-W. Lin, C.~M. Ko, B.-A. Li, B.~Zhang, and S.~Pal, {A Multi-phase transport
  model for relativistic heavy ion collisions}, \emph{Phys. Rev. C}. {\bf 72},
  \penalty0 064901  (2005).
\newblock \doi{10.1103/PhysRevC.72.064901}.

\bibitem{Bozek:2014cya}
P.~Bozek and W.~Broniowski, {Collective flow in ultrarelativistic $^3$He-Au
  collisions}, \emph{Phys. Lett. B}. {\bf 739}, \penalty0 308--312  (2014).
\newblock \doi{10.1016/j.physletb.2014.11.006}.

\bibitem{Shen:2022oyg}
C.~Shen and B.~Schenke, {Longitudinal dynamics and particle production in
  relativistic nuclear collisions}, \emph{Phys. Rev. C}. {\bf 105}\penalty0
  (6), \penalty0 064905  (2022).
\newblock \doi{10.1103/PhysRevC.105.064905}.

\bibitem{Zhao:2022ugy}
W.~Zhao, S.~Ryu, C.~Shen, and B.~Schenke, {3D structure of anisotropic flow in
  small collision systems at energies available at the BNL Relativistic Heavy
  Ion Collider}, \emph{Phys. Rev. C}. {\bf 107}\penalty0 (1), \penalty0 014904
  (2023).
\newblock \doi{10.1103/PhysRevC.107.014904}.

\bibitem{Carzon:2019qja}
P.~Carzon, M.~Martinez, M.~D. Sievert, D.~E. Wertepny, and J.~Noronha-Hostler,
  {Monte Carlo event generator for initial conditions of conserved charges in
  nuclear geometry}, \emph{Phys. Rev. C}. {\bf 105}\penalty0 (3), \penalty0
  034908  (2022).
\newblock \doi{10.1103/PhysRevC.105.034908}.

\bibitem{Du:2023msv}
L.~Du, C.~Shen, S.~Jeon, and C.~Gale.
\newblock {Constraints on initial baryon stopping and equation of state from
  directed flow}.
\newblock In \emph{{30th International Conference on Ultrarelativstic
  Nucleus-Nucleus Collisions}}  (12, 2023).

\bibitem{Pihan:2023dsb}
G.~Pihan, A.~Monnai, B.~Schenke, and C.~Shen.
\newblock {Tracing baryon and electric charge transport in isobar collisions}
  (12.
\newblock , 2023).

\bibitem{Arnold:2014jva}
P.~Arnold, P.~Romatschke, and W.~van~der Schee, {Absence of a local rest frame
  in far from equilibrium quantum matter}, \emph{JHEP}. {\bf 10}, \penalty0 110
   (2014).
\newblock \doi{10.1007/JHEP10(2014)110}.

\bibitem{Rougemont:2021qyk}
R.~Rougemont, J.~Noronha, W.~Barreto, G.~S. Denicol, and T.~Dore, {Violation of
  energy conditions and entropy production in holographic Bjorken flow},
  \emph{Phys. Rev. D}. {\bf 104}\penalty0 (12), \penalty0 126012  (2021).
\newblock \doi{10.1103/PhysRevD.104.126012}.

\bibitem{Rougemont:2021gjm}
R.~Rougemont, W.~Barreto, and J.~Noronha, {Hydrodynamization times of a
  holographic fluid far from equilibrium}, \emph{Phys. Rev. D}. {\bf
  105}\penalty0 (4), \penalty0 046009  (2022).
\newblock \doi{10.1103/PhysRevD.105.046009}.

\bibitem{Niemi:2014wta}
H.~Niemi and G.~S. Denicol, {How large is the Knudsen number reached in fluid
  dynamical simulations of ultrarelativistic heavy ion collisions?}  (4, 2014).

\bibitem{Noronha-Hostler:2015coa}
J.~Noronha-Hostler, J.~Noronha, and M.~Gyulassy, {Sensitivity of flow harmonics
  to subnucleon scale fluctuations in heavy ion collisions}, \emph{Phys. Rev.
  C}. {\bf 93}\penalty0 (2), \penalty0 024909  (2016).
\newblock \doi{10.1103/PhysRevC.93.024909}.

\bibitem{Inghirami:2022afu}
G.~Inghirami and H.~Elfner, {The applicability of hydrodynamics in heavy ion
  collisions at $\sqrt{s_\mathrm{NN}}$~=~2.4\textendash{}7.7~GeV}, \emph{Eur.
  Phys. J. C}. {\bf 82}\penalty0 (9), \penalty0 796  (2022).
\newblock \doi{10.1140/epjc/s10052-022-10718-x}.

\bibitem{Broniowski:2008qk}
W.~Broniowski, W.~Florkowski, M.~Chojnacki, and A.~Kisiel, {Free-streaming
  approximation in early dynamics of relativistic heavy-ion collisions},
  \emph{Phys. Rev. C}. {\bf 80}, \penalty0 034902  (2009).
\newblock \doi{10.1103/PhysRevC.80.034902}.

\bibitem{Liu:2015nwa}
J.~Liu, C.~Shen, and U.~Heinz, {Pre-equilibrium evolution effects on heavy-ion
  collision observables}, \emph{Phys. Rev. C}. {\bf 91}\penalty0 (6), \penalty0
  064906  (2015).
\newblock \doi{10.1103/PhysRevC.91.064906}.
\newblock [Erratum: Phys.Rev.C 92, 049904 (2015)].

\bibitem{Kurkela:2018vqr}
A.~Kurkela, A.~Mazeliauskas, J.-F. Paquet, S.~Schlichting, and D.~Teaney,
  {Effective kinetic description of event-by-event pre-equilibrium dynamics in
  high-energy heavy-ion collisions}, \emph{Phys. Rev. C}. {\bf 99}\penalty0
  (3), \penalty0 034910  (2019).
\newblock \doi{10.1103/PhysRevC.99.034910}.

\bibitem{Kurkela:2018wud}
A.~Kurkela, A.~Mazeliauskas, J.-F. Paquet, S.~Schlichting, and D.~Teaney,
  {Matching the Nonequilibrium Initial Stage of Heavy Ion Collisions to
  Hydrodynamics with QCD Kinetic Theory}, \emph{Phys. Rev. Lett.} {\bf
  122}\penalty0 (12), \penalty0 122302  (2019).
\newblock \doi{10.1103/PhysRevLett.122.122302}.

\bibitem{Arnold:2002zm}
P.~B. Arnold, G.~D. Moore, and L.~G. Yaffe, {Effective kinetic theory for high
  temperature gauge theories}, \emph{JHEP}. {\bf 01}, \penalty0 030  (2003).
\newblock \doi{10.1088/1126-6708/2003/01/030}.

\bibitem{Liyanage:2022nua}
D.~Liyanage, D.~Everett, C.~Chattopadhyay, and U.~Heinz, {Prehydrodynamic
  evolution and its impact on quark-gluon plasma signatures}, \emph{Phys. Rev.
  C}. {\bf 105}\penalty0 (6), \penalty0 064908  (2022).
\newblock \doi{10.1103/PhysRevC.105.064908}.

\bibitem{Jaiswal:2021uvv}
S.~Jaiswal, C.~Chattopadhyay, L.~Du, U.~Heinz, and S.~Pal, {Nonconformal
  kinetic theory and hydrodynamics for Bjorken flow}, \emph{Phys. Rev. C}. {\bf
  105}\penalty0 (2), \penalty0 024911  (2022).
\newblock \doi{10.1103/PhysRevC.105.024911}.

\bibitem{Kanakubo:2019ogh}
Y.~Kanakubo, Y.~Tachibana, and T.~Hirano, {Unified description of hadron yield
  ratios from dynamical core-corona initialization}, \emph{Phys. Rev. C}. {\bf
  101}\penalty0 (2), \penalty0 024912  (2020).
\newblock \doi{10.1103/PhysRevC.101.024912}.

\bibitem{Werner:2023jps}
K.~Werner, {Core-corona procedure and microcanonical hadronization to
  understand strangeness enhancement in proton-proton and heavy ion collisions
  in the EPOS4 framework}  (6, 2023).

\bibitem{Heller:2015dha}
M.~P. Heller and M.~Spalinski, {Hydrodynamics Beyond the Gradient Expansion:
  Resurgence and Resummation}, \emph{Phys. Rev. Lett.} {\bf 115}\penalty0 (7),
  \penalty0 072501  (2015).
\newblock \doi{10.1103/PhysRevLett.115.072501}.

\bibitem{Strickland:2017kux}
M.~Strickland, J.~Noronha, and G.~Denicol, {Anisotropic nonequilibrium
  hydrodynamic attractor}, \emph{Phys. Rev. D}. {\bf 97}\penalty0 (3),
  \penalty0 036020  (2018).
\newblock \doi{10.1103/PhysRevD.97.036020}.

\bibitem{Giacalone:2019ldn}
G.~Giacalone, A.~Mazeliauskas, and S.~Schlichting, {Hydrodynamic attractors,
  initial state energy and particle production in relativistic nuclear
  collisions}, \emph{Phys. Rev. Lett.} {\bf 123}\penalty0 (26), \penalty0
  262301  (2019).
\newblock \doi{10.1103/PhysRevLett.123.262301}.

\bibitem{Jankowski:2023fdz}
J.~Jankowski and M.~Spali\'nski, {Hydrodynamic attractors in ultrarelativistic
  nuclear collisions}, \emph{Prog. Part. Nucl. Phys.} {\bf 132}, \penalty0
  104048  (2023).
\newblock \doi{10.1016/j.ppnp.2023.104048}.

\bibitem{Monnai:2021kgu}
A.~Monnai, B.~Schenke, and C.~Shen, {QCD Equation of State at Finite Chemical
  Potentials for Relativistic Nuclear Collisions}, \emph{Int. J. Mod. Phys. A}.
  {\bf 36}\penalty0 (07), \penalty0 2130007  (2021).
\newblock \doi{10.1142/S0217751X21300076}.

\bibitem{EckartViscous}
C.~Eckart, The thermodynamics of irreversible processes {III}. {R}elativistic
  theory of the simple fluid, \emph{Physical Review}. {\bf 58}, \penalty0
  919--924  (1940).

\bibitem{PichonViscous}
G.~Pichon, {\'E}tude relativiste de fluides visqueux et charg{\'e}s,
  \emph{Annales de l'I.H.P. Physique th{\'e}orique}. {\bf 2}\penalty0 (1),
  \penalty0 21--85  (1965).
\newblock URL \url{http://www.numdam.org/item/AIHPA_1965__2_1_21_0}.

\bibitem{Hiscock_Lindblom_instability_1985}
W.~A. Hiscock and L.~Lindblom, Generic instabilities in first-order dissipative
  fluid theories, \emph{Phys. Rev. D}. {\bf 31}\penalty0 (4), \penalty0
  725--733  (1985).

\bibitem{Bemfica:2020zjp}
F.~S. Bemfica, M.~M. Disconzi, and J.~Noronha, {First-Order
  General-Relativistic Viscous Fluid Dynamics}, \emph{Phys. Rev. X}. {\bf
  12}\penalty0 (2), \penalty0 021044  (2022).
\newblock \doi{10.1103/PhysRevX.12.021044}.

\bibitem{Gavassino:2021owo}
L.~Gavassino, {Can We Make Sense of Dissipation without Causality?},
  \emph{Phys. Rev. X}. {\bf 12}\penalty0 (4), \penalty0 041001  (2022).
\newblock \doi{10.1103/PhysRevX.12.041001}.

\bibitem{BemficaDisconziNoronha}
F.~S. Bemfica, M.~M. Disconzi, and J.~Noronha, Causality and existence of
  solutions of relativistic viscous fluid dynamics with gravity, \emph{Phys.
  Rev.} {\bf D98}\penalty0 (10), \penalty0 104064 (26 pages)  (2018).
\newblock \doi{10.1103/PhysRevD.98.104064}.

\bibitem{Kovtun:2019hdm}
P.~Kovtun, First-order relativistic hydrodynamics is stable  (2019).

\bibitem{Bemfica:2019knx}
F.~S. Bemfica, F.~S. Bemfica, M.~M. Disconzi, M.~M. Disconzi, J.~Noronha, and
  J.~Noronha, {Nonlinear Causality of General First-Order Relativistic Viscous
  Hydrodynamics}, \emph{Phys. Rev. D}. {\bf 100}\penalty0 (10), \penalty0
  104020  (2019).
\newblock \doi{10.1103/PhysRevD.100.104020}.
\newblock [Erratum: Phys.Rev.D 105, 069902 (2022)].

\bibitem{Hoult:2020eho}
R.~E. Hoult and P.~Kovtun, {Stable and causal relativistic Navier-Stokes
  equations}, \emph{JHEP}. {\bf 06}, \penalty0 067  (2020).
\newblock \doi{10.1007/JHEP06(2020)067}.

\bibitem{Kovtun:2012rj}
K.~Kovtun, Lectures on hydrodynamic fluctuations in relativistic theories,
  \emph{J. Phys.} {\bf A45}, \penalty0 473001  (2012).
\newblock \doi{10.1088/1751-8113/45/47/473001}.

\bibitem{Pandya:2021ief}
A.~Pandya and F.~Pretorius, {Numerical exploration of first-order relativistic
  hydrodynamics}, \emph{Phys. Rev. D}. {\bf 104}\penalty0 (2), \penalty0 023015
   (2021).
\newblock \doi{10.1103/PhysRevD.104.023015}.

\bibitem{Pandya:2022pif}
A.~Pandya, E.~R. Most, and F.~Pretorius, {Conservative finite volume scheme for
  first-order viscous relativistic hydrodynamics}, \emph{Phys. Rev. D}. {\bf
  105}\penalty0 (12), \penalty0 123001  (2022).
\newblock \doi{10.1103/PhysRevD.105.123001}.

\bibitem{Pandya:2022sff}
A.~Pandya, E.~R. Most, and F.~Pretorius, {Causal, stable first-order viscous
  relativistic hydrodynamics with ideal gas microphysics}, \emph{Phys. Rev. D}.
  {\bf 106}\penalty0 (12), \penalty0 123036  (2022).
\newblock \doi{10.1103/PhysRevD.106.123036}.

\bibitem{Bantilan:2022ech}
H.~Bantilan, Y.~Bea, and P.~Figueras, {Evolutions in first-order viscous
  hydrodynamics}, \emph{JHEP}. {\bf 08}, \penalty0 298  (2022).
\newblock \doi{10.1007/JHEP08(2022)298}.

\bibitem{Rocha:2022ind}
G.~S. Rocha, G.~S. Denicol, and J.~Noronha, {Perturbative approaches in
  relativistic kinetic theory and the emergence of first-order hydrodynamics},
  \emph{Phys. Rev. D}. {\bf 106}\penalty0 (3), \penalty0 036010  (2022).
\newblock \doi{10.1103/PhysRevD.106.036010}.

\bibitem{MIS-6}
W.~Israel and J.~M. Stewart, Transient relativistic thermodynamics and kinetic
  theory, \emph{Ann. Phys.} {\bf 118}, \penalty0 341--372  (1979).

\bibitem{Baier:2007ix}
R.~Baier, P.~Romatschke, D.~T. Son, A.~O. Starinets, and M.~A. Stephanov,
  Relativistic viscous hydrodynamics, conformal invariance, and holography,
  \emph{JHEP}. {\bf 04}, \penalty0 100  (2008).
\newblock \doi{10.1088/1126-6708/2008/04/100}.

\bibitem{Alqahtani:2017mhy}
M.~Alqahtani, M.~Nopoush, and M.~Strickland, Relativistic anisotropic
  hydrodynamics, \emph{Prog. Part. Nucl. Phys.} {\bf 101}, \penalty0 204--248
  (2018).
\newblock \doi{10.1016/j.ppnp.2018.05.004}.

\bibitem{Luzum:2008cw}
M.~Luzum and P.~Romatschke, Conformal relativistic viscous hydrodynamics:
  Applications to rhic results at s(nn)**(1/2) = 200-gev, \emph{Phys. Rev.}
  {\bf C78}, \penalty0 034915  (2008).
\newblock \doi{10.1103/PhysRevC.78.034915, 10.1103/PhysRevC.79.039903}.
\newblock [Erratum: Phys. Rev.C79,039903(2009)].

\bibitem{Schenke:2010nt}
B.~Schenke, S.~Jeon, and C.~Gale, {(3+1)D hydrodynamic simulation of
  relativistic heavy-ion collisions}, \emph{Phys. Rev. C}. {\bf 82}, \penalty0
  014903  (2010).
\newblock \doi{10.1103/PhysRevC.82.014903}.

\bibitem{Schenke:2010rr}
B.~Schenke, S.~Jeon, and C.~Gale, Elliptic and triangular flow in
  event-by-event (3+1)d viscous hydrodynamics, \emph{Phys. Rev. Lett.} {\bf
  106}, \penalty0 042301  (2011).
\newblock \doi{10.1103/PhysRevLett.106.042301}.

\bibitem{Noronha-Hostler:2013gga}
J.~Noronha-Hostler, G.~S. Denicol, J.~Noronha, R.~P.~G. Andrade, and F.~Grassi,
  Bulk viscosity effects in event-by-event relativistic hydrodynamics,
  \emph{Phys. Rev.} {\bf C88}\penalty0 (4), \penalty0 044916  (2013).
\newblock \doi{10.1103/PhysRevC.88.044916}.

\bibitem{Noronha-Hostler:2014dqa}
J.~Noronha-Hostler, J.~Noronha, and F.~Grassi, Bulk viscosity-driven
  suppression of shear viscosity effects on the flow harmonics at energies
  available at the bnl relativistic heavy ion collider, \emph{Phys. Rev.} {\bf
  C90}\penalty0 (3), \penalty0 034907  (2014).
\newblock \doi{10.1103/PhysRevC.90.034907}.

\bibitem{Marrochio:2013wla}
H.~Marrochio, J.~Noronha, G.~S. Denicol, M.~Luzum, S.~Jeon, and C.~Gale,
  Solutions of conformal israel-stewart relativistic viscous fluid dynamics,
  \emph{Phys. Rev.} {\bf C91}\penalty0 (1), \penalty0 014903  (2015).
\newblock \doi{10.1103/PhysRevC.91.014903}.

\bibitem{brennen_2013}
C.~E. Brennen, \emph{Cavitation and Bubble Dynamics}. Cambridge University
  Press  (2013).
\newblock \doi{10.1017/CBO9781107338760}.

\bibitem{Torrieri:2007fb}
G.~Torrieri, B.~Tomasik, and I.~Mishustin, Bulk viscosity driven clusterization
  of quark-gluon plasma and early freeze-out in relativistic heavy-ion
  collisions, \emph{Phys. Rev.} {\bf C77}, \penalty0 034903  (2008).
\newblock \doi{10.1103/PhysRevC.77.034903}.

\bibitem{Rajagopal:2009yw}
K.~Rajagopal and N.~Tripuraneni, Bulk viscosity and cavitation in
  boost-invariant hydrodynamic expansion, \emph{JHEP}. {\bf 03}, \penalty0 018
  (2010).
\newblock \doi{10.1007/JHEP03(2010)018}.

\bibitem{Habich:2014tpa}
M.~Habich and P.~Romatschke, {Onset of cavitation in the quark-gluon plasma},
  \emph{JHEP}. {\bf 12}, \penalty0 054  (2014).
\newblock \doi{10.1007/JHEP12(2014)054}.

\bibitem{Denicol:2015bpa}
G.~S. Denicol, C.~Gale, and S.~Jeon, The domain of validity of fluid dynamics
  and the onset of cavitation in ultrarelativistic heavy ion collisions,
  \emph{PoS}. {\bf CPOD2014}, \penalty0 033  (2015).

\bibitem{Byres:2019xld}
M.~Byres, S.~H. Lim, C.~McGinn, J.~Ouellette, and J.~L. Nagle, {Bulk viscosity
  and cavitation in heavy ion collisions}, \emph{Phys. Rev. C}. {\bf
  101}\penalty0 (4), \penalty0 044902  (2020).
\newblock \doi{10.1103/PhysRevC.101.044902}.

\bibitem{Bemfica:2019cop}
F.~S. Bemfica, M.~M. Disconzi, and J.~Noronha, {Causality of the
  Einstein-Israel-Stewart Theory with Bulk Viscosity}, \emph{Phys. Rev. Lett.}
  {\bf 122}\penalty0 (22), \penalty0 221602  (2019).
\newblock \doi{10.1103/PhysRevLett.122.221602}.

\bibitem{Gavassino:2023mad}
L.~Gavassino, M.~M. Disconzi, and J.~Noronha, {Dispersion relations alone
  cannot guarantee causality}  (7, 2023).

\bibitem{Wang:2023csj}
D.-L. Wang and S.~Pu, {Stability and causality criteria in linear mode
  analysis: stability means causality}  (9, 2023).

\bibitem{Hoult:2023clg}
R.~E. Hoult and P.~Kovtun, {Causality and classical dispersion relations}  (9,
  2023).

\bibitem{Hiscock_Lindblom_stability_1983}
W.~A. Hiscock and L.~Lindblom, Stability and causality in dissipative
  relativistic fluids, \emph{Annals of Physics}. {\bf 151}\penalty0 (2),
  \penalty0 466--496  (1983).

\bibitem{Olson:1989ey}
T.~S. Olson, Stability and causality in the {I}srael-{S}tewart energy frame
  theory, \emph{Annals Phys.} {\bf 199}, \penalty0 18  (1990).
\newblock \doi{10.1016/0003-4916(90)90366-V}.

\bibitem{Denicol:2008ha}
G.~S. Denicol, T.~Kodama, T.~Koide, and P.~Mota, Stability and causality in
  relativistic dissipative hydrodynamics, \emph{J. Phys.} {\bf G35}, \penalty0
  115102  (2008).
\newblock \doi{10.1088/0954-3899/35/11/115102}.

\bibitem{Pu:2009fj}
S.~Pu, T.~Koide, and D.~H. Rischke, Does stability of relativistic dissipative
  fluid dynamics imply causality?, \emph{Phys. Rev.} {\bf D81}, \penalty0
  114039  (2010).
\newblock \doi{10.1103/PhysRevD.81.114039}.

\bibitem{Gavassino:2021cli}
L.~Gavassino, {Applying the Gibbs stability criterion to relativistic
  hydrodynamics}, \emph{Class. Quant. Grav.} {\bf 38}\penalty0 (21), \penalty0
  21LT02  (2021).
\newblock \doi{10.1088/1361-6382/ac2b0e}.

\bibitem{Gavassino:2021kjm}
L.~Gavassino, M.~Antonelli, and B.~Haskell, {Thermodynamic Stability Implies
  Causality}, \emph{Phys. Rev. Lett.} {\bf 128}\penalty0 (1), \penalty0 010606
  (2022).
\newblock \doi{10.1103/PhysRevLett.128.010606}.

\bibitem{Abboud:2023hos}
N.~Abboud, E.~Speranza, and J.~Noronha, {Causal and stable first-order chiral
  hydrodynamics}  (8, 2023).

\bibitem{Bemfica:2020xym}
F.~S. Bemfica, M.~M. Disconzi, V.~Hoang, J.~Noronha, and M.~Radosz, {Nonlinear
  Constraints on Relativistic Fluids Far from Equilibrium}, \emph{Phys. Rev.
  Lett.} {\bf 126}\penalty0 (22), \penalty0 222301  (2021).
\newblock \doi{10.1103/PhysRevLett.126.222301}.

\bibitem{Plumberg:2021bme}
C.~Plumberg, D.~Almaalol, T.~Dore, J.~Noronha, and J.~Noronha-Hostler,
  {Causality violations in realistic simulations of heavy-ion collisions},
  \emph{Phys. Rev. C}. {\bf 105}\penalty0 (6), \penalty0 L061901  (2022).
\newblock \doi{10.1103/PhysRevC.105.L061901}.

\bibitem{Chiu:2021muk}
C.~Chiu and C.~Shen, {Exploring theoretical uncertainties in the hydrodynamic
  description of relativistic heavy-ion collisions}, \emph{Phys. Rev. C}. {\bf
  103}\penalty0 (6), \penalty0 064901  (2021).
\newblock \doi{10.1103/PhysRevC.103.064901}.

\bibitem{Krupczak:2023jpa}
R.~Krupczak et~al., {Causality violations in simulations of large and small
  heavy-ion collisions}  (11, 2023).

\bibitem{Shen:2020gef}
C.~Shen, {Studying QGP with flow: A theory overview}, \emph{Nucl. Phys. A}.
  {\bf 1005}, \penalty0 121788  (2021).
\newblock \doi{10.1016/j.nuclphysa.2020.121788}.

\bibitem{Strickland:2014pga}
M.~Strickland, Anisotropic hydrodynamics: Three lectures, \emph{Acta Phys.
  Polon.} {\bf B45}\penalty0 (12), \penalty0 2355--2394  (2014).
\newblock \doi{10.5506/APhysPolB.45.2355}.

\bibitem{Strickland:2018exs}
M.~Strickland, {Small system studies: A theory overview}, \emph{Nucl. Phys. A}.
  {\bf 982}, \penalty0 92--98  (2019).
\newblock \doi{10.1016/j.nuclphysa.2018.09.071}.

\bibitem{Denicol:2014vaa}
G.~S. Denicol, S.~Jeon, and C.~Gale, {Transport Coefficients of Bulk Viscous
  Pressure in the 14-moment approximation}, \emph{Phys. Rev. C}. {\bf
  90}\penalty0 (2), \penalty0 024912  (2014).
\newblock \doi{10.1103/PhysRevC.90.024912}.

\bibitem{Rocha:2021zcw}
G.~S. Rocha, G.~S. Denicol, and J.~Noronha, {Novel Relaxation Time
  Approximation to the Relativistic Boltzmann Equation}, \emph{Phys. Rev.
  Lett.} {\bf 127}\penalty0 (4), \penalty0 042301  (2021).
\newblock \doi{10.1103/PhysRevLett.127.042301}.

\bibitem{Rocha:2022fqz}
G.~S. Rocha, M.~N. Ferreira, G.~S. Denicol, and J.~Noronha, {Transport
  coefficients of quasiparticle models within a new relaxation time
  approximation of the Boltzmann equation}, \emph{Phys. Rev. D}. {\bf
  106}\penalty0 (3), \penalty0 036022  (2022).
\newblock \doi{10.1103/PhysRevD.106.036022}.

\bibitem{deBrito:2023tgb}
C.~V.~P. de~Brito and G.~S. Denicol, {Third-order relativistic dissipative
  fluid dynamics from the method of moments}, \emph{Phys. Rev. D}. {\bf
  108}\penalty0 (9), \penalty0 096020  (2023).
\newblock \doi{10.1103/PhysRevD.108.096020}.

\bibitem{Nijs:2021clz}
G.~Nijs and W.~van~der Schee, {Predictions and postdictions for relativistic
  lead and oxygen collisions with the computational simulation code Trajectum},
  \emph{Phys. Rev. C}. {\bf 106}\penalty0 (4), \penalty0 044903  (2022).
\newblock \doi{10.1103/PhysRevC.106.044903}.

\bibitem{Romatschke:2017vte}
P.~Romatschke, {Relativistic Fluid Dynamics Far From Local Equilibrium},
  \emph{Phys. Rev. Lett.} {\bf 120}\penalty0 (1), \penalty0 012301  (2018).
\newblock \doi{10.1103/PhysRevLett.120.012301}.

\bibitem{Heller:2013fn}
M.~P. Heller, R.~A. Janik, and P.~Witaszczyk, Hydrodynamic gradient expansion
  in gauge theory plasmas, \emph{Phys. Rev. Lett.} {\bf 110}\penalty0 (21),
  \penalty0 211602  (2013).
\newblock \doi{10.1103/PhysRevLett.110.211602}.

\bibitem{Buchel:2016cbj}
A.~Buchel, M.~P. Heller, and J.~Noronha, Entropy production, hydrodynamics, and
  resurgence in the primordial quark-gluon plasma from holography, \emph{Phys.
  Rev.} {\bf D94}\penalty0 (10), \penalty0 106011  (2016).
\newblock \doi{10.1103/PhysRevD.94.106011}.

\bibitem{Denicol:2020eij}
G.~S. Denicol and J.~Noronha, {Connecting far-from-equilibrium hydrodynamics to
  resummed transport coefficients and attractors}, \emph{Nucl. Phys. A}. {\bf
  1005}, \penalty0 121748  (2021).
\newblock \doi{10.1016/j.nuclphysa.2020.121748}.

\bibitem{Blaizot:2017ucy}
J.-P. Blaizot and L.~Yan, {Fluid dynamics of out of equilibrium boost invariant
  plasmas}, \emph{Phys. Lett. B}. {\bf 780}, \penalty0 283--286  (2018).
\newblock \doi{10.1016/j.physletb.2018.02.058}.

\bibitem{Callen:1951vq}
H.~B. Callen and T.~A. Welton, {Irreversibility and generalized noise},
  \emph{Phys. Rev.} {\bf 83}, \penalty0 34--40  (1951).
\newblock \doi{10.1103/PhysRev.83.34}.

\bibitem{Kubo:1957mj}
R.~Kubo, {Statistical mechanical theory of irreversible processes. 1. General
  theory and simple applications in magnetic and conduction problems}, \emph{J.
  Phys. Soc. Jap.} {\bf 12}, \penalty0 570--586  (1957).
\newblock \doi{10.1143/JPSJ.12.570}.

\bibitem{Young:2014pka}
C.~Young, J.~I. Kapusta, C.~Gale, S.~Jeon, and B.~Schenke, {Thermally
  Fluctuating Second-Order Viscous Hydrodynamics and Heavy-Ion Collisions},
  \emph{Phys. Rev. C}. {\bf 91}\penalty0 (4), \penalty0 044901  (2015).
\newblock \doi{10.1103/PhysRevC.91.044901}.

\bibitem{Singh:2018dpk}
M.~Singh, C.~Shen, S.~McDonald, S.~Jeon, and C.~Gale, {Hydrodynamic
  Fluctuations in Relativistic Heavy-Ion Collisions}, \emph{Nucl. Phys. A}.
  {\bf 982}, \penalty0 319--322  (2019).
\newblock \doi{10.1016/j.nuclphysa.2018.10.061}.

\bibitem{Akamatsu:2016llw}
Y.~Akamatsu, A.~Mazeliauskas, and D.~Teaney, {A kinetic regime of hydrodynamic
  fluctuations and long time tails for a Bjorken expansion}, \emph{Phys. Rev.
  C}. {\bf 95}\penalty0 (1), \penalty0 014909  (2017).
\newblock \doi{10.1103/PhysRevC.95.014909}.

\bibitem{Akamatsu:2017rdu}
Y.~Akamatsu, A.~Mazeliauskas, and D.~Teaney, {Bulk viscosity from hydrodynamic
  fluctuations with relativistic hydrokinetic theory}, \emph{Phys. Rev. C}.
  {\bf 97}\penalty0 (2), \penalty0 024902  (2018).
\newblock \doi{10.1103/PhysRevC.97.024902}.

\bibitem{An:2019osr}
X.~An, G.~Basar, M.~Stephanov, and H.-U. Yee, {Relativistic Hydrodynamic
  Fluctuations}, \emph{Phys. Rev. C}. {\bf 100}\penalty0 (2), \penalty0 024910
  (2019).
\newblock \doi{10.1103/PhysRevC.100.024910}.

\bibitem{Martinez:2018wia}
M.~Martinez and T.~Sch\"afer, {Stochastic hydrodynamics and long time tails of
  an expanding conformal charged fluid}, \emph{Phys. Rev. C}. {\bf 99}\penalty0
  (5), \penalty0 054902  (2019).
\newblock \doi{10.1103/PhysRevC.99.054902}.

\bibitem{De:2022tkb}
A.~De, C.~Shen, and J.~I. Kapusta, {Stochastic hydrodynamics and
  hydro-kinetics: Similarities and differences}, \emph{Phys. Rev. C}. {\bf
  106}\penalty0 (5), \penalty0 054903  (2022).
\newblock \doi{10.1103/PhysRevC.106.054903}.

\bibitem{Stephanov:2017ghc}
M.~Stephanov and Y.~Yin, {Hydrodynamics with parametric slowing down and
  fluctuations near the critical point}, \emph{Phys. Rev. D}. {\bf 98}\penalty0
  (3), \penalty0 036006  (2018).
\newblock \doi{10.1103/PhysRevD.98.036006}.

\bibitem{Nahrgang:2018afz}
M.~Nahrgang, M.~Bluhm, T.~Schaefer, and S.~A. Bass, {Diffusive dynamics of
  critical fluctuations near the QCD critical point}, \emph{Phys. Rev. D}. {\bf
  99}\penalty0 (11), \penalty0 116015  (2019).
\newblock \doi{10.1103/PhysRevD.99.116015}.

\bibitem{Martinez:2019bsn}
M.~Martinez, T.~Sch\"afer, and V.~Skokov, {Critical behavior of the bulk
  viscosity in QCD}, \emph{Phys. Rev. D}. {\bf 100}\penalty0 (7), \penalty0
  074017  (2019).
\newblock \doi{10.1103/PhysRevD.100.074017}.

\bibitem{Rajagopal:2019xwg}
K.~Rajagopal, G.~Ridgway, R.~Weller, and Y.~Yin, {Understanding the
  out-of-equilibrium dynamics near a critical point in the QCD phase diagram},
  \emph{Phys. Rev. D}. {\bf 102}\penalty0 (9), \penalty0 094025  (2020).
\newblock \doi{10.1103/PhysRevD.102.094025}.

\bibitem{An:2019csj}
X.~An, G.~Ba\c{s}ar, M.~Stephanov, and H.-U. Yee, {Fluctuation dynamics in a
  relativistic fluid with a critical point}, \emph{Phys. Rev. C}. {\bf
  102}\penalty0 (3), \penalty0 034901  (2020).
\newblock \doi{10.1103/PhysRevC.102.034901}.

\bibitem{An:2020vri}
X.~An, G.~Ba\c{s}ar, M.~Stephanov, and H.-U. Yee, {Evolution of Non-Gaussian
  Hydrodynamic Fluctuations}, \emph{Phys. Rev. Lett.} {\bf 127}\penalty0 (7),
  \penalty0 072301  (2021).
\newblock \doi{10.1103/PhysRevLett.127.072301}.

\bibitem{Nahrgang:2020yxm}
M.~Nahrgang and M.~Bluhm, {Modeling the diffusive dynamics of critical
  fluctuations near the QCD critical point}, \emph{Phys. Rev. D}. {\bf
  102}\penalty0 (9), \penalty0 094017  (2020).
\newblock \doi{10.1103/PhysRevD.102.094017}.

\bibitem{Dore:2020jye}
T.~Dore, J.~Noronha-Hostler, and E.~McLaughlin, {Far-from-equilibrium search
  for the QCD critical point}, \emph{Phys. Rev. D}. {\bf 102}\penalty0 (7),
  \penalty0 074017  (2020).
\newblock \doi{10.1103/PhysRevD.102.074017}.

\bibitem{Du:2020bxp}
L.~Du, U.~Heinz, K.~Rajagopal, and Y.~Yin, {Fluctuation dynamics near the QCD
  critical point}, \emph{Phys. Rev. C}. {\bf 102}\penalty0 (5), \penalty0
  054911  (2020).
\newblock \doi{10.1103/PhysRevC.102.054911}.

\bibitem{Pradeep:2022mkf}
M.~Pradeep, K.~Rajagopal, M.~Stephanov, and Y.~Yin, {Freezing out fluctuations
  in Hydro+ near the QCD critical point}, \emph{Phys. Rev. D}. {\bf
  106}\penalty0 (3), \penalty0 036017  (2022).
\newblock \doi{10.1103/PhysRevD.106.036017}.

\bibitem{An:2022jgc}
X.~An, G.~Basar, M.~Stephanov, and H.-U. Yee, {Non-Gaussian fluctuation
  dynamics in relativistic fluid}  (12, 2022).

\bibitem{Kapusta:2011gt}
J.~I. Kapusta, B.~Muller, and M.~Stephanov, {Relativistic Theory of
  Hydrodynamic Fluctuations with Applications to Heavy Ion Collisions},
  \emph{Phys. Rev. C}. {\bf 85}, \penalty0 054906  (2012).
\newblock \doi{10.1103/PhysRevC.85.054906}.

\bibitem{Young:2013fka}
C.~Young, {Numerical integration of thermal noise in relativistic
  hydrodynamics}, \emph{Phys. Rev. C}. {\bf 89}\penalty0 (2), \penalty0 024913
  (2014).
\newblock \doi{10.1103/PhysRevC.89.024913}.

\bibitem{Sakai:2017rfi}
A.~Sakai, K.~Murase, and T.~Hirano, {Hydrodynamic fluctuations in Pb + Pb
  collisions at LHC}, \emph{Nucl. Phys. A}. {\bf 967}, \penalty0 445--448
  (2017).
\newblock \doi{10.1016/j.nuclphysa.2017.05.010}.

\bibitem{De:2020yyx}
A.~De, C.~Plumberg, and J.~I. Kapusta, {Calculating Fluctuations and
  Self-Correlations Numerically for Causal Charge Diffusion in Relativistic
  Heavy-Ion Collisions}, \emph{Phys. Rev. C}. {\bf 102}\penalty0 (2), \penalty0
  024905  (2020).
\newblock \doi{10.1103/PhysRevC.102.024905}.

\bibitem{Sakai:2020pjw}
A.~Sakai, K.~Murase, and T.~Hirano, {Rapidity decorrelation of anisotropic flow
  caused by hydrodynamic fluctuations}, \emph{Phys. Rev. C}. {\bf 102}\penalty0
  (6), \penalty0 064903  (2020).
\newblock \doi{10.1103/PhysRevC.102.064903}.

\bibitem{Kuroki:2023ebq}
K.~Kuroki, A.~Sakai, K.~Murase, and T.~Hirano, {Hydrodynamic fluctuations and
  ultra-central flow puzzle in heavy-ion collisions}  (5, 2023).

\bibitem{Calzetta:1997aj}
E.~Calzetta, {Relativistic fluctuating hydrodynamics}, \emph{Class. Quant.
  Grav.} {\bf 15}, \penalty0 653--667  (1998).
\newblock \doi{10.1088/0264-9381/15/3/015}.

\bibitem{Kovtun:2003vj}
P.~Kovtun and L.~G. Yaffe, {Hydrodynamic fluctuations, long time tails, and
  supersymmetry}, \emph{Phys. Rev. D}. {\bf 68}, \penalty0 025007  (2003).
\newblock \doi{10.1103/PhysRevD.68.025007}.

\bibitem{Dunkel:2008ngc}
J.~Dunkel and P.~H\"anggi, {Relativistic Brownian motion}, \emph{Phys. Rept.}
  {\bf 471}, \penalty0 1--73  (2009).
\newblock \doi{10.1016/j.physrep.2008.12.001}.

\bibitem{Kovtun:2011np}
P.~Kovtun, G.~D. Moore, and P.~Romatschke, {The stickiness of sound: An
  absolute lower limit on viscosity and the breakdown of second order
  relativistic hydrodynamics}, \emph{Phys. Rev. D}. {\bf 84}, \penalty0 025006
  (2011).
\newblock \doi{10.1103/PhysRevD.84.025006}.

\bibitem{Kumar:2013twa}
A.~Kumar, J.~R. Bhatt, and A.~P. Mishra, {Fluctuations in Relativistic Causal
  Hydrodynamics}, \emph{Nucl. Phys. A}. {\bf 925}, \penalty0 199--217  (2014).
\newblock \doi{10.1016/j.nuclphysa.2014.02.012}.

\bibitem{Murase:2016rhl}
K.~Murase and T.~Hirano, {Hydrodynamic fluctuations and dissipation in an
  integrated dynamical model}, \emph{Nucl. Phys. A}. {\bf 956}, \penalty0
  276--279  (2016).
\newblock \doi{10.1016/j.nuclphysa.2016.01.011}.

\bibitem{Kapusta:2014dja}
J.~I. Kapusta and C.~Young, {Causal Baryon Diffusion and Colored Noise},
  \emph{Phys. Rev. C}. {\bf 90}\penalty0 (4), \penalty0 044902  (2014).
\newblock \doi{10.1103/PhysRevC.90.044902}.

\bibitem{Martinez:2017jjf}
M.~Martinez and T.~Sch\"afer, {Hydrodynamic tails and a fluctuation bound on
  the bulk viscosity}, \emph{Phys. Rev. A}. {\bf 96}\penalty0 (6), \penalty0
  063607  (2017).
\newblock \doi{10.1103/PhysRevA.96.063607}.

\bibitem{Murase:2019cwc}
K.~Murase, {Causal hydrodynamic fluctuations in non-static and inhomogeneous
  backgrounds}, \emph{Annals Phys.} {\bf 411}, \penalty0 167969  (2019).
\newblock \doi{10.1016/j.aop.2019.167969}.

\bibitem{Calzetta:2020wzr}
E.~Calzetta, {Fully developed relativistic turbulence}, \emph{Phys. Rev. D}.
  {\bf 103}\penalty0 (5), \penalty0 056018  (2021).
\newblock \doi{10.1103/PhysRevD.103.056018}.

\bibitem{Torrieri:2020ezm}
G.~Torrieri, {Fluctuating Relativistic hydrodynamics from Crooks theorem},
  \emph{JHEP}. {\bf 02}, \penalty0 175  (2021).
\newblock \doi{10.1007/JHEP02(2021)175}.

\bibitem{Dore:2021xqq}
T.~Dore, L.~Gavassino, D.~Montenegro, M.~Shokri, and G.~Torrieri, {Fluctuating
  relativistic dissipative hydrodynamics as a gauge theory}, \emph{Annals
  Phys.} {\bf 442}, \penalty0 168902  (2022).
\newblock \doi{10.1016/j.aop.2022.168902}.

\bibitem{Petrosyan:2021lqi}
A.~Petrosyan and A.~Zaccone, {Relativistic Langevin equation derived from a
  particle-bath Lagrangian}, \emph{J. Phys. A}. {\bf 55}\penalty0 (1),
  \penalty0 015001  (2022).
\newblock \doi{10.1088/1751-8121/ac3a33}.

\bibitem{Abbasi:2022rum}
N.~Abbasi, A.~Davody, and S.~Tahery, {Correlation functions in stable
  first-order relativistic hydrodynamics}, \emph{Phys. Rev. D}. {\bf
  109}\penalty0 (3), \penalty0 036006  (2024).
\newblock \doi{10.1103/PhysRevD.109.036006}.

\bibitem{Chen:2022ryi}
Z.~Chen, D.~Teaney, and L.~Yan, {Hydrodynamic attractor of noisy plasmas}  (6,
  2022).

\bibitem{Crossley:2015evo}
M.~Crossley, P.~Glorioso, and H.~Liu, {Effective field theory of dissipative
  fluids}, \emph{JHEP}. {\bf 09}, \penalty0 095  (2017).
\newblock \doi{10.1007/JHEP09(2017)095}.

\bibitem{Sieberer:2015hba}
L.~M. Sieberer, A.~Chiocchetta, A.~Gambassi, U.~C. T\"auber, and S.~Diehl,
  {Thermodynamic Equilibrium as a Symmetry of the Schwinger-Keldysh Action},
  \emph{Phys. Rev. B}. {\bf 92}\penalty0 (13), \penalty0 134307  (2015).
\newblock \doi{10.1103/PhysRevB.92.134307}.

\bibitem{Haehl:2016pec}
F.~M. Haehl, R.~Loganayagam, and M.~Rangamani, {Schwinger-Keldysh formalism.
  Part I: BRST symmetries and superspace}, \emph{JHEP}. {\bf 06}, \penalty0 069
   (2017).
\newblock \doi{10.1007/JHEP06(2017)069}.

\bibitem{Glorioso:2017fpd}
P.~Glorioso, M.~Crossley, and H.~Liu, {Effective field theory of dissipative
  fluids (II): classical limit, dynamical KMS symmetry and entropy current},
  \emph{JHEP}. {\bf 09}, \penalty0 096  (2017).
\newblock \doi{10.1007/JHEP09(2017)096}.

\bibitem{Liu:2018kfw}
H.~Liu and P.~Glorioso, {Lectures on non-equilibrium effective field theories
  and fluctuating hydrodynamics}, \emph{PoS}. {\bf TASI2017}, \penalty0 008
  (2018).
\newblock \doi{10.22323/1.305.0008}.

\bibitem{Martin:1959jp}
P.~C. Martin and J.~S. Schwinger, {Theory of many particle systems. 1.},
  \emph{Phys. Rev.} {\bf 115}, \penalty0 1342--1373  (1959).
\newblock \doi{10.1103/PhysRev.115.1342}.

\bibitem{Grozdanov:2013dba}
S.~Grozdanov and J.~Polonyi, {Viscosity and dissipative hydrodynamics from
  effective field theory}, \emph{Phys. Rev. D}. {\bf 91}\penalty0 (10),
  \penalty0 105031  (2015).
\newblock \doi{10.1103/PhysRevD.91.105031}.

\bibitem{Kovtun:2014hpa}
P.~Kovtun, G.~D. Moore, and P.~Romatschke, {Towards an effective action for
  relativistic dissipative hydrodynamics}, \emph{JHEP}. {\bf 07}, \penalty0 123
   (2014).
\newblock \doi{10.1007/JHEP07(2014)123}.

\bibitem{Harder:2015nxa}
M.~Harder, P.~Kovtun, and A.~Ritz, {On thermal fluctuations and the generating
  functional in relativistic hydrodynamics}, \emph{JHEP}. {\bf 07}, \penalty0
  025  (2015).
\newblock \doi{10.1007/JHEP07(2015)025}.

\bibitem{Haehl:2015pja}
F.~M. Haehl, R.~Loganayagam, and M.~Rangamani, {Adiabatic hydrodynamics: The
  eightfold way to dissipation}, \emph{JHEP}. {\bf 05}, \penalty0 060  (2015).
\newblock \doi{10.1007/JHEP05(2015)060}.

\bibitem{Haehl:2015uoc}
F.~M. Haehl, R.~Loganayagam, and M.~Rangamani, {Topological sigma models \&
  dissipative hydrodynamics}, \emph{JHEP}. {\bf 04}, \penalty0 039  (2016).
\newblock \doi{10.1007/JHEP04(2016)039}.

\bibitem{Jensen:2017kzi}
K.~Jensen, N.~Pinzani-Fokeeva, and A.~Yarom, {Dissipative hydrodynamics in
  superspace}, \emph{JHEP}. {\bf 09}, \penalty0 127  (2018).
\newblock \doi{10.1007/JHEP09(2018)127}.

\bibitem{Jensen:2018hse}
K.~Jensen, R.~Marjieh, N.~Pinzani-Fokeeva, and A.~Yarom, {A panoply of
  Schwinger-Keldysh transport}, \emph{SciPost Phys.} {\bf 5}\penalty0 (5),
  \penalty0 053  (2018).
\newblock \doi{10.21468/SciPostPhys.5.5.053}.

\bibitem{Haehl:2018lcu}
F.~M. Haehl, R.~Loganayagam, and M.~Rangamani, {Effective Action for
  Relativistic Hydrodynamics: Fluctuations, Dissipation, and Entropy Inflow},
  \emph{JHEP}. {\bf 10}, \penalty0 194  (2018).
\newblock \doi{10.1007/JHEP10(2018)194}.

\bibitem{Jain:2020zhu}
A.~Jain and P.~Kovtun, {Late Time Correlations in Hydrodynamics: Beyond
  Constitutive Relations}, \emph{Phys. Rev. Lett.} {\bf 128}\penalty0 (7),
  \penalty0 071601  (2022).
\newblock \doi{10.1103/PhysRevLett.128.071601}.

\bibitem{Jain:2023obu}
A.~Jain and P.~Kovtun, {Schwinger-Keldysh effective field theory for stable and
  causal relativistic hydrodynamics}  (9, 2023).

\bibitem{Mullins:2023tjg}
N.~Mullins, M.~Hippert, and J.~Noronha, {Stochastic fluctuations in
  relativistic fluids: causality, stability, and the information current}  (6,
  2023).

\bibitem{Mullins:2023ott}
N.~Mullins, M.~Hippert, L.~Gavassino, and J.~Noronha, {Relativistic
  hydrodynamic fluctuations from an effective action: causality, stability, and
  the information current}  (9, 2023).

\bibitem{STAR:2016vqt}
L.~Adamczyk et~al., {Beam Energy Dependence of the Third Harmonic of Azimuthal
  Correlations in Au+Au Collisions at RHIC}, \emph{Phys. Rev. Lett.} {\bf
  116}\penalty0 (11), \penalty0 112302  (2016).
\newblock \doi{10.1103/PhysRevLett.116.112302}.

\bibitem{CMS:2023bvg}
A.~Tumasyan et~al., {Higher-order moments of the elliptic flow distribution in
  PbPb collisions at $\sqrt{s_\mathrm{NN}}$ = 5.02 TeV}  (11, 2023).

\bibitem{Cooper:1974mv}
F.~Cooper and G.~Frye, {Comment on the Single Particle Distribution in the
  Hydrodynamic and Statistical Thermodynamic Models of Multiparticle
  Production}, \emph{Phys. Rev. D}. {\bf 10}, \penalty0 186  (1974).
\newblock \doi{10.1103/PhysRevD.10.186}.

\bibitem{Teaney:2003kp}
D.~Teaney, {The Effects of viscosity on spectra, elliptic flow, and HBT radii},
  \emph{Phys. Rev. C}. {\bf 68}, \penalty0 034913  (2003).
\newblock \doi{10.1103/PhysRevC.68.034913}.

\bibitem{Dusling:2009df}
K.~Dusling, G.~D. Moore, and D.~Teaney, {Radiative energy loss and v(2) spectra
  for viscous hydrodynamics}, \emph{Phys. Rev. C}. {\bf 81}, \penalty0 034907
  (2010).
\newblock \doi{10.1103/PhysRevC.81.034907}.

\bibitem{Romatschke:2009im}
P.~Romatschke, New developments in relativistic viscous hydrodynamics,
  \emph{Int. J. Mod. Phys.} {\bf E19}, \penalty0 1--53  (2010).
\newblock \doi{10.1142/S0218301310014613}.

\bibitem{Wolff:2016vcm}
Z.~Wolff and D.~Molnar, {Flow harmonics from self-consistent particlization of
  a viscous fluid}, \emph{Phys. Rev. C}. {\bf 96}\penalty0 (4), \penalty0
  044909  (2017).
\newblock \doi{10.1103/PhysRevC.96.044909}.

\bibitem{JETSCAPE:2020avt}
J.-F. Paquet et~al., {Revisiting Bayesian constraints on the transport
  coefficients of QCD}, \emph{Nucl. Phys. A}. {\bf 1005}, \penalty0 121749
  (2021).
\newblock \doi{10.1016/j.nuclphysa.2020.121749}.

\bibitem{Oliinychenko:2019zfk}
D.~Oliinychenko and V.~Koch, {Microcanonical Particlization with Local
  Conservation Laws}, \emph{Phys. Rev. Lett.} {\bf 123}\penalty0 (18),
  \penalty0 182302  (2019).
\newblock \doi{10.1103/PhysRevLett.123.182302}.

\bibitem{Vovchenko:2020kwg}
V.~Vovchenko and V.~Koch, {Particlization of an interacting hadron resonance
  gas with global conservation laws for event-by-event fluctuations in
  heavy-ion collisions}, \emph{Phys. Rev. C}. {\bf 103}\penalty0 (4), \penalty0
  044903  (2021).
\newblock \doi{10.1103/PhysRevC.103.044903}.

\bibitem{Oliinychenko:2020cmr}
D.~Oliinychenko, S.~Shi, and V.~Koch, {Effects of local event-by-event
  conservation laws in ultrarelativistic heavy-ion collisions at
  particlization}, \emph{Phys. Rev. C}. {\bf 102}\penalty0 (3), \penalty0
  034904  (2020).
\newblock \doi{10.1103/PhysRevC.102.034904}.

\bibitem{Heinz:2019dbd}
U.~W. Heinz and J.~S. Moreland, {Hydrodynamic flow in small systems or:
  \textquotedblleft{}How the heck is it possible that a system emitting only a
  dozen particles can be described by fluid dynamics?\textquotedblright{}},
  \emph{J. Phys. Conf. Ser.} {\bf 1271}\penalty0 (1), \penalty0 012018  (2019).
\newblock \doi{10.1088/1742-6596/1271/1/012018}.

\bibitem{Shen:2015qba}
C.~Shen, J.~F. Paquet, G.~S. Denicol, S.~Jeon, and C.~Gale, {Thermal photon
  radiation in high multiplicity p+Pb collisions at the Large Hadron Collider},
  \emph{Phys. Rev. Lett.} {\bf 116}\penalty0 (7), \penalty0 072301  (2016).
\newblock \doi{10.1103/PhysRevLett.116.072301}.

\bibitem{Shen:2017fnn}
C.~Shen and B.~Schenke, {Initial state and hydrodynamic modeling of heavy-ion
  collisions at RHIC BES energies}, \emph{PoS}. {\bf CPOD2017}, \penalty0 006
  (2018).
\newblock \doi{10.22323/1.311.0006}.

\bibitem{Shen:2017ruz}
C.~Shen, G.~Denicol, C.~Gale, S.~Jeon, A.~Monnai, and B.~Schenke, {A hybrid
  approach to relativistic heavy-ion collisions at the RHIC BES energies},
  \emph{Nucl. Phys. A}. {\bf 967}, \penalty0 796--799  (2017).
\newblock \doi{10.1016/j.nuclphysa.2017.06.008}.

\bibitem{Zhao:2022ayk}
W.~Zhao, C.~Shen, and B.~Schenke, {Collectivity in Ultraperipheral Pb+Pb
  Collisions at the Large Hadron Collider}, \emph{Phys. Rev. Lett.} {\bf
  129}\penalty0 (25), \penalty0 252302  (2022).
\newblock \doi{10.1103/PhysRevLett.129.252302}.

\bibitem{Shen:2022daw}
C.~Shen, W.~Zhao, and B.~Schenke, {Collectivity in ultra-peripheral heavy-ion
  collisions}, \emph{EPJ Web Conf.} {\bf 276}, \penalty0 01002  (2023).
\newblock \doi{10.1051/epjconf/202327601002}.

\bibitem{Shen:2023awv}
C.~Shen, B.~Schenke, and W.~Zhao, {Viscosities of the Baryon-Rich Quark-Gluon
  Plasma from Beam Energy Scan Data}  (10, 2023).

\bibitem{Shen:2023pgb}
C.~Shen, B.~Schenke, and W.~Zhao.
\newblock {The effects of pseudorapidity-dependent observables on (3+1)D
  Bayesian Inference of relativistic heavy-ion collisions}.
\newblock In \emph{{30th International Conference on Ultrarelativstic
  Nucleus-Nucleus Collisions}}  (12, 2023).

\bibitem{CMS:2015xmx}
V.~Khachatryan et~al., {Evidence for transverse momentum and pseudorapidity
  dependent event plane fluctuations in PbPb and pPb collisions}, \emph{Phys.
  Rev. C}. {\bf 92}\penalty0 (3), \penalty0 034911  (2015).
\newblock \doi{10.1103/PhysRevC.92.034911}.

\bibitem{ATLAS:2017rij}
M.~Aaboud et~al., {Measurement of longitudinal flow decorrelations in Pb+Pb
  collisions at $\sqrt{s_{\text {NN}}}=2.76$ and 5.02 TeV with the ATLAS
  detector}, \emph{Eur. Phys. J. C}. {\bf 78}\penalty0 (2), \penalty0 142
  (2018).
\newblock \doi{10.1140/epjc/s10052-018-5605-7}.

\bibitem{Bozek:2017qir}
P.~Bozek and W.~Broniowski, {Longitudinal decorrelation measures of flow
  magnitude and event-plane angles in ultrarelativistic nuclear collisions},
  \emph{Phys. Rev. C}. {\bf 97}\penalty0 (3), \penalty0 034913  (2018).
\newblock \doi{10.1103/PhysRevC.97.034913}.

\bibitem{Pang:2018zzo}
L.-G. Pang, H.~Petersen, and X.-N. Wang, {Pseudorapidity distribution and
  decorrelation of anisotropic flow within the open-computing-language
  implementation CLVisc hydrodynamics}, \emph{Phys. Rev. C}. {\bf 97}\penalty0
  (6), \penalty0 064918  (2018).
\newblock \doi{10.1103/PhysRevC.97.064918}.

\bibitem{Li:2019eni}
H.~Li and L.~Yan, {Pseudorapidity dependent hydrodynamic response in heavy-ion
  collisions}, \emph{Phys. Lett. B}. {\bf 802}, \penalty0 135248  (2020).
\newblock \doi{10.1016/j.physletb.2020.135248}.

\bibitem{Franco:2019ihq}
R.~Franco and M.~Luzum, {Rapidity-dependent eccentricity scaling in
  relativistic heavy-ion collisions}, \emph{Phys. Lett. B}. {\bf 806},
  \penalty0 135518  (2020).
\newblock \doi{10.1016/j.physletb.2020.135518}.

\bibitem{Ryu:2021lnx}
S.~Ryu, V.~Jupic, and C.~Shen, {Probing early-time longitudinal dynamics with
  the \ensuremath{\Lambda} hyperon's spin polarization in relativistic
  heavy-ion collisions}, \emph{Phys. Rev. C}. {\bf 104}\penalty0 (5), \penalty0
  054908  (2021).
\newblock \doi{10.1103/PhysRevC.104.054908}.

\bibitem{Alzhrani:2022dpi}
S.~Alzhrani, S.~Ryu, and C.~Shen, {\ensuremath{\Lambda} spin polarization in
  event-by-event relativistic heavy-ion collisions}, \emph{Phys. Rev. C}. {\bf
  106}\penalty0 (1), \penalty0 014905  (2022).
\newblock \doi{10.1103/PhysRevC.106.014905}.

\bibitem{Lisa:2021zkj}
M.~A. Lisa, J.~a. G.~P. Barbon, D.~D. Chinellato, W.~M. Serenone, C.~Shen,
  J.~Takahashi, and G.~Torrieri, {Vortex rings from high energy central p+A
  collisions}, \emph{Phys. Rev. C}. {\bf 104}\penalty0 (1), \penalty0 011901
  (2021).
\newblock \doi{10.1103/PhysRevC.104.L011901}.

\bibitem{Serenone:2021zef}
W.~M. Serenone, J.~a. G.~P. Barbon, D.~D. Chinellato, M.~A. Lisa, C.~Shen,
  J.~Takahashi, and G.~Torrieri, {\ensuremath{\Lambda} polarization from
  thermalized jet energy}, \emph{Phys. Lett. B}. {\bf 820}, \penalty0 136500
  (2021).
\newblock \doi{10.1016/j.physletb.2021.136500}.

\bibitem{Ribeiro:2023waz}
V.~H. Ribeiro, D.~Dobrigkeit~Chinellato, M.~A. Lisa, W.~Matioli~Serenone,
  C.~Shen, J.~Takahashi, and G.~Torrieri, {$\Lambda$ polarization from vortex
  ring as medium response for jet thermalization}  (5, 2023).

\bibitem{Ollitrault:1992bk}
J.-Y. Ollitrault, {Anisotropy as a signature of transverse collective flow},
  \emph{Phys. Rev. D}. {\bf 46}, \penalty0 229--245  (1992).
\newblock \doi{10.1103/PhysRevD.46.229}.

\bibitem{Voloshin:1994mz}
S.~Voloshin and Y.~Zhang, {Flow study in relativistic nuclear collisions by
  Fourier expansion of Azimuthal particle distributions}, \emph{Z. Phys. C}.
  {\bf 70}, \penalty0 665--672  (1996).
\newblock \doi{10.1007/s002880050141}.

\bibitem{Poskanzer:1998yz}
A.~M. Poskanzer and S.~A. Voloshin, {Methods for analyzing anisotropic flow in
  relativistic nuclear collisions}, \emph{Phys. Rev. C}. {\bf 58}, \penalty0
  1671--1678  (1998).
\newblock \doi{10.1103/PhysRevC.58.1671}.

\bibitem{Luzum:2012da}
M.~Luzum and J.-Y. Ollitrault, {Eliminating experimental bias in
  anisotropic-flow measurements of high-energy nuclear collisions}, \emph{Phys.
  Rev. C}. {\bf 87}\penalty0 (4), \penalty0 044907  (2013).
\newblock \doi{10.1103/PhysRevC.87.044907}.

\bibitem{Bilandzic:2010jr}
A.~Bilandzic, R.~Snellings, and S.~Voloshin, {Flow analysis with cumulants:
  Direct calculations}, \emph{Phys. Rev. C}. {\bf 83}, \penalty0 044913
  (2011).
\newblock \doi{10.1103/PhysRevC.83.044913}.

\bibitem{Bilandzic:2013kga}
A.~Bilandzic, C.~H. Christensen, K.~Gulbrandsen, A.~Hansen, and Y.~Zhou,
  {Generic framework for anisotropic flow analyses with multiparticle azimuthal
  correlations}, \emph{Phys. Rev. C}. {\bf 89}\penalty0 (6), \penalty0 064904
  (2014).
\newblock \doi{10.1103/PhysRevC.89.064904}.

\bibitem{PHENIX:2017xrm}
C.~Aidala et~al., {Measurements of Multiparticle Correlations in
  $d+\mathrm{Au}$ Collisions at 200, 62.4, 39, and 19.6 GeV and $p+\mathrm{Au}$
  Collisions at 200 GeV and Implications for Collective Behavior}, \emph{Phys.
  Rev. Lett.} {\bf 120}\penalty0 (6), \penalty0 062302  (2018).
\newblock \doi{10.1103/PhysRevLett.120.062302}.

\bibitem{PHENIX:2018lia}
C.~Aidala et~al., {Creation of quark\textendash{}gluon plasma droplets with
  three distinct geometries}, \emph{Nature Phys.} {\bf 15}\penalty0 (3),
  \penalty0 214--220  (2019).
\newblock \doi{10.1038/s41567-018-0360-0}.

\bibitem{Citron:2018lsq}
Z.~Citron et~al., {Report from Working Group 5}: {Future physics opportunities
  for high-density QCD at the LHC with heavy-ion and proton beams}, \emph{CERN
  Yellow Rep. Monogr.} {\bf 7}, \penalty0 1159--1410  (2019).
\newblock \doi{10.23731/CYRM-2019-007.1159}.

\bibitem{ALICE:2019zfl}
S.~Acharya et~al., {Investigations of Anisotropic Flow Using Multiparticle
  Azimuthal Correlations in pp, p-Pb, Xe-Xe, and Pb-Pb Collisions at the LHC},
  \emph{Phys. Rev. Lett.} {\bf 123}\penalty0 (14), \penalty0 142301  (2019).
\newblock \doi{10.1103/PhysRevLett.123.142301}.

\bibitem{CMS:2016fnw}
V.~Khachatryan et~al., {Evidence for collectivity in pp collisions at the LHC},
  \emph{Phys. Lett. B}. {\bf 765}, \penalty0 193--220  (2017).
\newblock \doi{10.1016/j.physletb.2016.12.009}.

\bibitem{Zhou:2017wls}
M.~Zhou, {Measurement of multi-particle azimuthal correlations with the
  subevent cumulant method in pp and p +Pb collisions with the ATLAS detector},
  \emph{Nucl. Phys. A}. {\bf 967}, \penalty0 472--475  (2017).
\newblock \doi{10.1016/j.nuclphysa.2017.04.019}.

\bibitem{Zhao:2017rgg}
W.~Zhao, Y.~Zhou, H.~Xu, W.~Deng, and H.~Song, {Hydrodynamic collectivity in
  proton\textendash{}proton collisions at 13 TeV}, \emph{Phys. Lett. B}. {\bf
  780}, \penalty0 495--500  (2018).
\newblock \doi{10.1016/j.physletb.2018.03.022}.

\bibitem{Voloshin:2008dg}
S.~A. Voloshin, A.~M. Poskanzer, and R.~Snellings, {Collective phenomena in
  non-central nuclear collisions}, \emph{Landolt-Bornstein}. {\bf 23},
  \penalty0 293--333  (2010).
\newblock \doi{10.1007/978-3-642-01539-7_10}.

\bibitem{Borghini:2000sa}
N.~Borghini, P.~M. Dinh, and J.-Y. Ollitrault, {A New method for measuring
  azimuthal distributions in nucleus-nucleus collisions}, \emph{Phys. Rev. C}.
  {\bf 63}, \penalty0 054906  (2001).
\newblock \doi{10.1103/PhysRevC.63.054906}.

\bibitem{Jia:2014pza}
J.~Jia and S.~Radhakrishnan, {Limitation of multiparticle correlations for
  studying the event-by-event distribution of harmonic flow in heavy-ion
  collisions}, \emph{Phys. Rev. C}. {\bf 92}\penalty0 (2), \penalty0 024911
  (2015).
\newblock \doi{10.1103/PhysRevC.92.024911}.

\bibitem{ALICE:2011ab}
K.~Aamodt et~al., {Higher harmonic anisotropic flow measurements of charged
  particles in Pb-Pb collisions at $\sqrt{s_{NN}}$=2.76 TeV}, \emph{Phys. Rev.
  Lett.} {\bf 107}, \penalty0 032301  (2011).
\newblock \doi{10.1103/PhysRevLett.107.032301}.

\bibitem{Adam:2016izf}
J.~Adam et~al., {Anisotropic flow of charged particles in Pb-Pb collisions at
  $\sqrt{s_{\rm NN}}=5.02$ TeV}, \emph{Phys. Rev. Lett.} {\bf 116}\penalty0
  (13), \penalty0 132302  (2016).
\newblock \doi{10.1103/PhysRevLett.116.132302}.

\bibitem{Jia:2017hbm}
J.~Jia, M.~Zhou, and A.~Trzupek, {Revealing long-range multiparticle
  collectivity in small collision systems via subevent cumulants}, \emph{Phys.
  Rev. C}. {\bf 96}\penalty0 (3), \penalty0 034906  (2017).
\newblock \doi{10.1103/PhysRevC.96.034906}.

\bibitem{Huo:2017nms}
P.~Huo, K.~Gajdo\v{s}ov\'a, J.~Jia, and Y.~Zhou, {Importance of non-flow in
  mixed-harmonic multi-particle correlations in small collision systems},
  \emph{Phys. Lett. B}. {\bf 777}, \penalty0 201--206  (2018).
\newblock \doi{10.1016/j.physletb.2017.12.035}.

\bibitem{Dusling:2015gta}
K.~Dusling, W.~Li, and B.~Schenke, {Novel collective phenomena in high-energy
  proton\textendash{}proton and proton\textendash{}nucleus collisions},
  \emph{Int. J. Mod. Phys. E}. {\bf 25}\penalty0 (01), \penalty0 1630002
  (2016).
\newblock \doi{10.1142/S0218301316300022}.

\bibitem{Loizides:2016tew}
C.~Loizides, {Experimental overview on small collision systems at the LHC},
  \emph{Nucl. Phys. A}. {\bf 956}, \penalty0 200--207  (2016).
\newblock \doi{10.1016/j.nuclphysa.2016.04.022}.

\bibitem{Schlichting:2016sqo}
S.~Schlichting and P.~Tribedy, {Collectivity in Small Collision Systems: An
  Initial-State Perspective}, \emph{Adv. High Energy Phys.} {\bf 2016},
  \penalty0 8460349  (2016).
\newblock \doi{10.1155/2016/8460349}.

\bibitem{Zhao:2020pty}
W.~Zhao, Y.~Zhou, K.~Murase, and H.~Song, {Searching for small droplets of
  hydrodynamic fluid in proton\textendash{}proton collisions at the LHC},
  \emph{Eur. Phys. J. C}. {\bf 80}\penalty0 (9), \penalty0 846  (2020).
\newblock \doi{10.1140/epjc/s10052-020-8376-x}.

\bibitem{Kovner:2012jm}
A.~Kovner and M.~Lublinsky, {Angular and long range rapidity correlations in
  particle production at high energy}, \emph{Int. J. Mod. Phys. E}. {\bf 22},
  \penalty0 1330001  (2013).
\newblock \doi{10.1142/S0218301313300014}.

\bibitem{Dusling:2017dqg}
K.~Dusling, M.~Mace, and R.~Venugopalan, {Multiparticle collectivity from
  initial state correlations in high energy proton-nucleus collisions},
  \emph{Phys. Rev. Lett.} {\bf 120}\penalty0 (4), \penalty0 042002  (2018).
\newblock \doi{10.1103/PhysRevLett.120.042002}.

\bibitem{Kovchegov:2012nd}
Y.~V. Kovchegov and D.~E. Wertepny, {Long-Range Rapidity Correlations in
  Heavy-Light Ion Collisions}, \emph{Nucl. Phys. A}. {\bf 906}, \penalty0
  50--83  (2013).
\newblock \doi{10.1016/j.nuclphysa.2013.03.006}.

\bibitem{Schenke:2016lrs}
B.~Schenke, S.~Schlichting, P.~Tribedy, and R.~Venugopalan, {Mass ordering of
  spectra from fragmentation of saturated gluon states in high multiplicity
  proton-proton collisions}, \emph{Phys. Rev. Lett.} {\bf 117}\penalty0 (16),
  \penalty0 162301  (2016).
\newblock \doi{10.1103/PhysRevLett.117.162301}.

\bibitem{ALICE:2012mj}
B.~Abelev et~al., {Transverse momentum distribution and nuclear modification
  factor of charged particles in $p$-Pb collisions at $\sqrt{s_{NN}}=5.02$
  TeV}, \emph{Phys. Rev. Lett.} {\bf 110}\penalty0 (8), \penalty0 082302
  (2013).
\newblock \doi{10.1103/PhysRevLett.110.082302}.

\bibitem{CMS:2015ved}
V.~Khachatryan et~al., {Nuclear Effects on the Transverse Momentum Spectra of
  Charged Particles in pPb Collisions at $\sqrt{s_{_\mathrm {NN}}} =5.02$ TeV},
  \emph{Eur. Phys. J. C}. {\bf 75}\penalty0 (5), \penalty0 237  (2015).
\newblock \doi{10.1140/epjc/s10052-015-3435-4}.

\bibitem{ALICE:2018vuu}
S.~Acharya et~al., {Transverse momentum spectra and nuclear modification
  factors of charged particles in pp, p-Pb and Pb-Pb collisions at the LHC},
  \emph{JHEP}. {\bf 11}, \penalty0 013  (2018).
\newblock \doi{10.1007/JHEP11(2018)013}.

\bibitem{Eskola:2016oht}
K.~J. Eskola, P.~Paakkinen, H.~Paukkunen, and C.~A. Salgado, {EPPS16: Nuclear
  parton distributions with LHC data}, \emph{Eur. Phys. J. C}. {\bf
  77}\penalty0 (3), \penalty0 163  (2017).
\newblock \doi{10.1140/epjc/s10052-017-4725-9}.

\bibitem{Dong:2019byy}
X.~Dong, Y.-J. Lee, and R.~Rapp, {Open Heavy-Flavor Production in Heavy-Ion
  Collisions}, \emph{Ann. Rev. Nucl. Part. Sci.} {\bf 69}, \penalty0 417--445
  (2019).
\newblock \doi{10.1146/annurev-nucl-101918-023806}.

\bibitem{ALICE:2016yta}
J.~Adam et~al., {$D$-meson production in $p$-Pb collisions at $\sqrt{s_{\rm
  NN}}=$5.02 TeV and in pp collisions at $\sqrt{s}=$7 TeV}, \emph{Phys. Rev.
  C}. {\bf 94}\penalty0 (5), \penalty0 054908  (2016).
\newblock \doi{10.1103/PhysRevC.94.054908}.

\bibitem{CMS:2018loe}
A.~M. Sirunyan et~al., {Elliptic flow of charm and strange hadrons in
  high-multiplicity pPb collisions at $\sqrt{s_{_\mathrm{NN}}} =$ 8.16 TeV},
  \emph{Phys. Rev. Lett.} {\bf 121}\penalty0 (8), \penalty0 082301  (2018).
\newblock \doi{10.1103/PhysRevLett.121.082301}.

\bibitem{ATLAS:2016yzd}
M.~Aaboud et~al., {Measurements of long-range azimuthal anisotropies and
  associated Fourier coefficients for $pp$ collisions at $\sqrt{s}=5.02$ and
  $13$ TeV and $p$+Pb collisions at $\sqrt{s_{\mathrm{NN}}}=5.02$ TeV with the
  ATLAS detector}, \emph{Phys. Rev. C}. {\bf 96}\penalty0 (2), \penalty0 024908
   (2017).
\newblock \doi{10.1103/PhysRevC.96.024908}.

\bibitem{Pacik:2018gix}
V.~Pac\'\i{}k, {Elliptic flow of identified hadrons in small collisional
  systems measured with ALICE}, \emph{Nucl. Phys. A}. {\bf 982}, \penalty0
  451--454  (2019).
\newblock \doi{10.1016/j.nuclphysa.2018.09.020}.

\bibitem{ALICE:2016fzo}
J.~Adam et~al., {Enhanced production of multi-strange hadrons in
  high-multiplicity proton-proton collisions}, \emph{Nature Phys.} {\bf 13},
  \penalty0 535--539  (2017).
\newblock \doi{10.1038/nphys4111}.

\bibitem{ALICE:2013snk}
B.~B. Abelev et~al., {Long-range angular correlations of $\rm \pi$, K and p in
  p-Pb collisions at $\sqrt{s_{\rm NN}}$ = 5.02 TeV}, \emph{Phys. Lett. B}.
  {\bf 726}, \penalty0 164--177  (2013).
\newblock \doi{10.1016/j.physletb.2013.08.024}.

\bibitem{Zhao:2020wcd}
W.~Zhao, C.~M. Ko, Y.-X. Liu, G.-Y. Qin, and H.~Song, {Probing the Partonic
  Degrees of Freedom in High-Multiplicity $p-Pb$ collisions at $\sqrt {s_{NN}}$
  = 5.02 TeV}, \emph{Phys. Rev. Lett.} {\bf 125}\penalty0 (7), \penalty0 072301
   (2020).
\newblock \doi{10.1103/PhysRevLett.125.072301}.

\bibitem{Greco:2003xt}
V.~Greco, C.~M. Ko, and P.~Levai, {Parton coalescence and anti-proton / pion
  anomaly at RHIC}, \emph{Phys. Rev. Lett.} {\bf 90}, \penalty0 202302  (2003).
\newblock \doi{10.1103/PhysRevLett.90.202302}.

\bibitem{Han:2016uhh}
K.~C. Han, R.~J. Fries, and C.~M. Ko, {Jet Fragmentation via Recombination of
  Parton Showers}, \emph{Phys. Rev. C}. {\bf 93}\penalty0 (4), \penalty0 045207
   (2016).
\newblock \doi{10.1103/PhysRevC.93.045207}.

\bibitem{Wang:2013cia}
X.-N. Wang and Y.~Zhu, {Medium Modification of $\gamma$-jets in High-energy
  Heavy-ion Collisions}, \emph{Phys. Rev. Lett.} {\bf 111}\penalty0 (6),
  \penalty0 062301  (2013).
\newblock \doi{10.1103/PhysRevLett.111.062301}.

\bibitem{He:2015pra}
Y.~He, T.~Luo, X.-N. Wang, and Y.~Zhu, {Linear Boltzmann Transport for Jet
  Propagation in the Quark-Gluon Plasma: Elastic Processes and Medium Recoil},
  \emph{Phys. Rev. C}. {\bf 91}, \penalty0 054908  (2015).
\newblock \doi{10.1103/PhysRevC.91.054908}.
\newblock [Erratum: Phys.Rev.C 97, 019902 (2018)].

\bibitem{Cao:2017hhk}
S.~Cao, T.~Luo, G.-Y. Qin, and X.-N. Wang, {Heavy and light flavor jet
  quenching at RHIC and LHC energies}, \emph{Phys. Lett. B}. {\bf 777},
  \penalty0 255--259  (2018).
\newblock \doi{10.1016/j.physletb.2017.12.023}.

\bibitem{Sjostrand:2007gs}
T.~Sjostrand, S.~Mrenna, and P.~Z. Skands, {A Brief Introduction to PYTHIA
  8.1}, \emph{Comput. Phys. Commun.} {\bf 178}, \penalty0 852--867  (2008).
\newblock \doi{10.1016/j.cpc.2008.01.036}.

\bibitem{ALICE:2014jbq}
B.~B. Abelev et~al., {$K^*(892)^0$ and $\phi(1020)$ production in Pb-Pb
  collisions at $\sqrt{s{NN}}$ = 2.76 TeV}, \emph{Phys. Rev. C}. {\bf 91},
  \penalty0 024609  (2015).
\newblock \doi{10.1103/PhysRevC.91.024609}.

\bibitem{ALICE:2016sak}
J.~Adam et~al., {Production of K$^{*}$ (892)$^{0}$ and $\phi $ (1020) in
  p\textendash{}Pb collisions at $\sqrt{s_{{\text {NN}}}}$ = 5.02 TeV},
  \emph{Eur. Phys. J. C}. {\bf 76}\penalty0 (5), \penalty0 245  (2016).
\newblock \doi{10.1140/epjc/s10052-016-4088-7}.

\bibitem{ALICE:2018pal}
S.~Acharya et~al., {Multiplicity dependence of light-flavor hadron production
  in pp collisions at $\sqrt{s}$ = 7 TeV}, \emph{Phys. Rev. C}. {\bf
  99}\penalty0 (2), \penalty0 024906  (2019).
\newblock \doi{10.1103/PhysRevC.99.024906}.

\bibitem{ALICE:2013mez}
B.~Abelev et~al., {Centrality dependence of $\pi$, K, p production in Pb-Pb
  collisions at $\sqrt{s_{NN}}$ = 2.76 TeV}, \emph{Phys. Rev. C}. {\bf 88},
  \penalty0 044910  (2013).
\newblock \doi{10.1103/PhysRevC.88.044910}.

\bibitem{ALICE:2013wgn}
B.~B. Abelev et~al., {Multiplicity Dependence of Pion, Kaon, Proton and Lambda
  Production in p-Pb Collisions at $\sqrt{s_{NN}}$ = 5.02 TeV}, \emph{Phys.
  Lett. B}. {\bf 728}, \penalty0 25--38  (2014).
\newblock \doi{10.1016/j.physletb.2013.11.020}.

\bibitem{ALICE:2015mpp}
J.~Adam et~al., {Multi-strange baryon production in p-Pb collisions at
  $\sqrt{s_\mathbf{NN}}=5.02$ TeV}, \emph{Phys. Lett. B}. {\bf 758}, \penalty0
  389--401  (2016).
\newblock \doi{10.1016/j.physletb.2016.05.027}.

\bibitem{Bozek:2016yoj}
P.~Bozek, {Transverse-momentum\textendash{}flow correlations in relativistic
  heavy-ion collisions}, \emph{Phys. Rev. C}. {\bf 93}\penalty0 (4), \penalty0
  044908  (2016).
\newblock \doi{10.1103/PhysRevC.93.044908}.

\bibitem{Schenke:2020uqq}
B.~Schenke, C.~Shen, and D.~Teaney, {Transverse momentum fluctuations and their
  correlation with elliptic flow in nuclear collision}, \emph{Phys. Rev. C}.
  {\bf 102}\penalty0 (3), \penalty0 034905  (2020).
\newblock \doi{10.1103/PhysRevC.102.034905}.

\bibitem{Giacalone:2021clp}
G.~Giacalone, B.~Schenke, and C.~Shen, {Constraining the Nucleon Size with
  Relativistic Nuclear Collisions}, \emph{Phys. Rev. Lett.} {\bf 128}\penalty0
  (4), \penalty0 042301  (2022).
\newblock \doi{10.1103/PhysRevLett.128.042301}.

\bibitem{Gardim:2011xv}
F.~G. Gardim, F.~Grassi, M.~Luzum, and J.-Y. Ollitrault, {Mapping the
  hydrodynamic response to the initial geometry in heavy-ion collisions},
  \emph{Phys. Rev. C}. {\bf 85}, \penalty0 024908  (2012).
\newblock \doi{10.1103/PhysRevC.85.024908}.

\bibitem{Gardim:2014tya}
F.~G. Gardim, J.~Noronha-Hostler, M.~Luzum, and F.~Grassi, {Effects of
  viscosity on the mapping of initial to final state in heavy ion collisions},
  \emph{Phys. Rev. C}. {\bf 91}\penalty0 (3), \penalty0 034902  (2015).
\newblock \doi{10.1103/PhysRevC.91.034902}.

\bibitem{Betz:2016ayq}
B.~Betz, M.~Gyulassy, M.~Luzum, J.~Noronha, J.~Noronha-Hostler, I.~Portillo,
  and C.~Ratti, {Cumulants and nonlinear response of high $p_T$ harmonic flow
  at $\sqrt{s_{NN}}=5.02$ TeV}, \emph{Phys. Rev. C}. {\bf 95}\penalty0 (4),
  \penalty0 044901  (2017).
\newblock \doi{10.1103/PhysRevC.95.044901}.

\bibitem{Kovner:2010xk}
A.~Kovner and M.~Lublinsky, {Angular Correlations in Gluon Production at High
  Energy}, \emph{Phys. Rev. D}. {\bf 83}, \penalty0 034017  (2011).
\newblock \doi{10.1103/PhysRevD.83.034017}.

\bibitem{Kovner:2011pe}
A.~Kovner and M.~Lublinsky, {On Angular Correlations and High Energy
  Evolution}, \emph{Phys. Rev. D}. {\bf 84}, \penalty0 094011  (2011).
\newblock \doi{10.1103/PhysRevD.84.094011}.

\bibitem{Dumitru:2014dra}
A.~Dumitru and A.~V. Giannini, {Initial state angular asymmetries in high
  energy p+A collisions: spontaneous breaking of rotational symmetry by a color
  electric field and C-odd fluctuations}, \emph{Nucl. Phys. A}. {\bf 933},
  \penalty0 212--228  (2015).
\newblock \doi{10.1016/j.nuclphysa.2014.10.037}.

\bibitem{Dumitru:2014vka}
A.~Dumitru and V.~Skokov, {Anisotropy of the semiclassical gluon field of a
  large nucleus at high energy}, \emph{Phys. Rev. D}. {\bf 91}\penalty0 (7),
  \penalty0 074006  (2015).
\newblock \doi{10.1103/PhysRevD.91.074006}.

\bibitem{ATLAS-CONF-2021-001}
{Measurement of flow and transverse momentum correlations in Pb+Pb collisions
  at $\sqrt{s_{\mathrm{NN}}}=5.02$ TeV and Xe+Xe collisions at
  $\sqrt{s_{\mathrm{NN}}}=5.44$ TeV with the ATLAS detector}.
\newblock Technical report, CERN, Geneva  (2021).
\newblock URL \url{https://cds.cern.ch/record/2748818}.

\bibitem{ALICE:2021gxt}
S.~Acharya et~al., {Characterizing the initial conditions of heavy-ion
  collisions at the LHC with mean transverse momentum and anisotropic flow
  correlations}, \emph{Phys. Lett. B}. {\bf 834}, \penalty0 137393  (2022).
\newblock \doi{10.1016/j.physletb.2022.137393}.

\bibitem{STAR:2022pfn}
M.~I. Abdulhamid et~al., {Measurements of the Elliptic and Triangular Azimuthal
  Anisotropies in Central He3+Au, d+Au and p+Au Collisions at sNN=200\,\,GeV},
  \emph{Phys. Rev. Lett.} {\bf 130}\penalty0 (24), \penalty0 242301  (2023).
\newblock \doi{10.1103/PhysRevLett.130.242301}.

\bibitem{ALICE:2023gyf}
S.~Acharya et~al., {Measurements of long-range two-particle correlation over a
  wide pseudorapidity range in p$-$Pb collisions at $\sqrt{s_{\rm NN}}=5.0$
  TeV}  (8, 2023).

\bibitem{Ryu:2023bmx}
S.~Ryu, B.~Schenke, C.~Shen, and W.~Zhao.
\newblock {The role of longitudinal decorrelations for measurements of
  anisotropic flow in small collision systems} (12.
\newblock , 2023).

\bibitem{STAR:2023wmd}
{Measurement of flow coefficients in high-multiplicity $p$+Au, $d$+Au and
  $^{3}$He$+$Au collisions at $\sqrt{s_{_{\mathrm{NN}}}}$=200 GeV}  (12, 2023).

\bibitem{Shen:2016odt}
C.~Shen, {Electromagnetic Radiation from QCD Matter: Theory Overview},
  \emph{Nucl. Phys. A}. {\bf 956}, \penalty0 184--191  (2016).
\newblock \doi{10.1016/j.nuclphysa.2016.02.033}.

\bibitem{David:2019wpt}
G.~David, {Direct real photons in relativistic heavy ion collisions},
  \emph{Rept. Prog. Phys.} {\bf 83}\penalty0 (4), \penalty0 046301  (2020).
\newblock \doi{10.1088/1361-6633/ab6f57}.

\bibitem{Geurts:2022xmk}
F.~Geurts and R.-A. Tripolt, {Electromagnetic probes: Theory and experiment},
  \emph{Prog. Part. Nucl. Phys.} {\bf 128}, \penalty0 104004  (2023).
\newblock \doi{10.1016/j.ppnp.2022.104004}.

\bibitem{Shen:2016egw}
C.~Shen, C.~Park, J.-F. Paquet, G.~S. Denicol, S.~Jeon, and C.~Gale, {Direct
  photon production and jet energy-loss in small systems}, \emph{Nucl. Phys.
  A}. {\bf 956}, \penalty0 741--744  (2016).
\newblock \doi{10.1016/j.nuclphysa.2016.02.016}.

\bibitem{Shen:2016mmv}
C.~Shen, J.-F. Paquet, G.~S. Denicol, S.~Jeon, and C.~Gale, {Electromagnetic
  radiation and collectivity in small quark\textendash{}gluon droplets},
  \emph{Nucl. Part. Phys. Proc.} {\bf 289-290}, \penalty0 161--164  (2017).
\newblock \doi{10.1016/j.nuclphysbps.2017.05.034}.

\bibitem{Esha:2023ooh}
R.~Esha, {Thermal photon measurements at PHENIX}  (9, 2023).

\bibitem{Shen:2023aeg}
C.~Shen, A.~Noble, J.-F. Paquet, B.~Schenke, and C.~Gale.
\newblock {Illuminating early-stage dynamics of heavy-ion collisions through
  photons at RHIC BES energies}.
\newblock In \emph{{11th International Conference on Hard and Electromagnetic
  Probes of High-Energy Nuclear Collisions}: {Hard Probes 2023}}  (7, 2023).

\bibitem{ATLAS:2014qaj}
G.~Aad et~al., {Measurement of long-range pseudorapidity correlations and
  azimuthal harmonics in $\sqrt{s_{NN}}=5.02$ TeV proton-lead collisions with
  the ATLAS detector}, \emph{Phys. Rev. C}. {\bf 90}\penalty0 (4), \penalty0
  044906  (2014).
\newblock \doi{10.1103/PhysRevC.90.044906}.

\bibitem{ATLAS:2019vcm}
G.~Aad et~al., {Transverse momentum and process dependent azimuthal
  anisotropies in $\sqrt{s_{\mathrm{NN}}}=8.16$ TeV $p$+Pb collisions with the
  ATLAS detector}, \emph{Eur. Phys. J. C}. {\bf 80}\penalty0 (1), \penalty0 73
  (2020).
\newblock \doi{10.1140/epjc/s10052-020-7624-4}.

\bibitem{PHENIX:2015fgy}
A.~Adare et~al., {Centrality-dependent modification of jet-production rates in
  deuteron-gold collisions at $\sqrt{s_{NN}}$=200 GeV}, \emph{Phys. Rev. Lett.}
  {\bf 116}\penalty0 (12), \penalty0 122301  (2016).
\newblock \doi{10.1103/PhysRevLett.116.122301}.

\bibitem{Park:2016jap}
C.~Park, C.~Shen, S.~Jeon, and C.~Gale, {Rapidity-dependent jet energy loss in
  small systems with finite-size effects and running coupling}, \emph{Nucl.
  Part. Phys. Proc.} {\bf 289-290}, \penalty0 289--292  (2017).
\newblock \doi{10.1016/j.nuclphysbps.2017.05.066}.

\bibitem{JETSCAPE:2023xbc}
A.~Majumder et~al.
\newblock {A multistage framework for studying the evolution of jets and
  high-$p_T$ probes in small collision systems}.
\newblock In \emph{{11th International Conference on Hard and Electromagnetic
  Probes of High-Energy Nuclear Collisions}: {Hard Probes 2023}}  (8, 2023).

\bibitem{Soudi:2023epi}
I.~Soudi and A.~Majumder, {Azimuthal Anisotropy at high transverse momentum in
  $p$-$p$ and $p$-$A$ collisions}  (8, 2023).

\bibitem{ATLAS:2021jhn}
G.~Aad et~al., {Two-particle azimuthal correlations in photonuclear
  ultraperipheral Pb+Pb collisions at 5.02 TeV with ATLAS}, \emph{Phys. Rev.
  C}. {\bf 104}\penalty0 (1), \penalty0 014903  (2021).
\newblock \doi{10.1103/PhysRevC.104.014903}.

\bibitem{Bender:2003jk}
M.~Bender, P.-H. Heenen, and P.-G. Reinhard, {Self-consistent mean-field models
  for nuclear structure}, \emph{Rev. Mod. Phys.} {\bf 75}, \penalty0 121--180
  (2003).
\newblock \doi{10.1103/RevModPhys.75.121}.

\bibitem{Niel:2023zij}
E.~M. Niel, {SMOG at LHCb: experimental results}, \emph{PoS}. {\bf LHCP2022},
  \penalty0 150  (2023).
\newblock \doi{10.22323/1.422.0150}.

\bibitem{Giacalone:2023hwk}
G.~Giacalone, {Many-body correlations for nuclear physics across scales: from
  nuclei to quark-gluon plasmas to hadron distributions}, \emph{Eur. Phys. J.
  A}. {\bf 59}\penalty0 (12), \penalty0 297  (2023).
\newblock \doi{10.1140/epja/s10050-023-01200-7}.

\bibitem{Ryssens:2023fkv}
W.~Ryssens, G.~Giacalone, B.~Schenke, and C.~Shen, {Evidence of Hexadecapole
  Deformation in Uranium-238 at the Relativistic Heavy Ion Collider},
  \emph{Phys. Rev. Lett.} {\bf 130}\penalty0 (21), \penalty0 212302  (2023).
\newblock \doi{10.1103/PhysRevLett.130.212302}.

\bibitem{Gelis:2010nm}
F.~Gelis, E.~Iancu, J.~Jalilian-Marian, and R.~Venugopalan, {The Color Glass
  Condensate}, \emph{Ann. Rev. Nucl. Part. Sci.} {\bf 60}, \penalty0 463--489
  (2010).
\newblock \doi{10.1146/annurev.nucl.010909.083629}.

\bibitem{Krasnitz:2002ng}
A.~Krasnitz, Y.~Nara, and R.~Venugopalan, {Elliptic flow of colored glass in
  high-energy heavy ion collisions}, \emph{Phys. Lett. B}. {\bf 554}, \penalty0
  21--27  (2003).
\newblock \doi{10.1016/S0370-2693(02)03272-0}.

\bibitem{Gelis:2008ad}
F.~Gelis, T.~Lappi, and R.~Venugopalan, {High energy factorization in
  nucleus-nucleus collisions. II. Multigluon correlations}, \emph{Phys. Rev.
  D}. {\bf 78}, \penalty0 054020  (2008).
\newblock \doi{10.1103/PhysRevD.78.054020}.

\bibitem{Dumitru:2008wn}
A.~Dumitru, F.~Gelis, L.~McLerran, and R.~Venugopalan, {Glasma flux tubes and
  the near side ridge phenomenon at RHIC}, \emph{Nucl. Phys. A}. {\bf 810},
  \penalty0 91--108  (2008).
\newblock \doi{10.1016/j.nuclphysa.2008.06.012}.

\bibitem{Dusling:2018hsg}
K.~Dusling, M.~Mace, and R.~Venugopalan, {What does the matter created in high
  multiplicity proton-nucleus collisions teach us about the 3-D structure of
  the proton?}, \emph{PoS}. {\bf QCDEV2017}, \penalty0 039  (2018).
\newblock \doi{10.22323/1.308.0039}.

\bibitem{Greif:2020rhi}
M.~Greif, C.~Greiner, S.~Pl\"atzer, B.~Schenke, and S.~Schlichting,
  {Hadronization of correlated gluon fields}, \emph{Phys. Rev. D}. {\bf
  103}\penalty0 (5), \penalty0 054011  (2021).
\newblock \doi{10.1103/PhysRevD.103.054011}.

\bibitem{Bertulani:2005ru}
C.~A. Bertulani, S.~R. Klein, and J.~Nystrand, {Physics of ultra-peripheral
  nuclear collisions}, \emph{Ann. Rev. Nucl. Part. Sci.} {\bf 55}, \penalty0
  271--310  (2005).
\newblock \doi{10.1146/annurev.nucl.55.090704.151526}.

\bibitem{Baltz:2007kq}
A.~J. Baltz, {The Physics of Ultraperipheral Collisions at the LHC},
  \emph{Phys. Rept.} {\bf 458}, \penalty0 1--171  (2008).
\newblock \doi{10.1016/j.physrep.2007.12.001}.

\bibitem{Shi:2020djm}
Y.~Shi, L.~Wang, S.-Y. Wei, B.-W. Xiao, and L.~Zheng, {Exploring collective
  phenomena at the electron-ion collider}, \emph{Phys. Rev. D}. {\bf
  103}\penalty0 (5), \penalty0 054017  (2021).
\newblock \doi{10.1103/PhysRevD.103.054017}.

\bibitem{CMS:2023iam}
A.~Hayrapetyan et~al., {Observation of enhanced long-range elliptic
  anisotropies inside high-multiplicity jets in pp collisions at $\sqrt{s}$ =
  13 TeV}  (12, 2023).

\end{thebibliography}

\end{document}